\documentclass[fleqn,usenatbib]{aa}
\usepackage{txfonts}
\usepackage{graphicx}
\usepackage{natbib}
\usepackage{longtable}
\usepackage{subeqnarray}
\usepackage{cases}
\usepackage[colorlinks=true,citecolor=blue]{hyperref}
\usepackage{amsmath}
\usepackage[switch]{lineno}
\let\oldequation\equation 
\let\oldendequation\endequation 
\renewenvironment{equation}{\linenomathNonumbers\oldequation}{\oldendequation\endlinenomath} 

\begin{document}
\title{Molecular clouds at the edge of the Galaxy}
\subtitle{III. The first Galactic quadrant}
\author{C. S. Luo\inst{\ref{inst1},\ref{inst2}}
\and X. D. Tang\inst{\ref{inst1},\ref{inst2},\ref{inst3}}
\thanks{Corresponding author: tangxindi@xao.ac.cn}
\and C. Henkel\inst{\ref{inst4},\ref{inst1}}
\and Y. Sun\inst{\ref{inst5}}
\and Y. Gong\inst{\ref{inst5}}
\and X. W. Zheng\inst{\ref{inst6}}
\and T. Liu\inst{\ref{inst7}}
\and X. Lu\inst{\ref{inst7}}
\and Y. P. Ao\inst{\ref{inst5}}
\and X. P. Chen\inst{\ref{inst5}}
\and J. Esimbek\inst{\ref{inst1},\ref{inst2},\ref{inst3}}
\and D. L. Li\inst{\ref{inst1},\ref{inst2},\ref{inst3}}
\and Y. X. He\inst{\ref{inst1},\ref{inst2},\ref{inst3}}
\and K. Wang\inst{\ref{inst8}}
\and J. W. Wu\inst{\ref{inst9},\ref{inst2}}
\and J. J. Zhou\inst{\ref{inst1},\ref{inst2},\ref{inst3}}
\and G. Wu\inst{\ref{inst1},\ref{inst2},\ref{inst3}}
\and Y. X. Ma\inst{\ref{inst1},\ref{inst3}}
\and A. Sobolev\inst{\ref{inst1},\ref{inst3}}
\and W. A. Baan\inst{\ref{inst1},\ref{inst3}}
\and K. Tursun\inst{\ref{inst1},\ref{inst3}}
\and D. D. Zhou\inst{\ref{inst1},\ref{inst3}}
\and J. J. Qiu\inst{\ref{inst10}}
\and X. Zhao\inst{\ref{inst1},\ref{inst2}}
\and J. S. Li\inst{\ref{inst1},\ref{inst2}}
\and Q. Zhao\inst{\ref{inst1},\ref{inst2}}
\and L. D. Liu\inst{\ref{inst1},\ref{inst2}}
\and C. Y. Wang\inst{\ref{inst1},\ref{inst2}}
\and D. Ladeyschikov\inst{\ref{inst11},\ref{inst1}}
\and W. G. Ji\inst{\ref{inst1},\ref{inst3}}
\and T. Komesh\inst{\ref{inst1},\ref{inst12},\ref{inst13}}}

\titlerunning{Galactic edge clouds in the first Galactic quadrant}
\authorrunning{Luo et al.}

\institute{State Key Laboratory of Radio Astronomy and Technology, Xinjiang Astronomical Observatory, CAS, Urumqi 830011, PR China\label{inst1}
\and University of Chinese Academy of Sciences, Beijing 100080, PR China \label{inst2}
\and Xinjiang Key Laboratory of Radio Astrophysics, Urumqi 830011, PR China \label{inst3}
\and Max-Planck-Institut f\"{u}r Radioastronomie, Auf dem H\"{u}gel 69, 53121 Bonn, Germany \label{inst4}
\and Purple Mountain Observatory, Chinese Academy of Sciences, Nanjing 210008, PR China \label{inst5}
\and School of Astronomy and Space Science, Nanjing University, Nanjing 210093, PR China \label{inst6}
\and Shanghai Astronomical Observatory, Chinese Academy of Sciences, 80 Nandan Road, Shanghai 200030, PR China \label{inst7}
\and Kavli Institute for Astronomy and Astrophysics, Peking University, Beijing 100871, PR China \label{inst8}
\and National Astronomical Observatories, Chinese Academy of Sciences, Beijing 100101, PR China \label{inst9}
\and School of Mathematics and Physics, Jinggangshan University, Ji’an 343009, PR China \label{inst10}
\and Ural Federal University, 51 Lenin Str., 620051 Ekaterinburg, Russia \label{inst11}
\and Energetic Cosmos Laboratory, Nazarbayev University, Astana 010000, Kazakhstan \label{inst12}
\and Institute of Experimental and Theoretical Physics, Al-Farabi Kazakh National University, Almaty 050040, Kazakhstan \label{inst13}}

\date{Accepted 27/11/2026}
\abstract{Molecular clouds at the Galactic edge differ markedly from those in the solar neighborhood, being embedded in low-metallicity environments and exhibiting an enhanced fraction of CO-dark H$_2$ gas. 
In this work, we analyze 84 molecular clouds in the Outer Scutum-Centaurus arm in the first Galactic quadrant using $^{12}$CO\,$J$\,=\,1--0 and 3--2 data. 
Among the 69 Galactic edge clouds exhibiting significant CO\,(3--2) emission after smoothing to 50$^{\prime\prime}$, the integrated intensity ratio of CO\,$J$\,=\,3--2/1--0 ($R_{31}$) spans a range of 0.28 to 2.02, with a median value of 0.74\,$\pm$\,0.13. 
23 out of the 69 Galactic edge clouds exhibit $R_{31}$ ratios exceeding unity, suggesting a component of highly excited molecular gas within the Galactic edge. 
No significant dependence of the $R_{31}$ ratio variation on Galactocentric distance is found over $R_{\rm g}$\,$\sim$\,10--15\,kpc. 
A total of 110 CO\,(3--2) clumps are identified at $\sim$17$^{\prime\prime}$ resolution, and their physical properties are systematically evaluated. 
The derived clump properties span sizes of 0.6--5.9\,pc, velocity dispersions of 0.5--5.4\,km\,s$^{-1}$, masses of (0.5--213)\,$\times$\,$10^2$\,M$_\odot$, and surface densities of 7--666\,M$_{\odot}$\,pc$^{-2}$. 
These parameters show no systematic variation with Galactocentric distance over 10--15\,kpc, except for the virial parameter, which decreases exponentially as $\alpha_{\rm vir}$\,=\,28.1($\pm$\,16.2)e$^{-R_{\rm g}/8.4(\pm3.5)}$. 
Scaling relations reveal a velocity dispersion--size relation of $\sigma_{\rm v}$\,=\,0.98($\pm$0.06)$R_{\rm eff}^{0.35(\pm0.08)}$, indicating a suppressed growth of velocity dispersion at clump scales. 
The mass--size relation, $M_{\rm lum}$\,=\,165($\pm$22)$R_{\rm eff}^{1.60(\pm0.18)}$, is shallower than the constant--surface--density expectation. 
Virial parameters range from 1.6 to 31.7, with a median of 6.8\,$\pm$\,2.4, implying that the majority of the clumps are gravitationally unbound. 
The properties of the first-quadrant clumps, including their masses, virial states, radial trends, velocity dispersion--size relation, and highly excited molecular gas traced by CO\,(3--2), are broadly consistent with previous findings for the second Galactic quadrant at Galactocentric distances of $R_{\rm g}$\,$\sim$\,14--23\,kpc, albeit with modest variations in typical properties. } 

\maketitle

\section{Introduction}
\label{sect:Introduction}
Molecular clouds, referred to as stellar nurseries, are pivotal in the star formation process. 
The physical properties of molecular clouds are dependent on the local environment (e.g., \citealt{Fukui2010,Bolatto2013,Colombo2014,Heyer2015,Tang2017a,Tang2019,Tang2021,Henkel2022}), yet studying cloud properties in external galaxies poses a significant challenge due to limitations in spatial resolution. 
The environment at the edge of the Galaxy exhibits similarities to both irregular and dwarf galaxies \citep{Ruffle2007,Kobayashi2008}, as well as to the interstellar medium found at high redshift \citep{Ferguson1998}. 
It is characterized by low metallicity \citep{Smartt1997}, low gas density \citep{Heyer2015}, high gas-to-dust ratios \citep{Giannetti2017}, faint cosmic ray fluxes \citep{Bhat1984}, weak interstellar radiation fields \citep{Mathis1983}, little perturbation from spiral arms \citep{Monari2016}, and an absence of kinematic distance ambiguity \citep{Roman-Duval2009}. 
Therefore, the Galactic edge serves as an ideal location to investigate molecular clouds and star formation in extreme conditions. 

The initial conditions of star formation are influenced by the physical properties of molecular clouds (e.g., \citealt{Lada2003,Tang2017b,Tang2018a,Tang2018b,Zhao2024,Zhao2026,Gong2025}). 
Previous studies have found that star formation activity is present in the Galactic edge clouds, revealing stellar clusters (e.g., \citealt{Snell2002,Yasui2006,Yasui2008,Kobayashi2008,Izumi2014}), \ion{H}{II} regions (e.g., \citealt{deGeus1993,Rudolph1996,Anderson2015}), maser lines (e.g., \citealt{Wouterloot1988,Armentrout2017,Sun2018a}), protostellar outflows (e.g., \citealp{Izumi2024,Ikeda2025}), and hot molecular cores (e.g., \citealp{Shimonishi2021}). 
In the Galactic edge clouds located in the second quadrant, there is a notable prevalence of molecular gas with a high CO\,$J$=2--1/1--0 line ratio \citep{Luo2025}. 
This molecular gas is typically associated with compact clouds and areas characterized by star-forming activity. 
Radiation from $^{13}$CO\,(1--0), HCN\,(1--0), and HCO$^+$\,(1--0) has been detected in several molecular clouds \citep{Sun2015a,Braine2023}. 
Furthermore, a systematic decreasing trend of the virial parameter with increasing Galactocentric radius is observed ($R_{\rm g}$\,$\sim$\,9--24\,kpc), indicating that turbulent support is inadequate in low-metallicity molecular clouds \citep{Sun2024a,Lin2025,Luo2026}. 
Nevertheless, most molecular clouds are likely maintained in dynamical equilibrium by external pressure confinement, despite large uncertainties in the CO to H$_2$ conversion factor \citep{Brand1995}. 
The Galactic edge clouds identified in the second Galactic quadrant conform to Larson's scaling relations, and their large virial parameters indicate that they are not strongly gravitationally bound when external pressure confinement is not taken into account \citep{Luo2026}. 
To date, the sample of confirmed molecular clouds at the outskirts of the Milky Way remains small, and more comprehensive studies of their physical and chemical characteristics are required. 

\begin{figure}[t]
\centering
\includegraphics[width=0.45\textwidth]{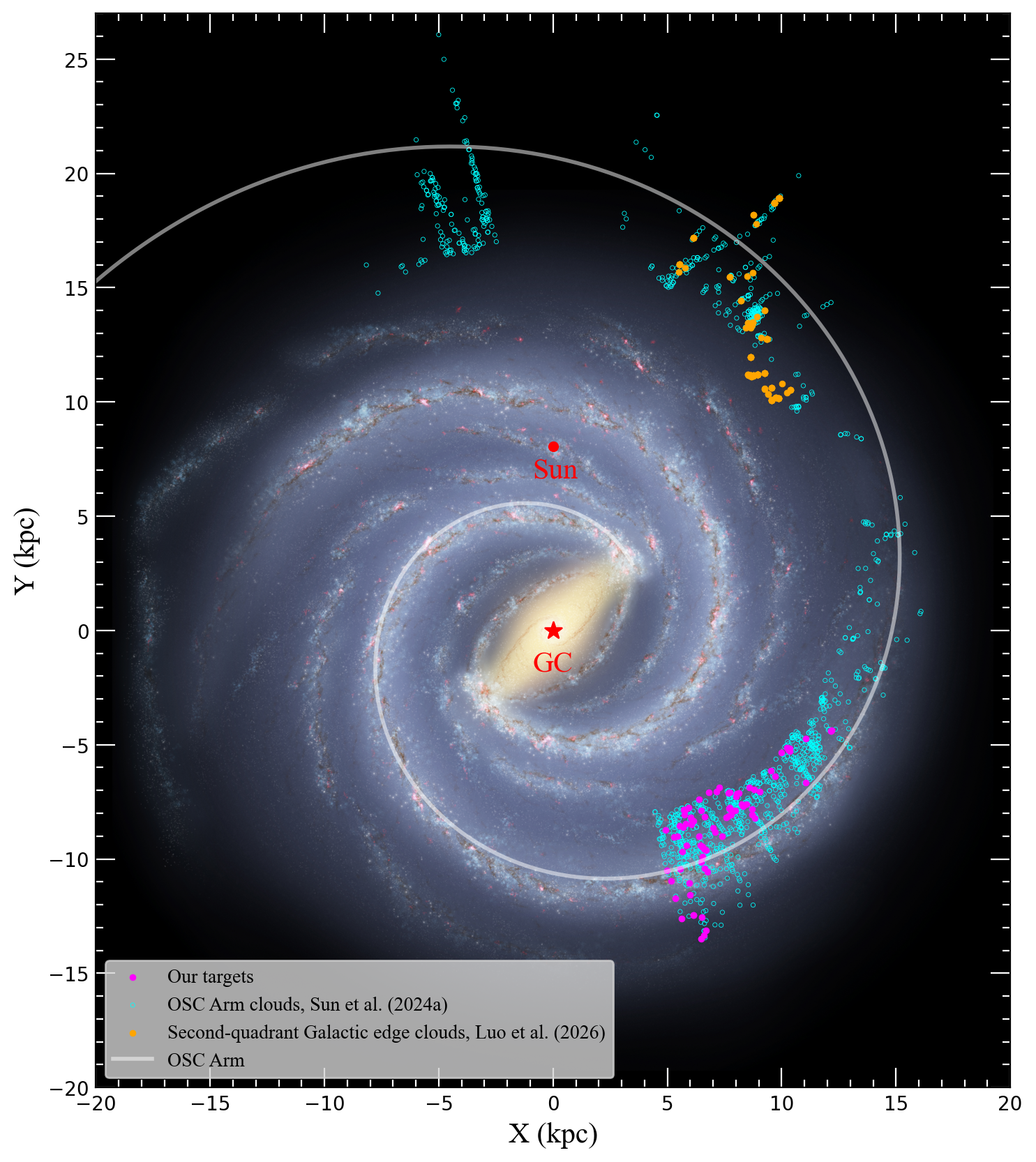}
\caption{Locations of 84 selected molecular clouds in the distant outer regions of the first Galactic quadrant (magenta circles). 
Orange circles, cyan circles, and white curve denote the Galactic edge clouds in the second Galactic quadrant \citep{Luo2026}, the OSC arm clouds identified by \citet{Sun2024b}, and the OSC arm, respectively. 
The background image is an artist's conception of the Milky Way, credited to R.Hurt: NASA/JPL-Caltech/SSC. }
\label{fig:target}
\end{figure}

Presently, an unbiased northern Galactic plane CO survey (10$^\circ$\,$\lesssim$\,$l$\,$\lesssim$\,230$^\circ$ and --5$^\circ$\,$\lesssim$\,$b$\,$\lesssim$\,5$^\circ$) is being conducted using the Delingha 13.7\,m telescope, referred to as the Milky Way Imaging Scroll Painting (MWISP) project\footnote{\tiny \url{http://www.radioast.nsdc.cn/english/mwisp.php}} \citep{Yang2026}. 
The MWISP survey identified a series of molecular clouds situated in the outer Galaxy (see Fig.\,\ref{fig:target}), with Galactocentric distances ranging from approximately 8 to 26\,kpc \citep{Sun2015a,Sun2017,Sun2024a,Sun2024b,Du2016,Su2016,Ma2021,Yan2021,Yuan2025}. 
It provides an ideal sample for studying molecular clouds and star formation at the outskirts of the Galaxy. 
Previous studies have focused on the second Galactic quadrant \citep{Luo2025,Luo2026}, and the inclusion of the first Galactic quadrant will provide a more comprehensive understanding of the properties of molecular clouds at the Galactic edge. 
Towards the Galactic edge in the first Galactic quadrant, observations have revealed the presence of the ammonia molecule \citep{Armentrout2017}, hydrogen radio recombination lines (RRL; \citealt{Anderson2015}), and maser emission from e.g., H$_2$O, CH$_3$OH, and OH \citep{Armentrout2017,Sun2018a}. 
Physical properties and scaling relations of first-quadrant Galactic edge clouds have been examined at a beam size of $\sim$\,50$^{\prime\prime}$ \citep{Sun2017}. 
Galactic edge clouds located in the first Galactic quadrant are more observationally challenging than those in the second quadrant due to their greater distances from the Sun. 
It is crucial to identify and analyze molecular clumps within first-quadrant Galactic edge clouds using appropriate resolution. 

Previous observations have shown a consistent relationship between the luminosity of $^{12}$C$^{16}$O\,$J$\,=\,3--2 (hereafter CO\,(3--2)) and the star formation rate (SFR) across various types of galaxies, ranging from quiescent spirals to intense starburst systems (e.g., \citealt{Iono2009, Wilson2012, Muraoka2016}). 
This correlation is evident at both global and resolved scales, supporting the hypothesis that CO\,(3--2) 
effectively traces the dense molecular gas directly involved in the process of star formation. In nearby galaxies, the emission of 
CO\,(3--2) aligns spatially with active star-forming regions, further supporting its utility as a proxy for the reservoir 
of molecular gas involved in star formation (e.g., \citealt{Yao2003,Komugi2007,Wilson2009,Leech2010,Mao2010,Ueda2012,Li2020a}). 
This paper aims to study the physical properties and scaling relations of the Galactic edge clouds in the first quadrant, as determined by CO\,(3--2) emission. 
Targets and data reduction are introduced in Sect.\,\ref{sect:Targets and data reduction}. 
Main results are presented in Sect.\,\ref{sect:Results}, while the CO\,$J$=3--2/1--0 line ratio ($R_{31}$) and dynamical states of the Galactic edge clouds are discussed in Sect.\,\ref{sect:discussion}. 
Our main conclusions are summarized in Sect.\,\ref{sect:summary}. 
This paper is part of the "Molecular Clouds at the Edge of the Galaxy" project, which focuses on the investigation of the physical and chemical properties of molecular clouds and star formation located in the outskirts of the Milky Way. 

\begin{figure*}[htb]
\centering
\includegraphics[width=1.0\textwidth]{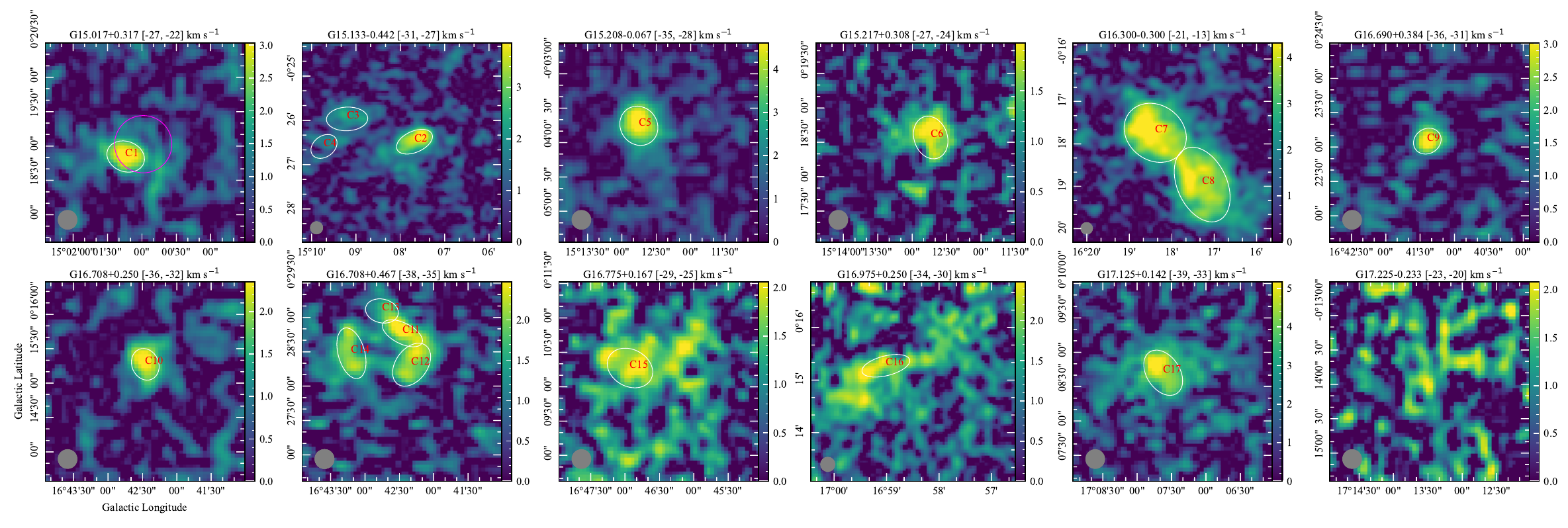}
\caption{CO\,(3--2) velocity-integrated intensity maps of 84 Galactic edge clouds. 
The color background shows $\int T_{\rm MB}dv$ with square-root scaling. 
The source names are located at the top of each panel, with a color bar in units of K\,km\,s$^{-1}$ situated on the right side of each panel. 
The source names correspond to the Table\,1 in \citet{Sun2024a}. 
White ellipses represent the fitted results of the identified CO clumps, with numbers ranging from 1 to 110, corresponding to the indices presented in Table\,\ref{table:clump parameters}. 
The magenta circle in the center of the first sub-figure illustrates the CO\,(1--0) emission peaks identified from the PMO\,13.7\,m telescope and its size corresponds to the telescope beam size $\sim$\,50$^{\prime\prime}$. 
The other maps are centered on the CO\,(1--0) emission peak in the same way. 
The gray filled circles in the lower-left corner of each map show the beam size of CO\,(3--2) observed with the JCMT\,15\,m telescope. 
Each map exhibits a distinct size. 
The velocity-integrated intensity maps for the remaining targets are presented in Fig.\,\ref{fig:map-continue}. 
} 
\label{fig:map} 
\end{figure*}

\section{Targets and data reduction}
\label{sect:Targets and data reduction}
\subsection{Targets}
\label{sect:Targets}
An extension of the Scutum-Centaurus Arm is found beyond the Outer Arm, known as the Outer Scutum-Centaurus (OSC) arm \citep{Dame2011,Sun2015a,Sun2024b}. 
A large number of CO molecular clouds were identified in this arm, despite its partial rise out of the Galactic plane due to the Galactic warp (see Fig.\,1, \citealt{Sun2024b}). 
Recently, the CO\,(3--2) High-Resolution Survey (COHRS) of the Galactic Plane was conducted using the JCMT\,15\,m (9.5$^\circ$\,$\le$\,$l$\,$\le$\,62.3$^\circ$ and --0.5$^\circ$\,$\lesssim$\,$b$\,$\lesssim$\,0.5$^\circ$; \citealt{Park2023}). 
The COHRS's data, combined with molecular cloud samples identified by the MWISP survey, offer a significant opportunity to study molecular clouds at the edge of the Galaxy. 
The COHRS survey encompasses the 313 OSC arm clouds identified by \citet{Sun2024b}, 84 of these clouds exhibit significant CO\,(3--2) emission at an angular resolution of 16.6$^{\prime\prime}$. 
Therefore, the targets of this work are comprised of these 84 OSC arm clouds, which span the Galactic longitude range of $15^\circ$\,<\,$l$\,<\,$50^\circ$ (refer to Fig.\,\ref{fig:target} and \citealt{Sun2024a}). 
The COHRS covers fewer molecular clouds as one moves in the direction where the Galactic longitude increases, due to the warped structure of the spiral arm in the far outer Galaxy. 
These Galactic edge clouds exhibit a wide range of Galactocentric radii, spanning approximately 10 to 15\,kpc, with a mean value of 12\,kpc \citep{Sun2024a}. 
Assuming that the $^{12}$CO\,(1--0) line is optically thick and that the beam-filling factors are near unity, the typical size, mass, velocity dispersion, volume density, and excitation temperature of these clouds are 6.4\,pc, 872\,M$_{\odot}$, 0.7\,km\,s$^{-1}$, 12\,cm$^{-3}$, and 5.9\,K \citep{Sun2024a}, respectively. 
The targets are listed in Table\,\ref{table:spectral parameters}. 

\subsection{Archival CO data} 
\label{sect:CO(3-2) data} 
$^{12}$CO\,(3--2) data were obtained as part of COHRS\footnote{\tiny \url{https://www.canfar.net/citation/landing?doi=22.0078}} \citep{Park2023}, conducted using the Heterodyne Array Receiver Program (HARP) on the JCMT\,15\,m telescope. 
The HARP, equipped with an array receiver comprising 4\,$\times$\,4 superconductor-insulator-superconductor heterodyne detectors spaced at intervals of 30$^{\prime\prime}$, possesses a beam size of 14$^{\prime\prime}$ and a main beam efficiency ($\eta_{\rm mb}$) of 0.61 at the frequency of 345.786\,GHz. 
The Auto-Correlation Spectral Imaging System (ACSIS) is configured with a 1\,GHz bandwidth and 2048 spectral channels, yielding a velocity coverage of --400\,km\,s$^{-1}$\,<\,$V_{\rm LSR}$\,<\,+400\,km\,s$^{-1}$ with a frequency resolution of 0.488\,MHz and a channel width of 0.42\,km\,s$^{-1}$ at 345.786\,GHz. 
The data released by COHRS encompasses an area of $\sim$\,52.8\,deg$^2$ (9.5$^\circ$\,$\le$\,$l$\,$\le$\,62.3$^\circ$ and --0.5$^\circ$\,$\le$\,$b$\,$\le$\,0.5$^\circ$), with a typical root-mean-square (RMS) noise level of 1.0\,K ($T_{\rm mb}$ scale) achieved by smoothing to an angular resolution of 16.6$^{\prime\prime}$ and a velocity resolution of 0.64\,km\,s$^{-1}$. 
The data were re-gridded into 6$^{\prime\prime}$\,$\times$\,6$^{\prime\prime}$ pixels, with a velocity range of --200\,km\,s$^{-1}$ to +300\,km\,s$^{-1}$ \citep{Park2023}. 
The main beam brightness temperature ($T_{\rm mb}$) of CO\,(3--2) is deduced using the formula $T_{\rm mb}$\,=\,$T_{\rm A}^{*}$/$\eta_{\rm mb}$, where $\eta_{\rm mb}$\,$\sim$\,0.61 is the main beam efficiency and $T_{\rm A}^{*}$ represents the antenna temperature. 

CO\,(1--0) data were obtained from the MWISP project \citep{Yang2026}, which surveyed an area of 2310\,deg$^2$ in the northern sky covering Galactic ranges of 10$^\circ$\,$\lesssim$\,$l$\,$\lesssim$\,230$^\circ$ and --5$^\circ$\,$\lesssim$\,$b$\,$\lesssim$\,5$^\circ$ with Local Standard of Rest (LSR) velocities from --200 to +300\,km s$^{-1}$. 
Spectral line data of the CO\,(1--0) line at a frequency of 115.271\,GHz have a beam size of approximately 50$^{\prime\prime}$, as observed by the Delingha 13.7\,m telescope.
All intensities were calibrated to the main beam brightness temperature ($T_{\rm mb}$) using the main beam efficiency ($\eta_{\rm mb}$\,$\sim$\,0.5), given by $T_{\rm mb}$\,=\,$T_{\rm A}^{*}$/$\eta_{\rm mb}$. 
The typical RMS noise level for the CO\,(1--0) data is 0.5\,K ($T_{\rm mb}$ scale) at a velocity resolution of $\sim$\,0.16\,km s$^{-1}$ \citep{Yang2026}. 

\begin{figure*}[t]
\centering
\includegraphics[width=0.95\textwidth]{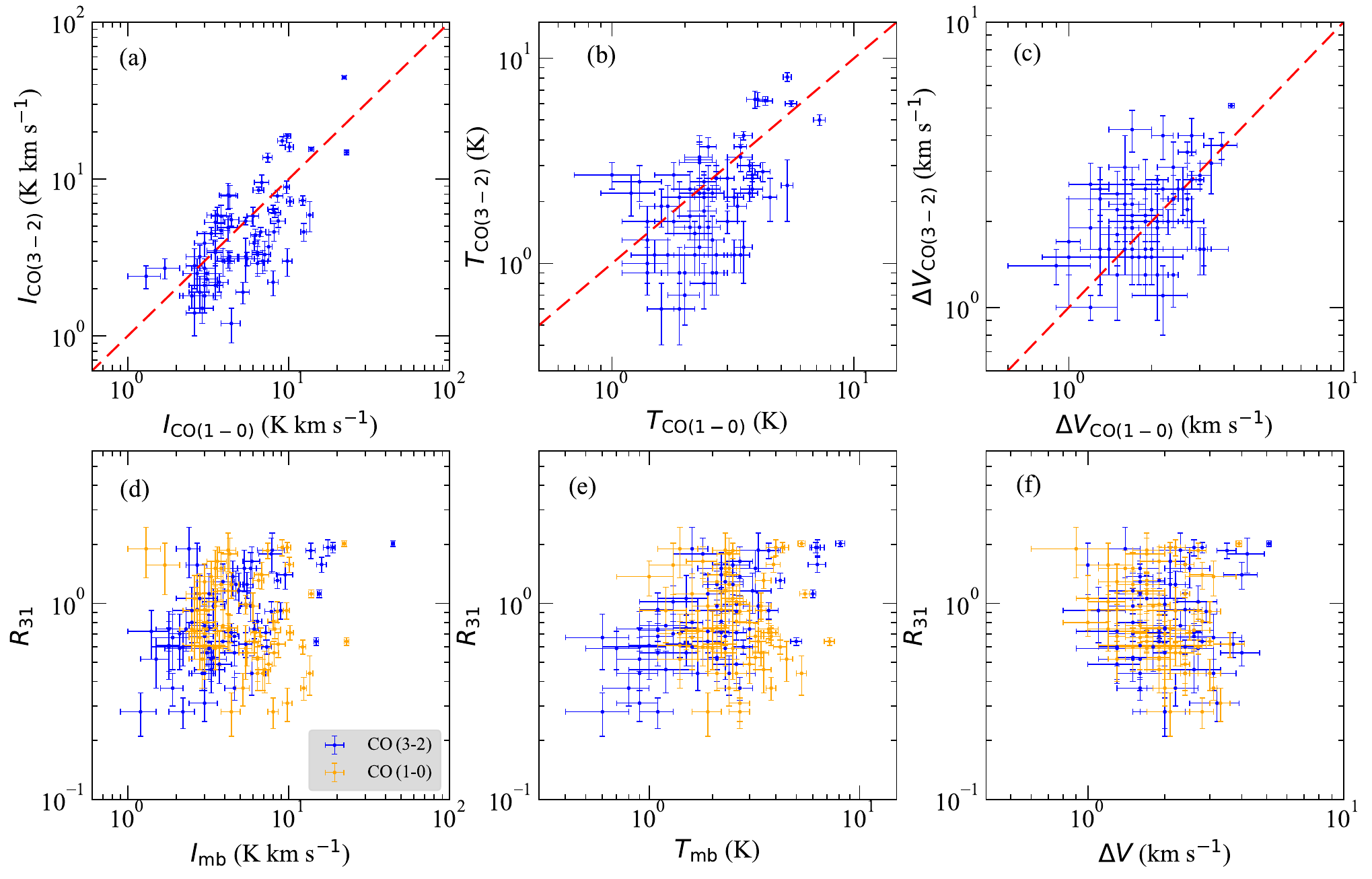}
\caption{Correlations of integrated intensities (a), peak temperatures (b), and line widths (c) of the CO\,(3--2) and CO\,(1--0) lines. 
The dashed red lines indicate Y\,=\,X. 
Correlations of $R_{31}$ with the CO integrated intensities (d), CO peak temperatures (e), and CO line widths (f) for the 69 Galactic edge clouds are detailed in Table\,\ref{table:spectral parameters}. } 
\label{fig:pra-r31} 
\end{figure*}

\subsection{Data reduction} 
\label{sect:data reduction} 
The processing of CO\,(3--2) data cubes was executed utilizing the GILDAS\footnote{\tiny \url{http://www.iram.fr/IRAMFR/GILDAS}} and {\tt Python}\footnote{\tiny \url{https://www.python.org/}} packages. 
As a first step, Flexible Image Transport System (FITS) cubes of 84 selected sources were cropped from COHRS. 
The cropping size, determined by the spatial extent covered by the CO\,(3--2) signal, displays a variation from 120$^{\prime\prime}$\,$\times$\,120$^{\prime\prime}$ to 360$^{\prime\prime}$\,$\times$\,360$^{\prime\prime}$. 
Furthermore, map-edge regions with large noise were systematically excluded. 
We generate RMS FITS files with a pixel scale of 6$^{\prime\prime}$\,$\times$\,6$^{\prime\prime}$ following baseline fitting. 

\section{Results}
\label{sect:Results}
\subsection{Overview}
\label{sect:Overview}
CO\,(3--2) integrated intensity maps of 84 Galactic edge clouds are depicted in Figs.\,\ref{fig:map} and \ref{fig:map-continue}. 
Approximately 28 CO\,(3--2) clouds exhibit compact and nearly round shapes, characterized by centrally concentrated emission peaks and relatively symmetric intensity distributions. 
Around 24 clouds show elongated morphologies, with emission preferentially extended along one direction. 
The remaining 32 clouds display irregular morphologies, multiple intensity peaks, or partially connected emission features, suggestive of complex internal substructures and heterogeneous gas distributions. 
These morphological characteristics are commonly observed in molecular clouds and may reflect different evolutionary stages or environmental effects. 
No significant morphological differences are apparent between the first- and second-quadrant clouds at the Galactic edge (see Fig.\,1 of \citealt{Luo2026}), suggesting they may share a similar gas structure despite their different locations with respect to the Galactic center. 

\subsection{$R_{31}$ line ratios}
\label{sect:R31 Line Ratios}
To ensure the derivation of reliable intensity ratios for CO\,(3--2) to CO\,(1--0), the CO\,(3--2) data cube was smoothed to an angular resolution of 50$^{\prime\prime}$ and regridded at 30$^{\prime\prime}$ intervals. 
Out of 84 CO\,(3--2) clouds by two-dimensional Gaussian smoothing\footnote{Although smoothing the data from 16.6$^{\prime\prime}$ to 50$^{\prime\prime}$ reduces the RMS noise level by a factor of $\sim$\,3, peak temperatures of marginally resolved clouds can decrease by a larger factor of $\sim$\,5, resulting in lower signal-to-noise (S/N) ratios.}, only 69 clouds displayed CO\,(3--2) emission due to beam dilution of clouds much less extended than 50$^{\prime\prime}$. 
For comparison, the CO\,(1--0) spectral lines were smoothed over four consecutive channels to achieve a velocity resolution of $\sim$\,0.64\,km\,s$^{-1}$, which closely aligns with the CO\,(3--2) velocity resolution. 
The CO\,(3--2) and (1--0) spectra of 69 Galactic edge clouds are depicted in Fig.\,\ref{fig:CO(3-2) spectra}, all smoothed to an angular resolution of 50$^{\prime\prime}$. 
The CO\,(3--2) spectra, which align with positions of the CO\,(1--0) peak emission, are fitted using Gaussian profiles. 
The fitting parameters include the local standard of rest velocity ($V_{\rm LSR}$), main beam brightness peak temperature ($T_{\rm mb}$), full width at half maximum line width ($\Delta V$), and velocity-integrated intensity ($\int$$T_{\rm mb}$d$v$). 
Similarly, the parameters of the CO\,(1--0) spectra are also fitted with a Gaussian profile. 
The fitted results are detailed in Table \ref{table:spectral parameters}. 
For CO\,(3--2), the line width, measured as the full width at half maximum (FWHM), ranges from 1.0 to 5.2\,km\,s$^{-1}$, with an average of 2.2\,$\pm$\,0.4\,km\,s$^{-1}$. 
This is comparable to that of the CO\,(1--0) line, which has a range of 0.9--3.9\,km\,s$^{-1}$ and an average of 2.0\,$\pm$\,0.3\,km\,s$^{-1}$. 
Both the CO\,(3--2) and CO\,(1--0) transitions display relatively narrow line widths, indicating quiescent molecular gas within the Galactic edge clouds. 
For the cloud with the broadest CO\,(3--2) line, the line widths of CO\,(3--2) and CO\,(1--0) are 5.1\,km\,s$^{-1}$ and 3.9\,km\,s$^{-1}$, respectively, while their peak brightness temperatures are 8.1\,K and 5.3\,K, respectively, yielding an $R_{31}$ ratio of 2.02. 
The broader CO\,(3--2) line may be preferentially caused by warmer and denser molecular gas with enhanced excitation. 
Nevertheless, correlations are positive in peak temperature, line width, and integrated intensity between CO\,(3--2) and (1--0) (see Fig.\,\ref{fig:pra-r31}). 
The observed agreement between the CO\,(3--2) and (1--0) lines suggests that these two transitions likely trace similar components of the molecular gas. 

\begin{figure}[t]
\centering
\includegraphics[width=0.45\textwidth]{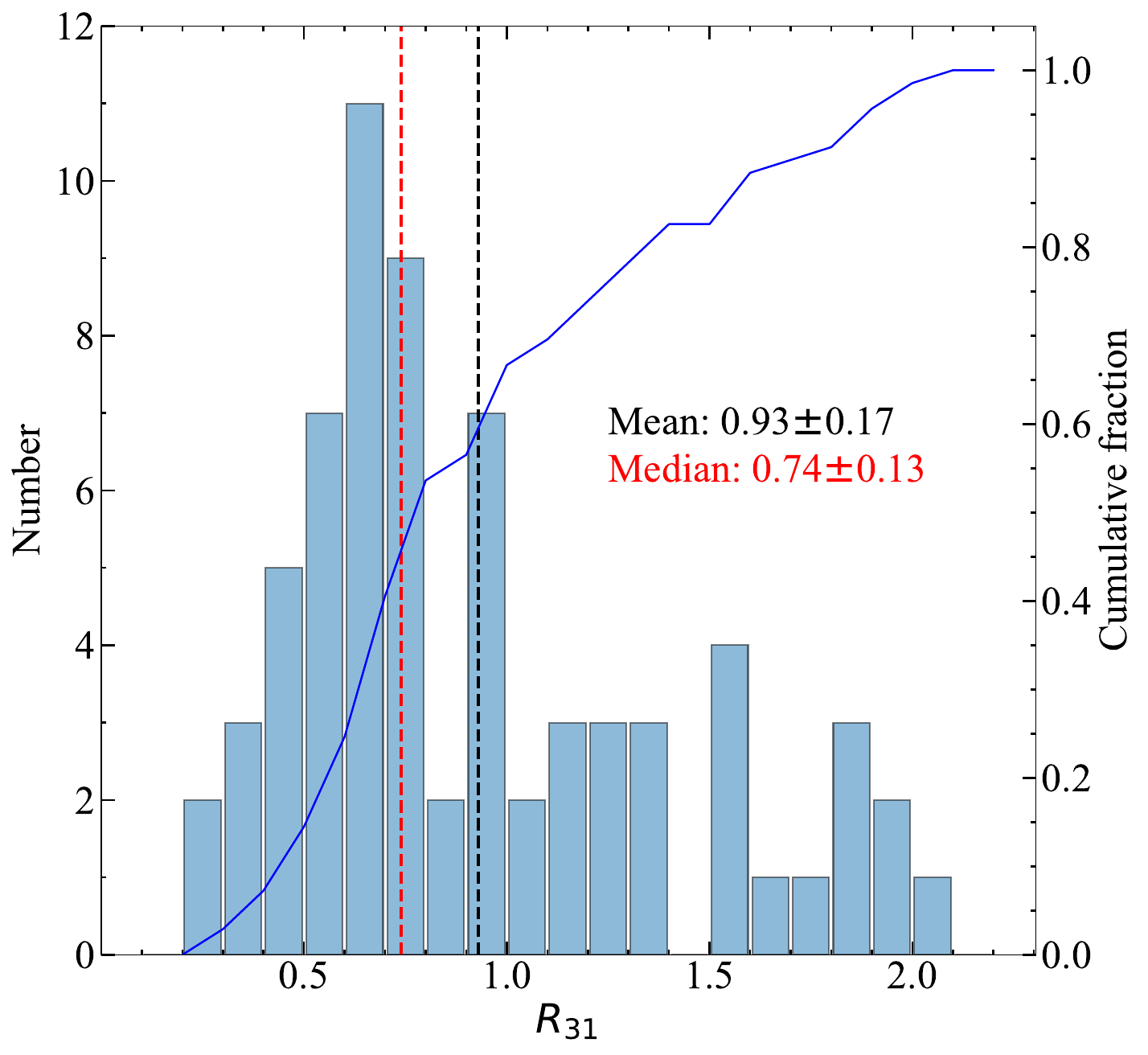}
\caption{Histogram of the $R_{31}$ ratios for the 69 Galactic edge clouds with significant CO\,(3--2) detections. 
The solid blue line shows the cumulative distribution function. 
The black and red dotted lines denote the mean and the median values, respectively. }
\label{fig:hist-r31}
\end{figure}

The $^{12}$CO\,$J$=3--2/1--0 integrated intensity ratios, denoted as $R_{31}$\,=\,$I_{\rm CO\,(3-2)}$/$I_{\rm CO\,(1-0)}$, were obtained for 69 of our 84 Galactic edge clouds (see Table\,\ref{table:spectral parameters}). 
For the remaining 15 clouds without significant CO\,(3--2) detections after smoothing to 50$^{\prime\prime}$, upper limits on $R_{31}$ ratios were derived and are also listed in Table\,\ref{table:spectral parameters}, all of which are below unity. 
Among the 69 Galactic edge clouds with significant CO\,(3--2) detections, 23 (33\%) have $R_{31}$\,>\,1.0, which corresponds to 27\% of the full sample of 84 clouds. 
Fifteen sources with no significant CO\,(3--2) detections were excluded from the statistical analysis. 
$R_{31}$ ratio ranges from 0.28 to 2.02, with a mean value of 0.93\,$\pm$\,0.17. 
The statistical histogram of the $R_{31}$ ratio is illustrated in Fig.\,\ref{fig:hist-r31}. 
The cumulative distribution of the $R_{31}$ ratio reaches 50\% at a median value of 0.74\,$\pm$\,0.13 and rises steeply within the 0.6--0.8 interval, where the bulk of the ratios is concentrated. 
The correlations between $R_{31}$ ratios and both CO peak temperatures and integrated intensities are stronger than those with line width (see Fig.\,\ref{fig:pra-r31}). 

\subsection{Clump identification and physical properties}
\label{sect:Clump identification} 
Clumps within molecular clouds are extracted with the Python package, {\tt quickclump}\footnote{\tiny \url{https://github.com/vojtech-sidorin/quickclump/}} \citep{Sidorin2017}, which has been widely utilized for identifying clumps (e.g., \citealt{Finn2021,Finn2022,Finn2024a,Luo2025}). 
This algorithm is akin to the non-hierarchical clump identification method known as CLUMPFIND \citep{Williams1994}. 
The {\tt quickclump} determines the spatial boundary of a clump employing a threshold-based approach within the three-dimensional position-position-velocity (3D-PPV) space. 
A recognized clump must satisfy four criteria: minimum intensity threshold $T_{\rm cutoff}$, minimum peak intensity $I_{\rm minpk}$\footnote{$I_{\rm minpk}$ was introduced by \citet{Indebetouw2020}, refer to \tiny \url{https://github.com/indebetouw/quickclump}.}, minimum number of voxels $N_{\rm pix}$, and minimum intensity difference between the local maximum and the highest adjacent saddle point $dT_{\rm leaf}$. 
In this work, clump identification was performed on the original 16.6$^{\prime\prime}$ CO\,(3--2) cubes, rather than on the 50$^{\prime\prime}$-smoothed cubes used for the $R_{31}$ analysis. 
CO clumps were identified within the FITS cube using the following parameters: $T_{\rm cutoff}$\,=\,3$\sigma_{\rm rms}$, $dT_{\rm leaf}$\,=\,2$\sigma_{\rm rms}$, $I_{\rm minpk}$\,=\,5$\sigma_{\rm rms}$, $N_{\rm pix}$\,=\,2\,beams, where $\sigma_{\rm rms}$ is derived from the median of the RMS FITS file. 
The velocity resolution of the CO\,(3-2) data is 0.64\,km\,s$^{-1}$. 
Clouds with very narrow line widths ($\sim$\,0.5--1.0\,km\,s$^{-1}$) are poorly resolved, as their emission is confined to only one or two velocity channels and may be difficult to distinguish from noise or artifacts. 
Consequently, our sample may be incomplete for extremely narrow linewidth clouds. 
Clumps are primarily located within compact clouds. 
While some dense regions exhibit CO\,(3--2) signals, our spatial resolution is insufficient for discerning the structure of the clumps, necessitating a higher resolution for accurate identification. 
Considering these limitations, 110 CO clumps have been identified within the 84 Galactic edge clouds (see Fig.\,\ref{fig:map}). 

Pixels used to calculate the velocity dispersion, effective radius, CO luminosity and other physical properties are defined by the 110 CO clump mask as described below. 
The observed velocity dispersion ($\sigma_{\rm obs}$) is obtained by Gaussian fitting\footnote{\tiny \url{https://scipy.org/}} of the intensity-weighted mean profile of each identified clump. 
The local standard of rest velocity ($V_{\rm LSR}$), main beam brightness temperature ($T_{\rm mb}$), and integrated intensity ($I_{\rm CO}$) of a clump were determined simultaneously (see Table\,\ref{table:clump parameters}). 
The intrinsic velocity dispersion ($\sigma_{\rm v}$) has been deconvolved using the velocity resolution ($\Delta v$, 0.64\,km\,s$^{-1}$): 
\begin{equation}
\sigma_{\rm v} = \sqrt{\sigma_{\rm obs}^2-(\frac{\Delta v}{\sqrt{8\ln2}})^2} \hspace{2mm}{\rm km}\,{\rm s}^{-1}. 
\label{eq:Standard deviation of velocity}
\end{equation}
Errors of the parameters propagate from the uncertainty in the fitting method. 
The full width at half maximum ($\Delta V$) of the CO\,(3--2) line is approximately 2.355 times $\sigma_{\rm v}$. 
For the 110 CO clumps, the $\sigma_{\rm v}$ ranges from 0.5 to 5.4\,km\,s$^{-1}$, with a median of 1.2\,$\pm$\,0.1\,km\,s$^{-1}$ (see Tables\,\ref{table:statistics of physical parameters} and \ref{table:clump parameters}). 

\begin{table*}[t]
\setlength{\tabcolsep}{4.8pt}
\caption{Statistical parameters of the Galactic edge clumps.}
\centering
\begin{tabular}
{lcccccccccccc}
\hline\hline
\smash{\raisebox{-1.3ex}{Properties}}
& \multicolumn{4}{c}{First Galactic quadrant (this work)} & & \multicolumn{4}{c}{Second Galactic quadrant \citep{Luo2026}}\\
\cline{2-5} \cline{7-10}
& Min& Max& Mean& Median& & Min& Max& Mean& Median\\
\hline
$R_{\rm g}$ (kpc)                           &9.7\,$\pm$\,0.3  &14.9\,$\pm$\,0.5  &11.5\,$\pm$\,0.3  &11.4\,$\pm$\,0.3 & &14.2\,$\pm$\,0.1 &22.9\,$\pm$\,0.7  &16.7\,$\pm$\,0.3  &15.4\,$\pm$\,0.3 \\
$D_{\rm s}$ (kpc)                           &16.7\,$\pm$\,0.1 &22.5\,$\pm$\,0.5  &18.1\,$\pm$\,0.4  &17.5\,$\pm$\,0.3 & &9.3 \,$\pm$\,0.3 &16.2\,$\pm$\,0.6  &11.2\,$\pm$\,0.4  &10.3\,$\pm$\,0.4 \\
$R_{\rm eff}$ (pc)                          &0.6\,$\pm$\,0.1  &5.9\,$\pm$\,0.2   &2.0 \,$\pm$\,0.1  &2.0\,$\pm$\,0.1  & &0.6 \,$\pm$\,0.1 &3.4 \,$\pm$\,0.2  &1.2 \,$\pm$\,0.1  &1.2 \,$\pm$\,0.1 \\
$\sigma_{\rm v}$ (km\,s$^{-1}$)             &0.5\,$\pm$\,0.1  &5.4 \,$\pm$\,0.2  &1.3 \,$\pm$\,0.1  &1.2 \,$\pm$\,0.1 & &0.3 \,$\pm$\,0.1 &1.7 \,$\pm$\,0.1  &0.8 \,$\pm$\,0.1  &0.7 \,$\pm$\,0.1 \\
$L_{\rm CO}$ (K\,km\,s$^{-1}$\,pc$^{2}$)    &8.4\,$\pm$\,1.7  &2682\,$\pm$\,595  &149\,$\pm$\,29    &49.3\,$\pm$\,9.6 & &3.0 \,$\pm$\,0.4 &285 \,$\pm$\,31   &32.5\,$\pm$\,4.1  &16.8\,$\pm$\,2.3 \\
$M_{\rm H_2}$ ($10^{2}$\,M$_\odot$)         &0.5\,$\pm$\,0.2  &213 \,$\pm$\,64   &10.1\,$\pm$\,3.1  &3.3 \,$\pm$\,0.9 & &0.3 \,$\pm$\,0.1 &82.5\,$\pm$\,10.0 &5.2 \,$\pm$\,0.9  &2.2 \,$\pm$\,0.4 \\
$\Sigma_{\rm H_2}$ (M$_{\odot}$\,pc$^{-2}$) &7.4\,$\pm$\,2.8  &666 \,$\pm$\,198  &65.1\,$\pm$\,21.0 &42.7\,$\pm$\,12.2& &11.7\,$\pm$\,3.0 &1025\,$\pm$\,261  &92.9\,$\pm$\,18.7 &62.6\,$\pm$\,13.0\\
$N_{\rm H_2}$ ($10^{21}$\,cm$^{-2}$)        &0.3\,$\pm$\,0.1  &31.0\,$\pm$\,9.2  &3.0\,$\pm$\,1.0   &2.0 \,$\pm$\,0.6 & &0.5 \,$\pm$\,0.2 &47.7\,$\pm$\,12.1 &4.3 \,$\pm$\,0.9  &2.9 \,$\pm$\,0.6 \\
$n_{\rm H_2}$ ($10^{2}$\,cm$^{-3}$)         &0.3\,$\pm$\,0.1  &44.8\,$\pm$\,14.0 &5.0\,$\pm$\,1.6   &2.5 \,$\pm$\,0.9 & &1.0 \,$\pm$\,0.3 &161 \,$\pm$\,30   &9.8 \,$\pm$\,1.8  &6.3 \,$\pm$\,1.1 \\
$\alpha_{\rm vir}$                          &1.6\,$\pm$\,0.6  &31.7\,$\pm$\,12.7 &8.7\,$\pm$\,3.0   &6.8 \,$\pm$\,2.4 & &0.6 \,$\pm$\,0.1 &15.3\,$\pm$\,3.5  &3.4 \,$\pm$\,0.8  &2.8 \,$\pm$\,0.6 \\
\hline
\end{tabular}
\label{table:statistics of physical parameters}\\ 
\tablefoot{The $\pm$ uncertainties are derived through error propagation.} 
\end{table*}

Kinematic distances are used to estimate clump properties in this work, owing to the absence of other reliable distance constraints for molecular clouds in the outer Galaxy. 
The Galactocentric distance ($R_{\rm g}$) and heliocentric distance ($D_{\rm s}$) are adopted from \citet{Sun2024a}, which utilizes the Galactic rotation curve from \cite{Reid2019}. 
The heliocentric distances of 84 Galactic edge clouds range from approximately 16.7 to 22.5\,kpc, with a median of $\sim$\,17.5\,kpc. 
Correspondingly, the linear scales range from 1.3 to 1.8\,pc at a beam size of 16.6$^{\prime\prime}$, with a median of $\sim$\,1.4\,pc. 
An ellipse is fitted to each clump to determine the effective radius, using the full-width at half-maximum for both major ($\delta_{\rm maj}$) and minor ($\delta_{\rm min}$) axes. 
Approximating each clump as a two-dimensional Gaussian, the deconvolved values of the major ($\delta_{\rm maj}^{\prime}$) and minor ($\delta_{\rm min}^{\prime}$) axes are obtained by deconvolving the telescope beam: 
\begin{align}
\delta_{\rm maj}^{\prime} = \sqrt{(\delta_{\rm maj}^2-\theta_{\rm beam}^2)/(8\ln2)} \hspace{2mm}{\rm arcsec}, 
\label{eq:major}
\\
\delta_{\rm min}^{\prime} = \sqrt{(\delta_{\rm min}^2-\theta_{\rm beam}^2)/(8\ln2)} \hspace{2mm}{\rm arcsec}. 
\label{eq:minor}
\end{align}
The angular size in arcsec ($^{\prime\prime}$), measured from the major and minor axes, is converted to physical units (pc) using the heliocentric distance. 
Following previous studies (e.g., \citealt{Wong2022,Luo2026}), the effective radius ($R_{\rm eff}$) is derived by 
\begin{equation}
R_{\rm eff} = 1.91\sqrt{\delta_{\rm maj}^{\prime} \delta_{\rm min}^{\prime}} \hspace{2mm}{\rm pc}. 
\label{eq:effective radius}
\end{equation}
For the 110 clumps, the median diameter (2$R_{\rm eff}$) is $\sim$\,4.0\,pc (see Table\,\ref{table:statistics of physical parameters}), corresponding to approximately three times the median linear resolution, indicating that most clumps are reasonably resolved. 
Only a few clumps have diameters comparable to the linear resolution (see Table\,\ref{table:clump parameters}), indicating that they are marginally resolved. 

CO luminosities ($L_{\rm CO}$), referring to CO\,(1--0) and defined as the sum of the integrated CO intensities over each clump, are determined using the CO\,$J$=3--2/1--0 integrated intensity ratio, following 
\begin{equation}
L_{\rm CO} = A_{\rm pix} \frac{\Sigma I_{\rm CO(3-2)}}{R_{31}} \hspace{2mm}{\rm K}\,{\rm km}\,{\rm s}^{-1}\,{\rm pc}^{2},
\label{eq:luminosities}
\end{equation}
where $A_{\rm pix}$ (pc$^2$) is the area of an individual pixel and $\Sigma I_{\rm CO(3-2)}$ (K\,km\,s$^{-1}$) is the total integrated intensity of CO\,(3--2) summed over all pixels within a clump. 
Clumps within each cloud consistently correspond to a single $R_{31}$ value (see Table \ref{table:spectral parameters}). 
Small clouds, when diluted by a large beam size, lack $R_{31}$ ratios. 
In these cases we utilize the median value. 
The clump masses ($M_{\rm H_2}$) are calculated by the conversion factor from CO luminosity to H$_2$ mass ($\alpha_{\rm CO}$), according to 
\begin{equation}
M_{\rm H_2} = \alpha_{\rm CO} L_{\rm CO} \hspace{2mm}{\rm M}_\odot. 
\label{eq:luminous mass}
\end{equation}
The CO-to-H$_2$ conversion factor depends on metallicity. 
In this work, the relationship $\alpha_{\rm CO}$\,$\propto$\,$Z^{-1.0}$ \citep{Bolatto2013} has been employed, where $Z$ signifies metallicity (${\rm log(}Z{\rm )}$\,=\,$-0.056R_{\rm g}-1.176$; for details, see \citealt{Giannetti2017}). 
Over the Galactocentric distance range spanned by our sample, this corresponds to metallicities of $Z$\,$\simeq $\,0.010--0.019, and hence to $\alpha_{\rm CO}$ values of 5.3--10.3\,M$_{\odot}$\,(K\,km\,s$^{-1}$\,pc$^2$)$^{-1}$. 
CO luminosities and masses range from 8 to 2682\,K\,km\,s$^{-1}$\,pc$^{2}$ and 50 to 21300\,M$_\odot$, respectively. 
The corresponding median values are approximately 49\,K\,km\,s$^{-1}$\,pc$^{2}$ and 330\,M$_\odot$, respectively (see Tables\,\ref{table:statistics of physical parameters} and \ref{table:clump parameters}). 

\begin{figure*}[t]
\centering
\includegraphics[width=0.90\textwidth]{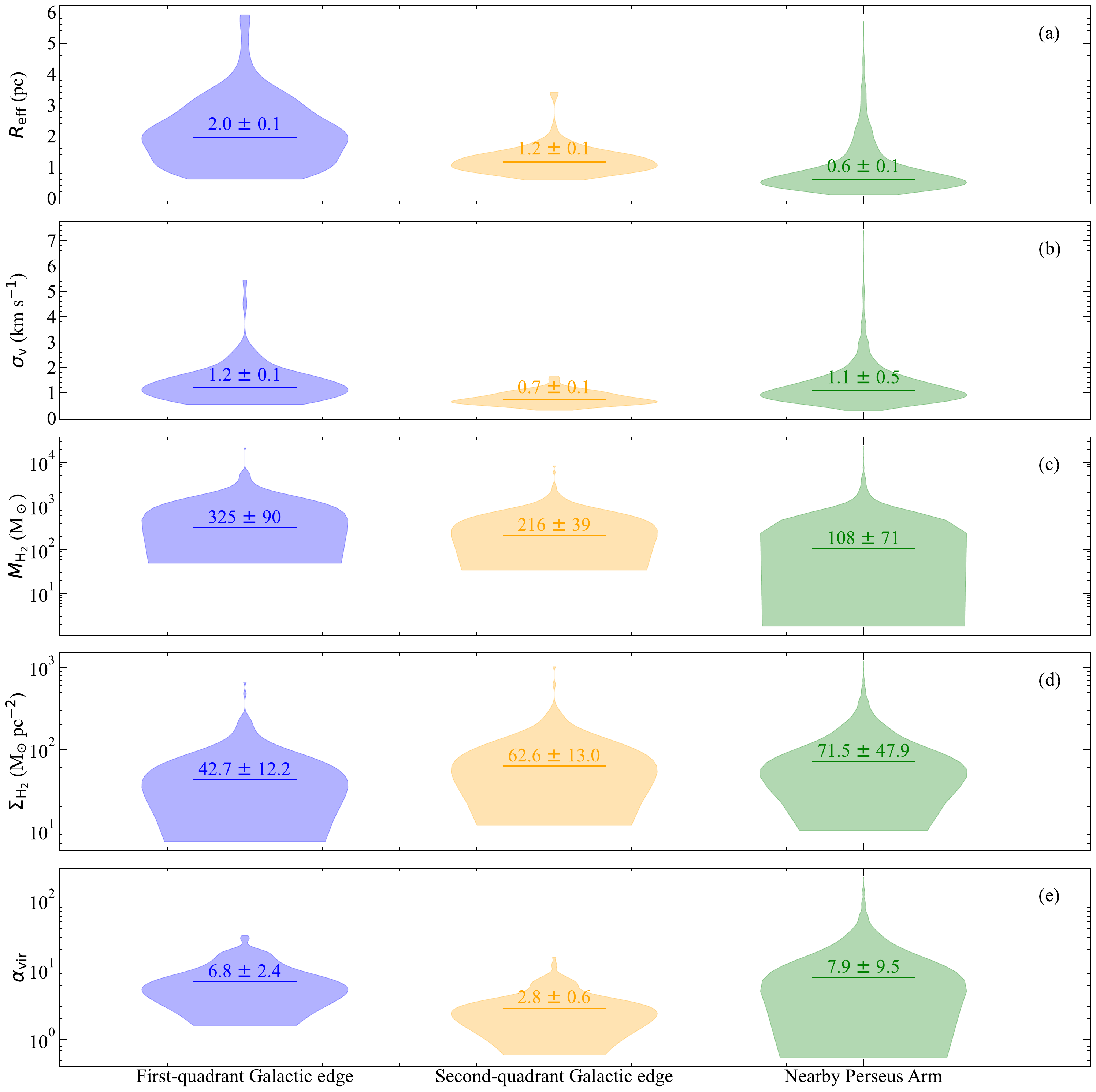}
\caption{Violin plots of CO clump effective radii (a), velocity dispersions (b), masses (c), surface densities (d), and virial parameters (e). 
This includes 110 first-quadrant Galactic edge clumps (blue; this work), 112 second-quadrant Galactic edge clumps (orange; \citealt{Luo2026}), and 978 nearby Perseus Arm clumps (green; \citealt{Rigby2025}). 
The median values are marked by horizontal colored lines, with their corresponding values labeled above these lines. } 
\label{fig:PAMS}
\end{figure*}

\begin{figure*}[t]
\centering
\includegraphics[width=0.95\textwidth]{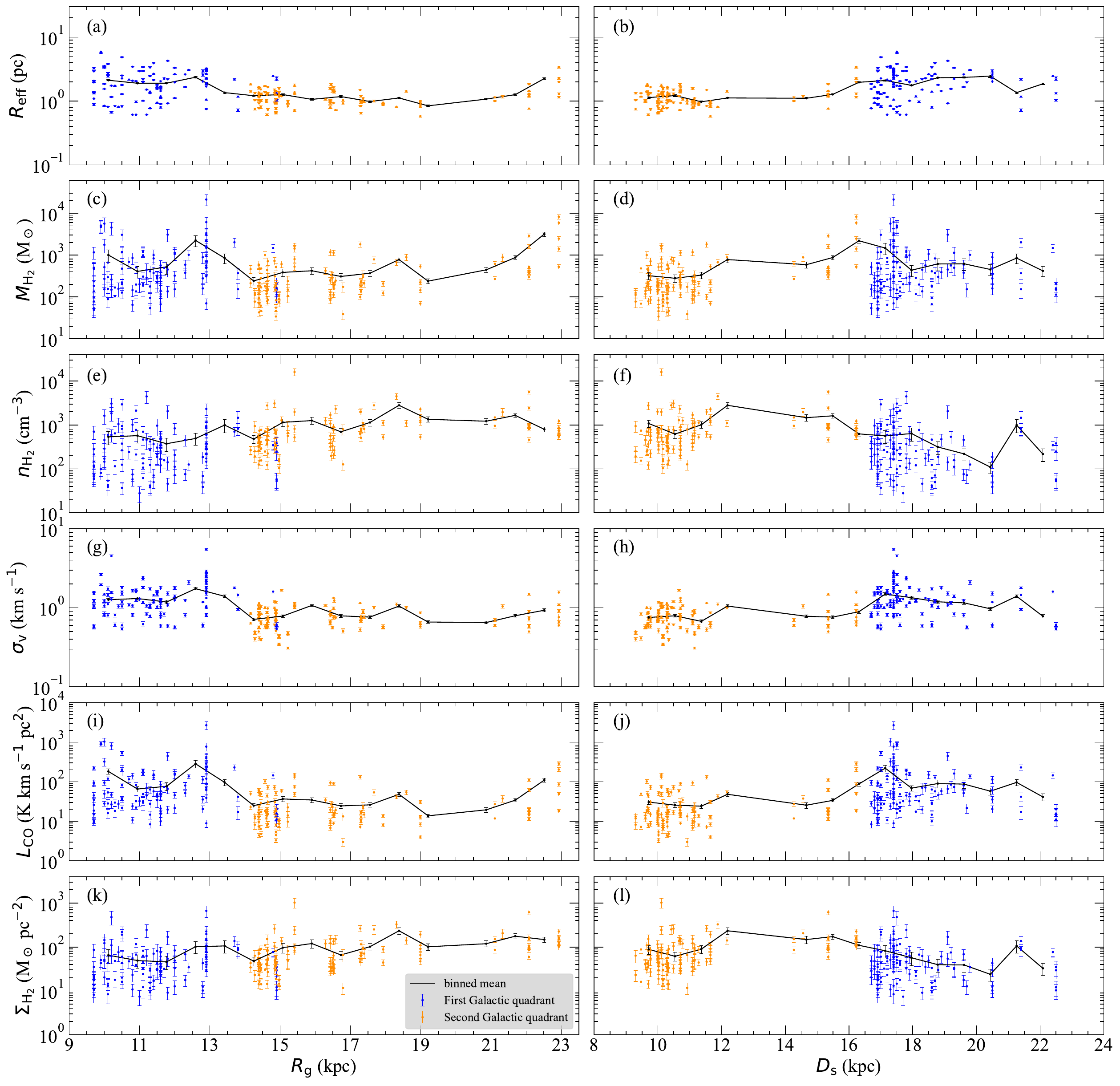}
\caption{Variation in CO clump properties with Galactocentric (\emph{left column}) and heliocentric distances 
(\emph{right column}): effective radius (a--b), luminous mass (c--d), volume density (e--f), velocity dispersion (g--h), 
CO luminosity (i--j), and surface density (k--l). The solid black lines connecting the data points represent the mean 
value within a bin, with each bin spanning 0.5\,kpc. 
Galactic edge clumps in the first Galactic quadrant (blue; this work) and second Galactic quadrant (orange; \citealt{Luo2026}). } 
\label{fig:pra_dis} 
\end{figure*}

The H$_2$ column densities ($N_{\rm H_2}$), H$_2$ volume densities ($n_{\rm H_2}$), and surface densities ($\Sigma_{\rm H_2}$) of the clumps are derived using the following equations: 
\begin{align}
N_{\rm H_2} &= X_{\rm CO} L_{\rm CO} / (\pi R_{\rm eff}^2) \hspace{2mm}{\rm cm}^{-2}, 
\label{eq:column density}
\\
n_{\rm H_2} &= 3 M_{\rm H_2} / (4 \pi \mu m_{\rm H} R_{\rm eff}^3) \hspace{2mm}{\rm cm}^{-3}, 
\label{eq:volume density}
\\
\Sigma_{\rm H_2} &= M_{\rm H_2} / (\pi R_{\rm eff}^2) \hspace{2mm}{\rm M}_{\odot}\,{\rm pc}^{-2}, 
\label{eq:surface density}
\end{align}
where $X_{\rm CO}$ represents the CO-to-H$_2$ conversion factor, corresponding to $\alpha_{\rm CO}$\,=\,$\mu$$m_{\rm H}$$X_{\rm CO}$(1\,pc)$^2$/M$_{\odot}$. 
$m_{\rm H}$ and $\mu$ denote the atomic hydrogen mass ($\sim$\,1.67\,$\times$\,10$^{-27}$\,kg) and the average molecular weight per H$_2$ molecule ($\sim$\,2.72) in the interstellar medium, respectively. 
The median values for H$_2$ column density, volume density, and surface density of the CO clumps are 2.0\,$\times$\,$10^{21}$\,cm$^{-2}$, 250\,cm$^{-3}$, and 42.7\,M$_{\odot}$\,pc$^{-2}$, respectively. 
Additionally, column densities vary between 3.4\,$\times$\,$10^{20}$\,cm$^{-2}$ and 3.1\,$\times$\,$10^{22}$\,cm$^{-2}$, with a mean value of 3.0\,$\times$\,$10^{21}$\,cm$^{-2}$.  
The volume densities range from 30 to 4480\,cm$^{-3}$, with a mean value of 500\,$\pm$\,160\,cm$^{-3}$ (see Tables\,\ref{table:statistics of physical parameters} and \ref{table:clump parameters}). 

The virial parameter ($\alpha_{\rm vir}$), a measure of the equilibrium between gravitational and kinetic energy within a clump \citep{Bertoldi1992,McKee2007}, can be calculated by 
\begin{equation}
\alpha_{\rm vir} = 5\sigma_{\rm v}^2 R_{\rm eff}/{\rm G}M_{\rm H_2} \,, 
\label{eq:virial parameter}
\end{equation}
where G denotes the gravitational constant. 
The measured $\alpha_{\rm vir}$ in our clumps span from 1.6 to 31.7, with a mean of 8.7\,$\pm$\,3.0 and a median of 6.8\,$\pm$\,2.4 (see Tables\,\ref{table:statistics of physical parameters} and \ref{table:clump parameters}). 
The uncertainty in derived parameters stems from error propagation, which is determined by the uncertainties in kinematic distance, velocity dispersion, and CO luminosity. 
Note that our clump sample is subject to several observational biases. 
Clumps are considered unresolved if their angular size is less than 16.6$^{\prime\prime}$, their line width falls below 0.64\,km\,s$^{-1}$, or their peak intensity is below the 5\,$\sigma_{\rm rms}$ detection limit. 
The derived $X_{\rm CO}$ factor may also be affected by spatial resolution, in addition to metallicity. 
These limits partially account for the systematic differences observed between the first and second quadrant samples (see Sect.\,\ref{sect:Comparison of Galactic edge clump properties between the first and second Galactic quadrants}). 

\section{Discussion}
\label{sect:discussion}
\subsection{Comparison of clump properties in the first and second Galactic quadrants}
\label{sect:Comparison of Galactic edge clump properties between the first and second Galactic quadrants} 
To investigate the universality of molecular cloud and clump properties in low-metallicity environments at the Galactic outskirts, we compare clouds in the first and second Galactic quadrants. 
The first-quadrant sample probes the OSC arm at $R_{\rm g}$\,$\sim$\,10--15\,kpc \citep{Sun2024a,Sun2024b}, while the second-quadrant sample extends to $R_{\rm g}$\,$\sim$\,14--23\,kpc \citep{Sun2015a,Luo2026}. 
Both samples reside in the outer Galactic disk along the same spiral arm, but span different Galactocentric radii and azimuths and may experience different ambient radiation fields. 
A comparison of two samples therefore provides a valuable test of whether clump properties and scaling relations depend on these environmental factors. 

The distributions of physical parameters for Galactic edge clumps in the first and second Galactic quadrants are shown in Fig.\,\ref{fig:PAMS}. 
The Galactic edge clumps in the first quadrant have statistically larger sizes and higher velocity dispersions than those in the second Galactic quadrant. 
In addition, the clumps in the first quadrant extend to higher masses and CO luminosities, up to 10$^{4}$\,M$_{\odot}$ and  10$^{3}$\,K\,km\,s$^{-1}$\,pc$^{2}$ (see Table\,\ref{table:statistics of physical parameters}). 
While the overall distributions of properties are broadly similar between the two quadrants in Fig.\,\ref{fig:PAMS}, the statistical average values for surface density, column density, and volume density tend to be higher in the second Galactic quadrant (see Table\,\ref{table:statistics of physical parameters}). 
The lower-mass but higher-density clumps may be attributed to the significant proportion of CO-dark gas at large Galactocentric distances, which constrains CO emission to regions with particularly high (column) density \citep{Pineda2013,Luo2025}. 
Another non-negligible factor is that the average linear spatial scale corresponding to the telescope beam sizes for the Galactic edge clouds in the first Galactic quadrant ($\sim$\,1.4\,pc) is larger than that of the second Galactic quadrant ($\sim$\,0.6\,pc). 
As a result, compact and low-mass clumps are more likely to be missed in the first quadrant, while blending of adjacent structures may increase measured sizes and velocity dispersions. 
Furthermore, the virial parameter exhibits a broader distribution in the first quadrant, encompassing a significant proportion of elevated values, whereas it is predominantly concentrated below 5 in the second quadrant  (see Fig.\,\ref{fig:PAMS}). 

Clump properties as a function of the Galactocentric distance are investigated by comparing the first and second Galactic quadrant samples in Fig.\,\ref{fig:pra_dis}. 
Over the Galactocentric distance range of 10--23 kpc, we find no compelling evidence for systematic radial trends in clump size, mass, luminosity, velocity dispersion, volume density, or surface density. 
Spiral density wave perturbations become weaker with increasing Galactocentric distance, corresponding to a decrease in stellar density. 
The clump properties in the OSC arm may also be determined by the azimuthal environment, in addition to the Galactocentric radius. 
These molecular cloud samples, all located in the Scutum-Centaurus Arm, are subject to stronger perturbations in the first quadrant than in the second quadrant \citep{Dame2011,Sun2015a}. 
These perturbations, which include spiral arms, velocity shear, and stellar feedback, may lead to greater velocity dispersion and larger virial parameters. 

The Galactic edge clumps in the first quadrant are distributed at greater distances from the Sun ($D_{\rm s}$\,$\sim$\,16--23\,kpc), whereas those in the second quadrant are located at closer distances ($D_{\rm s}$\,$\sim$\,9--16\,kpc). 
The distributions of clump mass and velocity dispersion across varying heliocentric distances display significant scatter, with no discernible systematic variation (see Fig.\,\ref{fig:pra_dis}). 
Clump size and CO luminosity increase with heliocentric distance, while the volume density and surface density tend to decrease. 
These trends are consistent with observational biases such as sensitivity limits and beam dilution, which preferentially select larger and brighter clumps at greater distances. 
Therefore, the systematically larger sizes and masses, as well as the lower densities observed in the first Galactic quadrant, may be partially influenced by distance-related biases rather than intrinsic physical differences. 
In addition, differences in physical properties likely reflect environmental effects, including metallicity gradients and the increasing contribution of CO-dark gas in the outer Galaxy (see Sect.\,\ref{sect:The impact of metallicity and CO-dark gas}). 

To compare clump properties at comparable Galactocentric radii ($R_{\rm g}$\,$\sim$\,10\,kpc), we selected nearby Perseus Arm clumps identified in the Perseus Arm Molecular Survey (PAMS; \citealt{Rigby2025}), which covers star-forming regions such as W5, S157, and NGC\,7538. 
As illustrated in Fig.\,\ref{fig:PAMS}, first-quadrant Galactic edge clumps have statistically larger sizes and masses but lower surface densities compared with nearby Perseus Arm clumps. 
In addition, the typical velocity dispersions of first-quadrant Galactic edge clumps (1.2\,$\pm$\,0.1\,km\,s$^{-1}$) are comparable to those of nearby Perseus Arm clumps (1.1\,$\pm$\,0.5\,km\,s$^{-1}$). 
Both samples show relatively large virial parameters, with median values of $\alpha_{\rm vir}$\,$\sim$\,7--8, indicating that most clumps are not gravitationally bound. 
This suggests that the Galactic edge clumps in the first Galactic quadrant may not be significantly different from nearby Perseus arm clumps, and that the observed differences could be due to lower physical resolution of the first quadrant data. 
Future observations with higher angular resolution and uniform sensitivity will be essential to disentangle the intrinsic clump properties from distance-related observational biases. 

One should note that the first and second Galactic quadrant samples are non-overlapping in heliocentric distance, and that the finite beam introduces a distance-dependent bias in clump sizes and masses. 
The comparison presented here is statistical. 
We focus on the Galactocentric trends, intensity ratios, and scaling relation slopes, verifying that our main conclusions are robust to smoothing the data to a common physical resolution and/or restricting it to the overlapping range in Galactocentric distance. 

\subsection{Variation in $R_{31}$ ratios}
\label{Sec:Variation in R31 ratios}
CO is frequently used as a tracer for H$_2$, with its mass typically calculated using the CO-to-H$_2$ conversion factor. 
Estimates of molecular gas mass, based on the assumption of a constant $R_{31}$ value, may lead to misinterpretation due to the significant variations that the $R_{31}$ exhibits in different environments. 
The $R_{31}$ ratio has been observed across diverse environments, including the Milky Way (e.g., \citealt{Oka2007,Oka2012,Enokiya2018,Tokuyama2019,Park2023}) and external galaxies (e.g., \citealt{Devereux1994,Meier2001,Yao2003,Muraoka2007,Nakanishi2007,Mao2010,Hunt2017,Li2019,Li2020a,Lamperti2020,Leroy2022}). 
Within the Milky Way, the $R_{31}$ ratios of the Galactic center and spiral arms are approximately 0.7 and 0.4, respectively \citep{Oka2012}. 
The median $R_{31}$ value is 0.7 for the Large Magellanic Cloud (LMC) \citep{Minamidani2008,Celis2019} and 1.0 for the Small Magellanic Cloud (SMC) \citep{Nikolic2007,Muller2014}. 
Furthermore, $R_{31}$ ratios of $\gtrsim$\,1.0 are observed in metal-poor galaxies ($Z/Z_{\odot}$\,$\sim$\,0.3; \citealt{Hunt2017}). 
The average $R_{31}$ values for galaxies categorized as starbursts and barred spirals are 0.89\,$\pm$\,0.11 and 0.88\,$\pm$\,0.15, respectively \citep{Mao2010}. 
At high redshift ($z$\,>\,2), the median $R_{31}$ ratio is 0.55\,$\pm$\,0.05 in luminous infrared ($L_{\rm IR}$\,$\sim$\,10$^{12}$\,$L_{\odot}$) galaxies, which is consistent with local galaxies ($L_{\rm IR}$\,>\,10$^{11}$\,$L_{\odot}$) \citep{Ivison2011}. 
A median value of 0.97\,$\pm$\,0.17 is observed in submillimeter galaxies that host quasars \citep{Riechers2011}. 
These findings suggest that the variation in $R_{31}$ is influenced by local molecular gas conditions. 

\begin{figure}[t]
\centering
\includegraphics[width=0.45\textwidth]{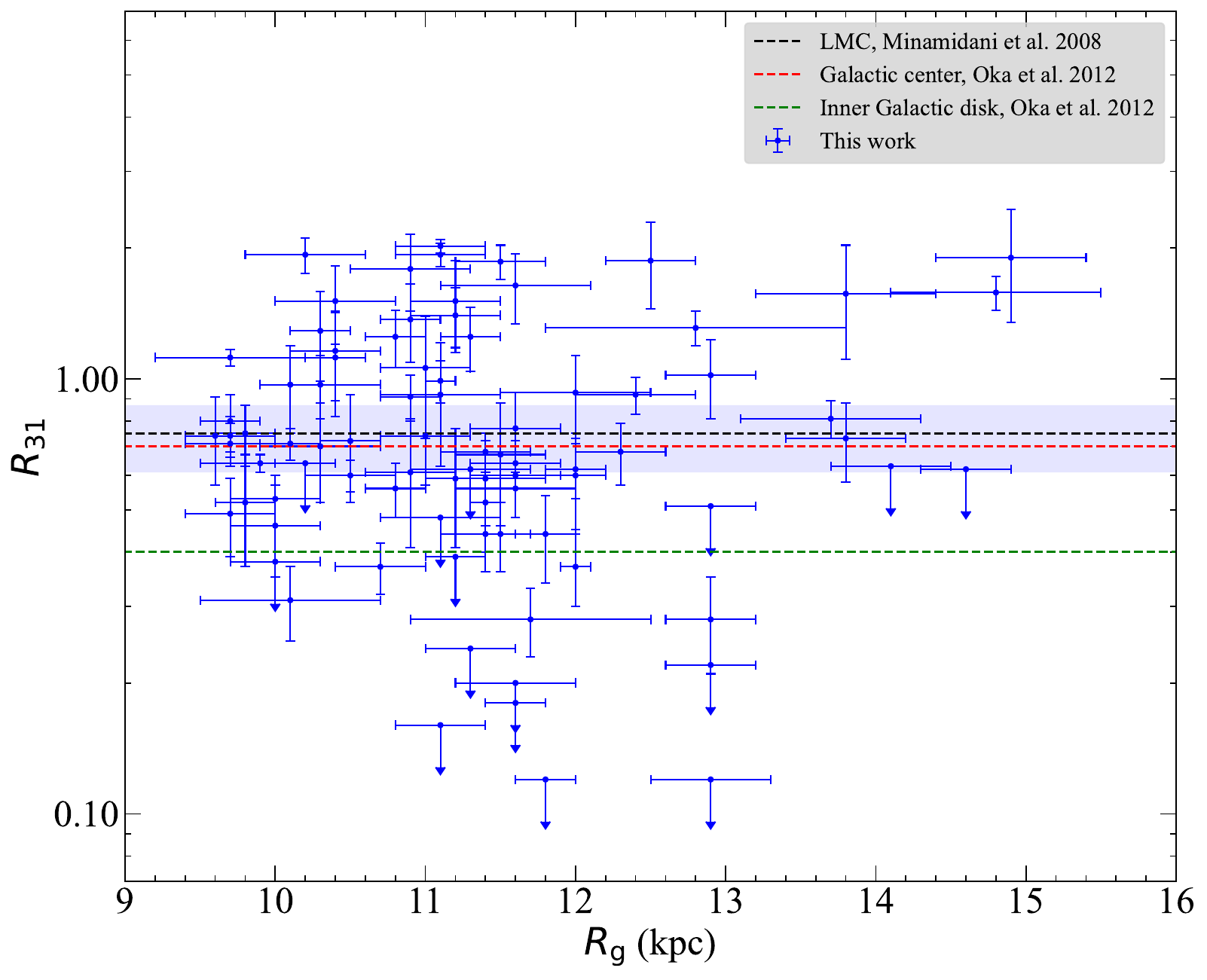}
\caption{Variation in the $R_{31}$ ratios with Galactocentric distance, $R_{\rm g}$. 
Downward arrows denote the upper limits of $R_{31}$ ratios derived from CO\,(3--2) signal-to-noise ratios below 3 after smoothing to 50$^{\prime\prime}$. 
The gray-shaded region illustrates the median and error range of $R_{31}$ (0.74\,$\pm$\,0.13). 
The black, red, and green dotted lines represent typical $R_{31}$ values of the LMC (0.75; \citealt{Minamidani2008}), Galactic center (0.7; \citealt{Oka2012}), and inner Galactic disk (0.4; \citealt{Oka2012}), respectively. } 
\label{fig:Rg-R31}
\end{figure}

\begin{figure*}[t]
\centering
\includegraphics[width=0.95\textwidth]{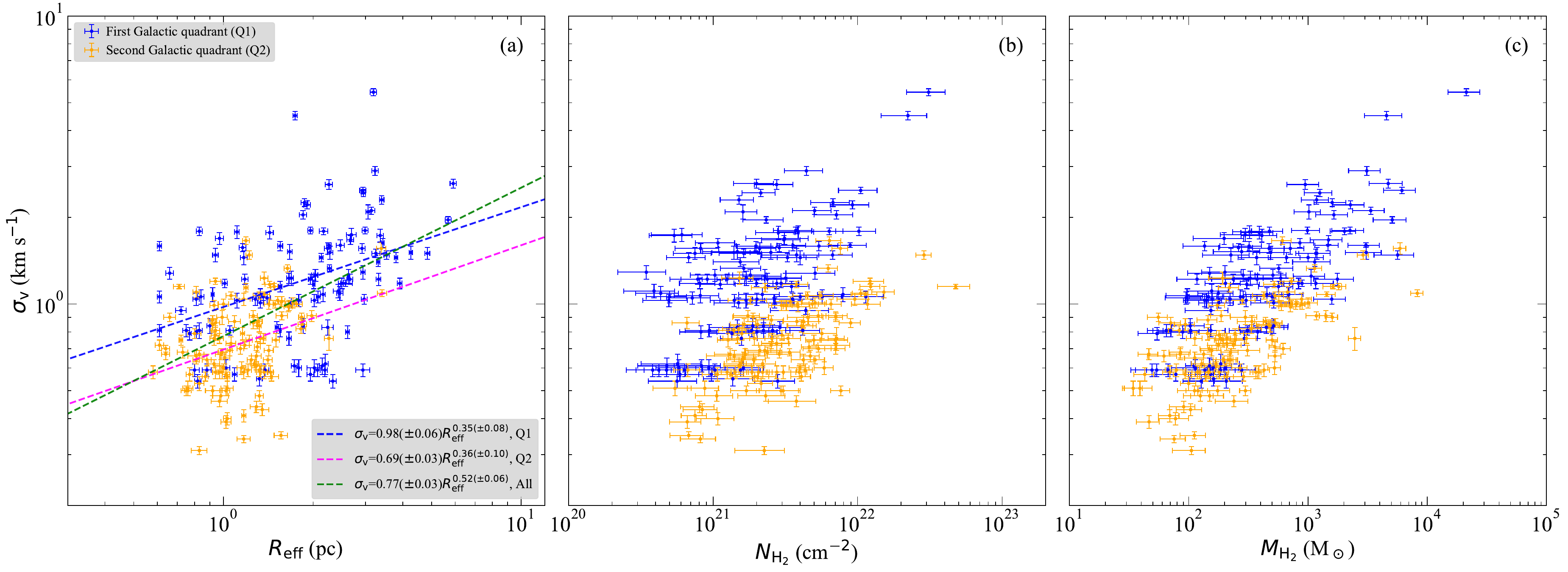}
\caption{(a) The $\sigma_v$--$R$ relation for Galactic edge clumps, located in the first Galactic quadrant (blue, this work) and second Galactic quadrant (orange, \citealt{Luo2026}). 
The blue, magenta, and green dashed lines represent the fits obtained for the Galactic edge clumps in the first quadrant, second quadrant, and both combined, respectively. 
(b) Variation in the velocity dispersion with the column density of the clumps. 
(c) Variation in the velocity dispersion with the mass of the clumps. 
} 
\label{fig:Scaling-Relation} 
\end{figure*}

Among the 69 Galactic edge clouds with significant CO\,(3--2) detections after smoothing to 50$^{\prime\prime}$, the $R_{31}$ ratios show a significant dispersion (0.28--2.02) (see Sect.\,\ref{sect:R31 Line Ratios}). 
23 clouds out of the 69 Galactic edge clouds have $R_{31}$\,>\,1.0 (see Table\,\ref{table:spectral parameters}). 
High $R_{31}$ ratios of 2.02 and 1.93 are associated with the dense clouds G26.608--0.208 and G25.833+0.400, respectively (see Fig.\,\ref{fig:map}). 
A high $R_{31}$ ratio could suggest the presence of molecular gas characterized by increased temperature, density, and/or reduced optical depth (e.g., \citealt{Hunt2017,Li2019,Li2020a}). 
However, a value of $R_{31}$\,>\,1.0 is not uniquely indicative of highly excited gas as the observed ratio can be affected by optical depth effects, self absorption, multiple velocity components, and calibration uncertainties. 
The median $R_{31}$ of 0.74\,$\pm$\,0.13 in the Galactic edge clouds is comparable to that of the Galactic center, in spite of their different environmental boundary conditions. 
Similarly, such a ratio is characteristic of the LMC, which likely demonstrates comparable molecular excitation to that of the Galactic edge. 
One potential cause is the increase in the fraction of CO-dark H$_2$ relative to the total H$_2$ mass as the metallicity decreases \citep{Pineda2013,Langer2014}. 
CO emission may preferentially trace the well-shielded, high (column) density regions of molecular clouds, which could influence the observed $R_{31}$ distribution. 
Previous observations indicate that enhanced CO line ratios are typically associated with compact clouds and regions with star forming activity \citep{Park2023,Luo2025}. 
In other environments, the $R_{31}$ ratio exhibits a positive correlation with H$\alpha$ flux \citep{Minamidani2008}, dust temperature \citep{Li2020a}, star formation efficiency \citep{Lamperti2020}, and star formation rate \citep{Li2020a,Leroy2022}. 
Excitation conditions of the molecular gas at the Galactic edge, as traced by $R_{31}$ in the first Galactic quadrant, are consistent with previous findings based on $^{12}$CO\,$J$=2--1/1--0 integrated intensity ratios ($R_{21}$) in the second Galactic quadrant \citep{Luo2025}. 
Possible mechanisms for the enhanced CO line ratios include ultraviolet (UV) heating associated with massive stars, interstellar shocks, and dynamical effects related to the spiral arm, although these possibilities cannot be distinguished with the present data. 

The $R_{31}$ profile exhibits fluctuations across the Galactic longitude range of $15^\circ$\,<\,$l$\,<\,$50^\circ$ \citep{Park2023}. 
The $R_{31}$ profile in Galactic star-forming regions (e.g., W43, W49A, and W51) is relatively enhanced compared with the overall profile, which has a median value of 0.27. 
In the outer Galaxy, there is no clear trend in the $R_{31}$ ratio with Galactocentric distances ranging from 10 to 15\,kpc (see Fig.\,\ref{fig:Rg-R31}). 
Nevertheless, the $R_{31}$ ratios show typical values of 0.7 at the Galactic center, 0.4 within the inner Galactic disk, and 0.77 in the outer Galaxy. 
This variation aligns with the CO\,$J$=2--1/1--0 line ratios observed in the Milky Way \citep{Luo2025}. 
A similar trend is also observed in the M31 disk, where the $R_{31}$ ratio increases slightly with galactocentric radius from $\sim$\,0.14 at 2--9\,kpc to $\sim$\,0.27 in the 10\,kpc ring, while reaching 0.8 in the 1\,kpc nuclear ring \citep{Li2020a}. 

\subsection{Scaling relations}
\label{sect:Scaling Relations}
\subsubsection{Velocity dispersion--size relation}
\label{sect:Line width-size relation}
The power-law relation between the molecular cloud size and velocity dispersion was first discovered by \citet{Larson1981}. 
The relation is further confirmed by examining molecular clouds located in the first quadrant of the Galactic disk \citep{Solomon1987}. 
Considering the equivalence between gravitational and kinetic energy, this power-law relation demonstrates a correlation of the velocity dispersion with the mass surface density \citep{Heyer2009}. 
The velocity dispersion--size relation (hereafter $\sigma_{\rm v}$--$R$ relation) has been categorized into four distinct types \citep{Goodman1998}: 
Type 1 involves multiple tracers and multiple clouds; Type 2 is characterized by a single tracer and multiple clouds; Type 3 is denoted by multiple tracers within a single cloud; and Type 4 represents a single tracer in a single cloud. 
Overall, the $\sigma_{\rm v}$--$R$ relation of different Types adheres to the formula $\sigma_{\rm v}$\,=\,$v_0 R_{\rm eff}^{\,a}$ as observed in the Milky Way (e.g., \citealt{Heyer2015,Rice2016,Miville-Deschenes2017,Sun2017,Sun2024a,He2026}) and nearby galaxies (e.g., \citealt{Wong2017,Wong2019,Wong2022,Saldano2023}). 

In this work, we investigate the $\sigma_{\rm v}$--$R$ relation of Type 2 by analyzing 110 Galactic edge clumps in the first quadrant using CO\,(3--2) as a single molecular tracer (see Fig.\,\ref{fig:Scaling-Relation}). 
The relation was fitted by the equation $\sigma_{\rm v}$\,=\,0.98($\pm$0.06)$R_{\rm eff}^{0.35(\pm0.08)}$, indicating a weak positive correlation as evidenced by a Pearson correlation coefficient of 0.39. 
A $p$-value of less than $10^{-3}$ indicates high significance (see Table\,\ref{table:Fit results for the scaling relations}). 
The fitted power-law index is consistent, within the uncertainty, with the value of 0.38 reported by  \citet{Larson1981}. 
The index is also somewhat lower than the value ($\sim$\,0.5) commonly reported for some samples of large-scale molecular clouds ($R_{\rm eff}$\,$\gtrsim$\,1\,pc; \citealt{Solomon1987,Sun2017,Sun2024a}). 
At the clump scale, the power-law index is notably less than 0.5, indicating a large scatter in the velocity dispersion \citep{Heyer2001,Falgarone2009}. 
This may indicate that the growth of velocity dispersion slows on clump scales, as turbulence injected from large clouds is dissipated at small-scale clumps \citep{Falgarone2009}. 
As shown in Fig.\,\ref{fig:Scaling-Relation}, the power-law index derived from the Galactic edge clumps in the first quadrant is close to that in the second quadrant ($a$\,=\,0.36), but both are shallower than that of the combined sample ($a$\,=\,0.52). 
The latter is consistent with the value of 0.5 found for large-scale molecular clouds by \citet{Solomon1987}. 
The coefficient $v_0$ of the $\sigma_{\rm v}$--$R$ relation, which is 0.98 in the first-quadrant Galactic edge clumps (see Fig.\,\ref{fig:Scaling-Relation}), is higher than that in the second Galactic quadrant (0.69; \citealt{Luo2026}). 
The two samples have different distributions in the $\sigma_{\rm v}$--$R$ parameter space, where the second-quadrant sample contains more clumps with low size and low velocity dispersion, while the first-quadrant sample is dominated by relatively high velocity dispersion clumps. 

As star formation activity increases, there is a corresponding increase in the $v_0$ value \citep{Wong2017,Wong2019}. 
In regions of high density, large velocity dispersion may be triggered by the energetic feedback from recent star formation, or by the collapse caused by gravity-driven motions. 
The two anomalously large velocity dispersions of 4.5 and 5.4\,km\,s$^{-1}$ correspond to clumps with column densities of 2.2\,$\times$\,$10^{22}$ and 3.1\,$\times$\,$10^{22}$\,cm$^{-2}$, respectively, in G20.442+0.350 and G44.500+0.340 (see Table\,\ref{table:clump parameters}), identified from CO\,(3--2) data at a resolution of 16.6$^{\prime\prime}$. 
The correlations indicate that regions of high column density exhibit large velocity dispersions (see Fig.\,\ref{fig:Scaling-Relation}). 
Nevertheless, the two clumps with anomalously high velocity dispersions have virial parameters exceeding 5 (see Table\,\ref{table:clump parameters}), suggesting that they may not be gravitationally bound and require external pressure. 
In addition, velocity dispersion increases with the mass of the clump (see Fig.\,\ref{fig:Scaling-Relation}), suggesting that more massive clumps exhibit stronger internal support against turbulent motions. 
At a given clump mass or column density, first-quadrant clumps exhibit larger velocity dispersions on average than second-quadrant clumps, indicating a potential difference in their dynamical properties.  
Note that the first-quadrant sample is incomplete for clouds with line widths below the velocity resolution (0.64\,km\,s$^{-1}$; see Sect.\,\ref{sect:CO(3-2) data}), while opacity broadening may also contribute to the measured line widths. 
This could result in the clump velocity dispersion of the first Galactic quadrant being systematically larger than that of the second Galactic quadrant. 
In addition, the difference in heliocentric distance between the two quadrants may introduce an observational bias such that larger and more turbulent clumps are preferentially discerned in the first Galactic quadrant. 

\subsubsection{Mass--size relation}
\label{sect:Mass-size relation}
The relationship between mass and size in molecular clouds can also be described by a power law, represented as $M$\,=\,$m_0 R_{\rm eff}^b$ (hereafter $M$--$R$ relation). 
The $M$--$R$ relation has been observed in molecular clouds within both the Milky Way (e.g., \citealt{Kauffmann2010a,Kauffmann2010b,Kauffmann2010c}) and nearby galaxies (e.g., \citealt{Faesi2018,Lada2024}). 
Previous studies suggest that the power-law index $b$ commonly lies at 1--3. 
The $b$\,=\,2 implies that the mean column density remains nearly constant, whereas $b$\,=\,3 indicates that the volume density is uniform for different cloud radii \citep{Larson1981}. 
Mean column density escalates with increasing radius when $b$\,>\,2, implying that larger molecular clouds exhibit greater compactness. 
In contrast, mean column density decreases as the radius increases when $b$\,<\,2 \citep{Kauffmann2010a,Kauffmann2010b}. 

\begin{figure}[t]
\centering
\includegraphics[width=0.45\textwidth]{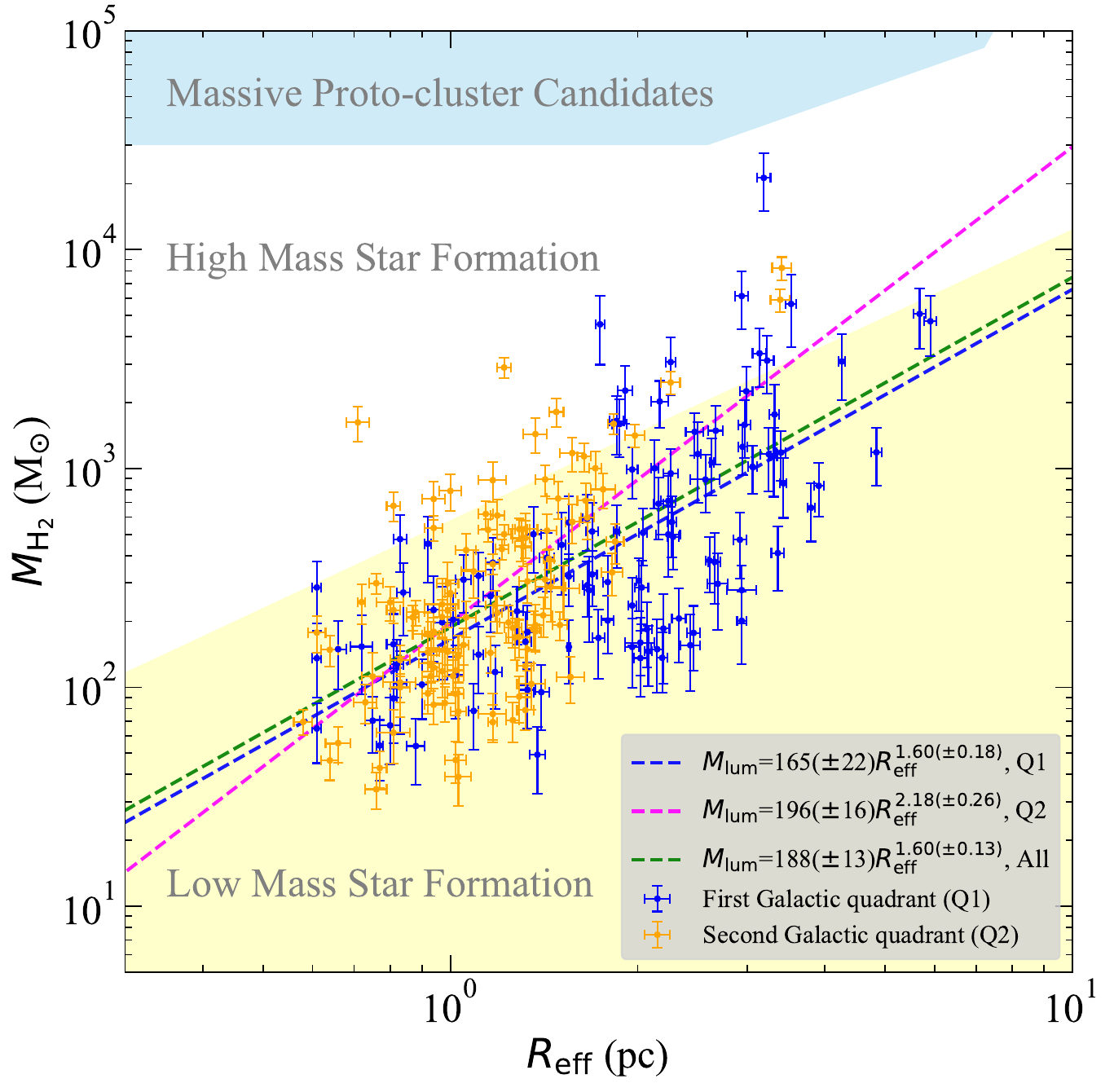}
\caption{The $M$--$R$ relation for CO clumps at the Galactic edge. 
The yellow-shaded region indicates the parameter space below the empirical massive-star-formation criterion, where $M$\,$\leq$\,580\,M$_{\odot}$\,($R_{\rm eff}$/pc)$^{1.33}$ \citep{Kauffmann2010c}. 
The cyan-shaded region represents the parameter space associated with massive proto-cluster candidates \citep{Bressert2012}. 
The blue, magenta, and green dashed lines represent the fits obtained for the Galactic edge clumps in the first quadrant (this work), second quadrant \citep{Luo2026}, and both combined, respectively. 
} 
\label{fig:mass-size ralation} 
\end{figure}

For Galactic edge clumps in the first Galactic quadrant (see Fig.\,\ref{fig:mass-size ralation}), the $M$--$R$ relation is fitted to $M_{\rm lum}$\,=\,165($\pm$22)$R_{\rm eff}^{1.60(\pm0.18)}$, indicating a strong correlation as evidenced by the Pearson correlation coefficient of 0.67. 
A $p$-value less than $10^{-3}$ indicates high significance (see Table\,\ref{table:Fit results for the scaling relations}). 
The value of $b$ is 1.60\,$\pm$\,0.18, indicating that the column density shows only a weak dependence on clump size. 
This may be attributed to the difficulty in probing the outer layer of the molecular clouds with CO\,(3--2), which results in the underestimation of the clump mass in low-metallicity environments. 
In this environment, CO is readily photodissociated by UV radiation, given the lack of dust obstruction \citep{ONeill2022a}. 
When the $M$--$R$ relation follows $M$\,=\,$m_0 R_{\rm eff}^2$, the corresponding average surface density is $m_0$/$\pi$. 
For the first-quadrant Galactic edge clumps, the $m_0$/$\pi$ value (53\,M$_{\odot}$\,pc$^{-2}$) is comparable to the typical clump surface density (43\,M$_{\odot}$\,pc$^{-2}$; see Table\,\ref{table:statistics of physical parameters}). 
The outer Galactic clouds exhibit smaller size, lower mass, and narrower line width compared to those found within the inner Galaxy \citep{May1997}. 
The empirical threshold for massive star formation typically conforms to the relation $M(R)$\,>\,580\,M$_{\odot}$\,($R$/pc)$^{1.33}$ \citep{Kauffmann2010c}. 
The normalization of 580\,M$_{\odot}$ corresponds to the threshold mass at a physical size of $R$\,=\,1\,pc. 
Most of the Galactic edge clumps lie below this relation (see Fig.\,\ref{fig:mass-size ralation}), indicating that they do not satisfy this empirical criterion for massive star formation. 
However, the presence of a significant amount of CO-dark molecular gas in the outskirts of the Galaxy may lead to an underestimation of their total molecular masses and therefore some clumps might actually surpass this threshold. 

As shown in Fig.\,\ref{fig:mass-size ralation}, the $M$--$R$ relation of the Galactic edge clumps shows a shallower power-law index in the first quadrant (1.60) compared to the second quadrant (2.18). 
This suggests that smaller clumps in the first-quadrant sample tend to have higher surface densities, whereas the surface density shows little dependence on clump size in the second-quadrant sample. 
In contrast, the second-quadrant Galactic edge clumps are closer to a constant surface density. 
In the second-quadrant Galactic edge ($R_{\rm g}$\,$\sim$\,14--23\,kpc), as compared to the first-quadrant Galactic edge ($R_{\rm g}$\,$\sim$\,10--15\,kpc), the significance of CO-dark gas increases \citep{Wolfire2010,Pineda2013}. 
Therefore, CO preferentially traces dense and well-shielded regions, whereas extended low-density structures are more likely to be missed \citep{Bolatto2013}. 
This incompleteness can bias the observed mass–size relation by preferentially excluding low-mass clumps with large size. 
This results in a steeper $M$--$R$ relation for the second-quadrant Galactic edge clumps. 
In addition, the power-law index derived from the combined sample (1.60) is nearly identical to that of the first quadrant alone, despite the much steeper index obtained for the second quadrant (see Fig.\,\ref{fig:mass-size ralation}). 
This suggests that the global $M$--$R$ relation is dominated by the first quadrant sample, which covers a wider range in size and mass. 

\begin{figure*}[t]
\centering
\includegraphics[width=0.95\textwidth]{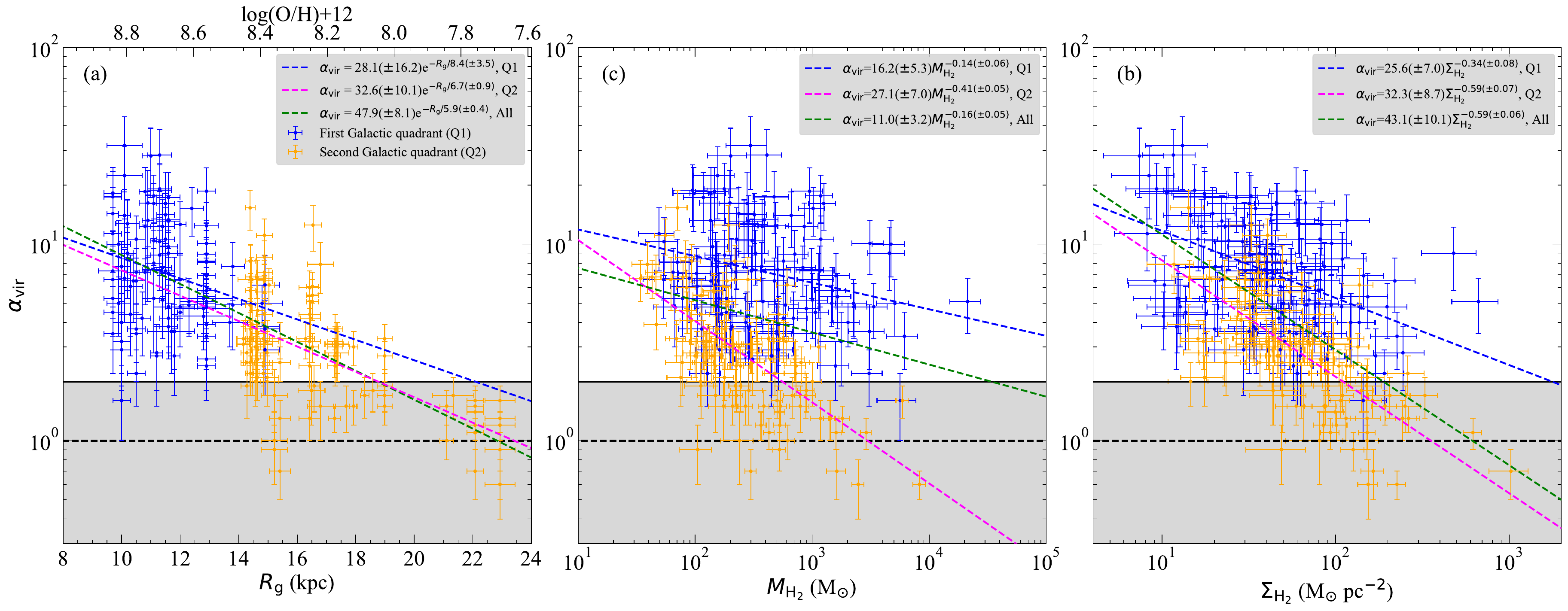}
\caption{Variation of the clump virial parameter with Galactocentric distance (a), mass (b), and surface density (c). 
The blue, magenta, and green dashed lines represent the fits obtained for the Galactic edge clumps in the first quadrant (this work), second quadrant \citep{Luo2026}, and both combined, respectively. 
The horizontal blue dashed and solid lines denote $\alpha_{\rm vir}$\,=\,1 and 2, respectively. 
} 
\label{fig:Rg-alpha} 
\end{figure*}

\subsection{Dynamical state of the Galactic edge clumps}
\label{sect:Dynamical state of the clumps}
\subsubsection{Virial parameter}
\label{sect:Virial parameter}
The dynamical state of molecular clouds is normally evaluated by the virial parameter (see Eq.\,\ref{eq:virial parameter}; e.g., \citealt{Kauffmann2013,Lin2025}). 
A critical virial parameter ($\alpha_{\rm cri}$) is used to distinguish molecular clouds as either gravitationally bound or unbound. 
Subcritical clouds, defined by $\alpha_{\rm vir}$\,<\,$\alpha_{\rm cri}$, are gravitationally bound and will collapse upon perturbation. 
Supercritical clouds, defined by $\alpha_{\rm vir}$\,>\,$\alpha_{\rm cri}$, are unbound and thus subject to dissipation. 
However, the critical virial parameter, depending on the cloud's internal structure and its external environment (e.g., density structure, magnetic fields, and external pressure), is not constant. 
The $\alpha_{\rm cri}$ is equated to 1, taking into account only the internal pressure and self-gravity in a scenario involving a homogeneous and spherical density sphere. 
$\alpha_{\rm cri}$ is 2, taking into account comprehensive perturbations in both external pressure and density gradients \citep{Kauffmann2013}. 
The observed virial parameter in our Galactic edge clumps varies from 1.6 to 31.7, with the majority concentrated between 2 and 8 (see Tables\,\ref{table:statistics of physical parameters} and \ref{table:clump parameters}, or Sect.\,\ref{sect:Clump identification}). 
The median virial parameter is 6.8\,$\pm$\,2.4, with a mean of 8.7\,$\pm$\,3.0 (see Table\,\ref{table:statistics of physical parameters}), which is larger than the typical value for that in the second Galactic quadrant (2.8\,$\pm$\,0.6; \citealt{Luo2026}). 
Most $\alpha_{\rm vir}$ values lie above the commonly adopted critical range of $\sim$\,1--2, suggesting that the majority of clumps located at the Galactic periphery are gravitationally unbound. 
However, the presence of CO-dark H$_2$ gas can lead to an underestimation of the molecular gas mass and a consequent overestimation of the virial parameter. 

Radial gradients in the virial parameter as a function of Galactocentric distance have been reported in previous studies of the outer Galaxy (e.g., \citealt{Sun2024a,Lin2025,Luo2025}). 
In the first Galactic quadrant, $\alpha_{\rm vir}$ exhibits a decrease from Galactocentric distance of 10 to 15\,kpc (see Fig.\,\ref{fig:Rg-alpha}), as represented by the exponential function $\alpha_{\rm vir}$\,=\,28.1($\pm$\,16.2)e$^{-R_{\rm g}/8.4(\pm3.5)}$. 
The Pearson correlation coefficient of $-0.27$ indicates a weak negative correlation, while the corresponding $p$-value of 0.004 demonstrates that the correlation is very significant. 
In the second Galactic quadrant, $\alpha_{\rm vir}$ decreases exponentially with increasing Galactocentric distance from 14 to 23\,kpc, which can be expressed as $\alpha_{\rm vir}$\,=\,32.6($\pm$\,10.1)e$^{-R_{\rm g}/6.7(\pm0.9)}$ \citep{Luo2026}. 
Combined with the first and second Galactic quadrant samples (see Fig.\,\ref{fig:Rg-alpha}), the exponential relation is expressed as $\alpha_{\rm vir}$\,=\,47.9($\pm$\,8.1)e$^{-R_{\rm g}/5.9(\pm0.4)}$, exhibiting a strong (Pearson's coefficient of $-0.72$) and highly significant correlation ($p$-value\,<\,10$^{-3}$). 
Analogously, there is a gradual decrease in the virial parameters, transitioning from the metal-rich inner Galaxy to the metal-poor outer Galaxy (3\,<\,$R_{\rm g}$\,<\,25\,kpc; \citealt{Lin2025}). 
This implies that turbulence may decline at the periphery of the Galaxy, which is characterized by a metal-deficient environment. 

\begin{figure}[t]
\centering
\includegraphics[width=0.45\textwidth]{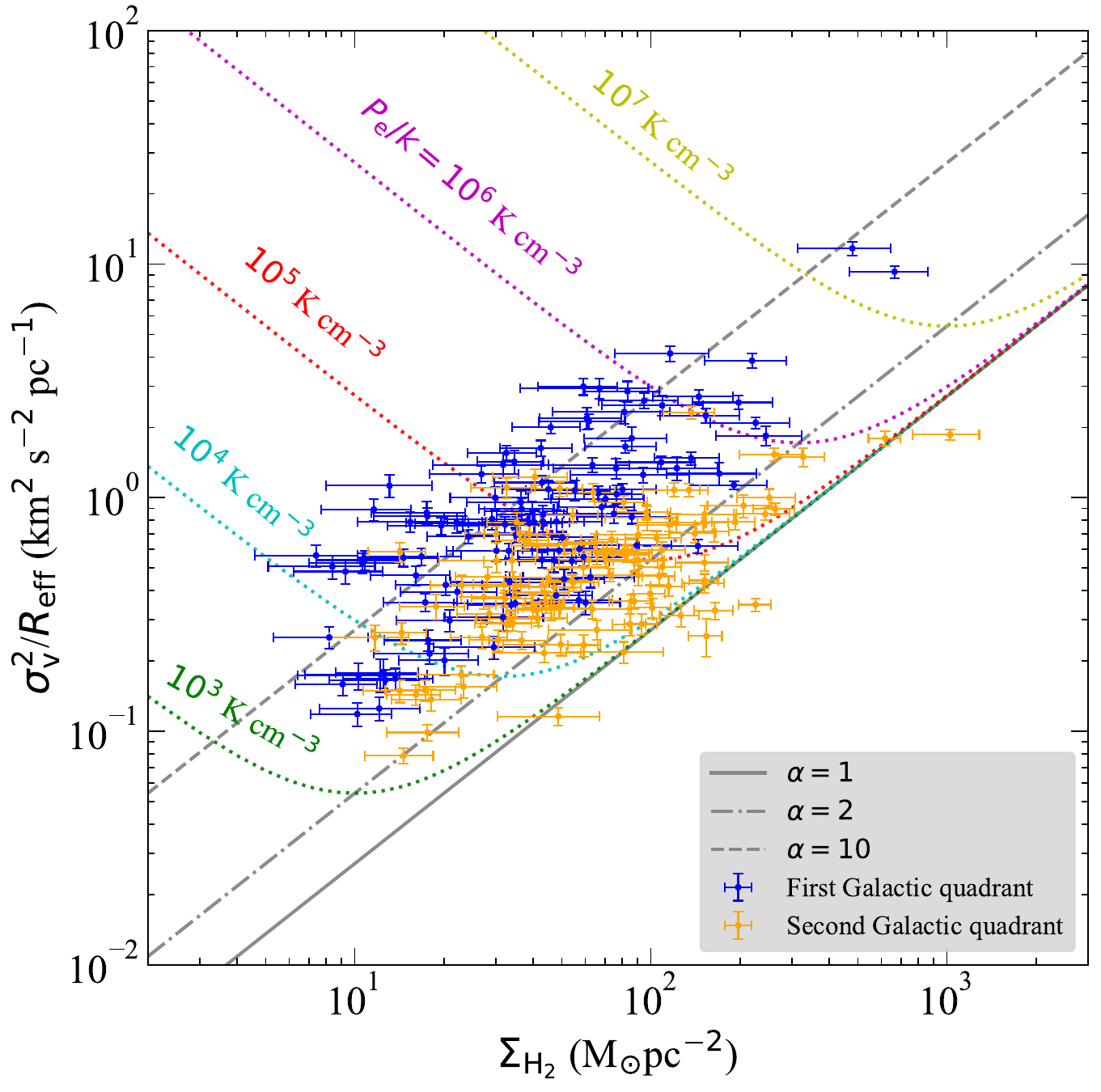}
\caption{Scaling coefficient ($\sigma_{\rm v}^2$/$R_{\rm eff}$) as a function of the mass surface density ($\Sigma_{\rm H_2}$). 
Results are obtained from the Galactic edge clumps of the first Galactic quadrant (blue, this work) and second Galactic quadrant (orange, \citealt{Luo2026}). 
The linear solid line represents virialized clouds in the absence of pressure ($\alpha_{\rm vir}$\,=\,1). The linear dash-dotted and dashed lines correspond to $\alpha_{\rm vir}$ values of 2 and 10, respectively. The dotted colored curves indicate gravitational virial conditions for clouds confined by external pressures ($P_{\rm e}/k$) from $10^3$ to $10^7$\,K\,cm$^{-3}$ (details see Sect.\,\ref{sect:Dynamical state of the clumps}). } 
\label{fig:size-linewidth-surface density} 
\end{figure}

\subsubsection{Gravitational boundedness}
\label{sect:Gravitational boundedness}
Massive clumps are often more strongly gravitationally bound than lower-mass clumps \citep{Kauffmann2013,Urquhart2014}. 
Prior studies suggest that the virial parameter decreases as the clump mass increases, following a power-law equation $\alpha_{\rm vir}$\,=\,$iM_{\rm H_2}^{j}$ (--1\,<\,$j$\,<\,0; e.g., \citealt{Kauffmann2013,Urquhart2014}). 
A weak correlation (Pearson's coefficient of $-0.19$) is observed in our first Galactic quadrant sample ($\alpha_{\rm vir}$\,=\,16.2($\pm$5.3)$M_{\rm H_2}^{-0.14(\pm0.06)}$; Fig.\,\ref{fig:Rg-alpha}), with statistical significance ($p$-value $\sim$\,0.05). 
In contrast, the second Galactic quadrant sample follows a steeper relation, $\alpha_{\rm vir}$\,=\,27.1($\pm$7.0)$M_{\rm H_2}^{-0.41(\pm0.05)}$. 
When the two samples are combined, the overall correlation becomes substantially weaker ($\alpha_{\rm vir}$\,=\,11.0($\pm$3.2)$M_{\rm H_2}^{-0.16(\pm0.05)}$), resembling that of the first-quadrant sample. 
This may result from the systematic differences between the two samples, including their environmental conditions, spatial resolutions, and parameter distributions, which partially obscure the underlying trend. 
Clumps with masses above $\sim$\,10$^3$\,M$_\odot$ are more likely to be gravitationally bound in the second Galactic quadrant, whereas first-quadrant clumps may require higher masses to reach a comparable degree of gravitational boundedness (see Fig.\,\ref{fig:Rg-alpha}). 
Additionally, as shown in Fig.\,\ref{fig:Rg-alpha}, a power-law correlation between the virial parameter and surface density is also found for the Galactic edge clumps in both the first ($\alpha_{\rm vir}$\,=\,25.6($\pm$7.0)$\Sigma_{\rm H_2}^{-0.34(\pm0.08)}$) and second ($\alpha_{\rm vir}$\,=\,32.3($\pm$8.7)$\Sigma_{\rm H_2}^{-0.59(\pm0.07)}$) Galactic quadrant samples, as well as in the combined sample ($\alpha_{\rm vir}$\,=\,43.1($\pm$10.1)$\Sigma_{\rm H_2}^{-0.59(\pm0.06)}$). 
The correlation is highly significant, with a $p$-value less than 10$^{-3}$ and a Pearson's coefficient greater than 0.4. 
Low surface density clumps ($\lesssim$\,100\,M$_{\odot}$\,pc$^{-2}$) are characterized by internal kinetic energy that is more significant relative to their self-gravity, rendering them less gravitationally bound. 
In high-density clumps ($\gtrsim$\,100\,M$_{\odot}$\,pc$^{-2}$) with virial parameters typically near the critical value of 1--2, self-gravity is dynamically more significant \citep{Hunt2026}. 
This indicates that more massive and denser clumps tend to exhibit a lower virial parameter, implying a higher degree of gravitational boundedness. 
Furthermore, only dense cores are observed in metal-poor environments due to the significant presence of CO-dark H$_2$ gas \citep{ONeill2022a}. 
This could potentially lead to a decrease in the luminosity-based mass as derived from the CO tracer. 
Consequently, it may appear that the virial parameter increases as the luminous mass decreases \citep{Luo2026}. 

\subsubsection{Pressure-confined equilibrium}
\label{sect:Pressure-confined equilibrium}
In the context of an isolated, self-gravitating, and isothermal spherical cloud, the pressure-bounded virial equilibrium \citep{Field2011} adheres to the equation 
\begin{align}
v_0^2 = \frac{\sigma_{\rm v}^{2}}{R_{\rm eff}} = \frac{1}{3}(\frac{3 \pi \rm G \Sigma_{\rm H_2}}{5}+\frac{4P_{\rm e}}{\Sigma_{\rm H_2}}) \,,
\label{eq:size-mass-linewidth-pressure} 
\end{align} 
where $P_{\rm e}$ denotes external pressure, and $v_0$ is the coefficient of the $\sigma_{\rm v}$--$R$ relation (see Sect.\,\ref{sect:Line width-size relation}). 
Both the first- and second-quadrant samples exhibit a positive correlation between $\sigma_{\rm v}^{2}$/$R_{\rm eff}$ and $\Sigma_{\rm H_2}$ (see Fig.\,\ref{fig:size-linewidth-surface density}), suggesting that molecular clumps with higher surface densities tend to have larger internal dynamical pressure terms. 
Relative to the second-quadrant sample, first-quadrant clumps exhibit larger $v_0$ values at a given surface density, indicating potentially enhanced internal motions. 
As illustrated in Fig.\,\ref{fig:size-linewidth-surface density}, the external pressures mainly fall within the range of $10^4$--$10^6$\,K\,cm$^{-3}$ for our first-quadrant Galactic edge clumps. 
These pressures are comparable in magnitude to those reported for Galactic edge clumps in the second quadrant \citep{Luo2026} and star-forming regions in the LMC \citep{Wong2022}, although they are evidently lower than those found in molecular clouds in the Galactic center \citep{Battersby2025} and the overlap regions of interacting galaxies \citep{Krahm2024}. 
For our Galactic edge clumps, the virial state may depend on the level of external pressure. 
The equilibrium state might be met when the data points align with the external pressure curve. 

\begin{figure}[t]
\centering
\includegraphics[width=0.45\textwidth]{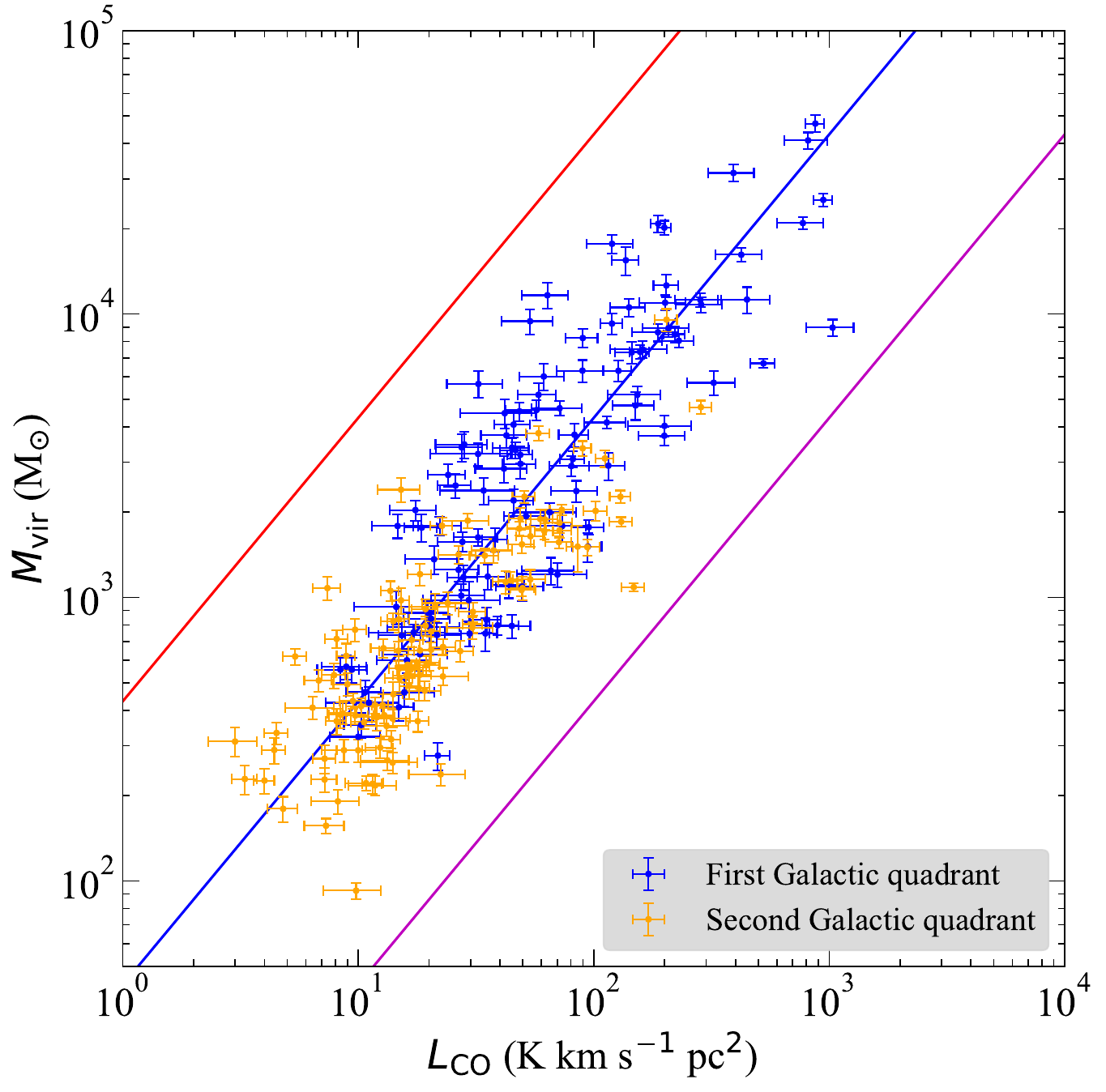}
\caption{Virial mass ($M_{\rm vir}$) of molecular clouds as a function of CO luminosity. 
Results are obtained from molecular clouds located in the first Galactic quadrant (blue, this work) and second Galactic quadrant (orange; \citealt{Luo2026}).
The magenta, blue, and red lines denote the $X_{\rm CO}$ factors of $\sim$\,2, 20, and 200\,$\times$\,10$^{20}$\,cm$^{-2}$\,(K\,km s$^{-1}$)$^{-1}$, corresponding to $\alpha_{\rm CO}$ values of $\sim$\,4.3, 43, and 430\,M$_{\odot}$\,(K\,km\,s$^{-1}$\,pc$^2$)$^{-1}$). 
This comparison assumes that $M_{\rm vir}$ is a reasonable proxy for the true molecular cloud mass, which is expected if the clouds are approximately in virial equilibrium. }
\label{fig:lum-Mvir} 
\end{figure}

\subsection{Factors affecting the CO-to-H$_2$ conversion factor}
\label{sect:The impact of metallicity and CO-dark gas}
Metallicity plays a crucial role in determining molecular cloud properties, as it regulates CO abundance, dust shielding, and gas cooling \citep{Leroy2011,Bolatto2013}. 
The oxygen abundance gradient across the Galactic disk exhibits a slope of --0.052($\pm$0.004)\,dex\,kpc$^{-1}$ \citep{Wenger2019}. 
Consequently, molecular clouds in the outer Galaxy reside in a relatively low-metallicity environment.
In metal-poor clouds, the reduced dust abundance weakens shielding against UV radiation, allowing photons to penetrate deeper into molecular gas \citep{Amorin2016}. 
As a result, CO is more susceptible to photodissociation than H$_2$, while H$_2$ can remain abundant owing to its efficient self-shielding \citep{Bolatto2013,Shi2016,Rodriguez-Ambler2026}. 
This leads to a layered cloud structure consisting of a CO-bright central region surrounded by an extended diffuse H$_2$ envelope that emits weakly or remains undetected in CO, commonly referred to as CO-dark H$_2$ gas \citep{Wolfire2010,Madden2020}. 
As the metallicity decreases, the fraction of CO-dark H$_2$ increases and CO emission traces an increasingly smaller fraction of the total molecular gas reservoir. 
Consequently, molecular cloud masses inferred from CO observations alone may be systematically underestimated \citep{Elmegreen2013,Rubio2015,Bisbas2025}. 
This effect is reflected in the metallicity dependence of the CO-to-H$_2$ conversion factor ($X_{\rm CO}$), which tends to increase toward lower metallicities. 

Across the Galactic disk, the metallicity decreases by a factor of $\sim$\,3 between Galactocentric radii of 8 and 16\,kpc, resulting in a comparable increase in $X_{\rm CO}$ (see Sect.\,\ref{sect:Clump identification}). 
In this work, the metallicity-corrected $X_{\rm CO}$ factor ranges from (2.4\,$\pm$\,0.7)\,$\times$\,10$^{20}$\,cm$^{-2}$\,(K\,km s$^{-1}$)$^{-1}$ to (4.8\,$\pm$\,0.8)\,$\times$\,10$^{20}$\,cm$^{-2}$\,(K\,km s$^{-1}$)$^{-1}$ for Galactocentric distances of 10 to 15\,kpc, with a mean of (3.1\,$\pm$\,0.7)\,$\times$\,10$^{20}$\,cm$^{-2}$\,(K\,km s$^{-1}$)$^{-1}$ and a median of (3.0\,$\pm$\,0.7)\,$\times$\,10$^{20}$\,cm$^{-2}$\,(K\,km s$^{-1}$)$^{-1}$. 
In addition, total mass ($M_{\rm total}$) of molecular clouds can be corrected using the relation $M_{\rm total}$\,=\,$M_{\rm CO}$/(1-$f_{\rm dark}$), where $M_{\rm CO}$ is the CO-derived mass and $f_{\rm dark}$ represents the fraction of CO-dark H$_2$ gas. 
Previous observations indicate that the $f_{\rm dark}$ values increase with Galactocentric distance, rising from $\sim$\,0.2 at 4\,kpc to $\sim$\,0.8 at 10\,kpc \citep{Pineda2013}. 
This corresponds to mass correction factors (1/(1-$f_{\rm dark}$)) of $\sim$\,1.25 at a Galactocentric distance of 4\,kpc, increasing to $\sim$\,5 at 10\,kpc. 
The $f_{\rm dark}$ value is approximately 0.6 in the Solar neighborhood \citep{Paradis2012} and is expected to be further enhanced in the metal-poor environments of the Galactic outskirts \citep{Langer2014}. 
When the $f_{\rm dark}$ value reaches 0.96 at the edge of the Galaxy, the corresponding mass correction factor is $\sim$\,25, which is ten times larger than 2.5 in the Solar neighborhood. 
This result is consistent with the $X_{\rm CO}$ factor in Galactic edge clumps being approximately an order of magnitude higher than that in the Solar neighborhood (see Fig.\,\ref{fig:lum-Mvir}). 
The presence of CO-dark H$_2$ gas leads to a systematic underestimation of cloud sizes based on CO identification. 
In contrast, velocity dispersion primarily originates from the dense central regions and is less affected, resulting in larger coefficient $v_0$ for the $\sigma_{\rm v}$--$R$ relation \citep{ONeill2022a}. 
The CO-dark gas effect becomes more prevalent at larger Galactocentric distances, resulting in steeper $M$--$R$ relations and biased molecular surface densities (see Sect.\,\ref{sect:Mass-size relation}). 

The CO-to-H$_2$ conversion factor depends not only on metallicity but also on the beam size, particularly when the CO-emitting structures are not fully spatially resolved \citep{Rubio1991,Rubio1993,Schruba2017,Henkel2022}. 
Beam dilution reduces the observed CO intensity, resulting in an increased virial-based estimate $X_{\rm CO}$ without necessarily implying a genuinely higher conversion factor. 
As illustrated in Fig.\,\ref{fig:lum-Mvir}, the first-quadrant samples extend to significantly higher CO luminosities and virial masses than the second-quadrant samples, with both samples exhibiting a strong positive correlation. 
This indicates that the first-quadrant samples exhibit a larger population of massive molecular clumps and that their greater heliocentric distances preferentially limit the detection of faint and low-mass clumps. 
Therefore, the $X_{\rm CO}$ factor may be underestimated for clumps at the Galactic edge or these clumps are highly gravitationally unbound. 
Furthermore, underestimating the gas mass results in inflated virial parameters, which may lead to misidentification of the dynamical state of molecular clouds. 
Therefore, properly accounting for the impact of metallicity, CO-dark gas and beam size is essential to understanding the physical properties of molecular clouds at the edge of the Galaxy. 
The large virial parameters inferred for the Galactic edge clumps may reflect genuinely supervirial dynamics, an underestimate of the molecular mass associated with CO emission, or a combination of both. 
The present data do not uniquely distinguish these possibilities. 

\section{Summary}
\label{sect:summary}
We investigated properties of 84 Galactic edge clouds situated in the OSC arm within the first Galactic quadrant at 
Galactocentric distances ranging from 10 to 15\,kpc using CO\,$J$\,=\,1--0 and 3--2 lines obtained from the 
MWISP and COHRS surveys. The main results are as follows: 
\begin{enumerate}
\item 
84 Galactic edge clouds, as resolved by CO\,(3--2), exhibit compact, elongated, and irregular morphological structures.
110 clumps identified in these Galactic edge clouds have sizes, velocity dispersions, masses and surface densities in the ranges of 0.6--5.9\,pc, 0.5--5.4\,km\,s$^{-1}$, (0.5--213)\,$\times$\,$10^2$\,M$_\odot$, and 7--666\,M$_{\odot}$\,pc$^{-2}$, respectively. 

\item 
Over the Galactocentric distance range of 10--15\,kpc, no systematic variations are observed in the physical parameters of clumps at the Galactic edge in the first Galactic quadrant, including size, mass, volume density, velocity dispersion, CO luminosity, and surface density. 
The virial parameters show a decreasing trend with increasing Galactocentric distances, described by an exponential relation $\alpha_{\rm vir}$\,=\,28.1($\pm$\,16.2)e$^{-R_{\rm g}/8.4(\pm3.5)}$, similar to previous findings in the second Galactic quadrant. 

\item
For the 69 Galactic edge clouds with significant CO\,(3--2) emission after smoothing to 50$^{\prime\prime}$, the $R_{31}$ ratio ranges from 0.28 to 2.02, with a mean of 0.93\,$\pm$\,0.17 and a median of 0.74\,$\pm$\,0.13. 
A substantial fraction of Galactic edge clouds (23/69) exhibit $R_{31}$\,>\,1.0, suggesting that highly excited gas may be present at the Galactic edge. 
This is consistent with observations in the second-quadrant Galactic edge clouds. 
Nevertheless, within the range of $R_{\rm g}$\,$\sim$\,10--15\,kpc, no discernible trend is observed in $R_{31}$. 

\item
The $\sigma_v$--$R$ relation for first-quadrant Galactic edge clumps is fitted by $\sigma_{\rm v}$\,=\,0.98($\pm$0.06)$R_{\rm eff}^{0.35(\pm0.08)}$. 
This power-law index is similar to that of the second quadrant sample, implying a decreasing rate of velocity dispersion growth on clump scales. 
The $M$--$R$ relation of CO clumps is fitted by $M_{\rm lum}$\,=\,165($\pm$22)$R_{\rm eff}^{1.60(\pm0.18)}$, exhibiting a shallower power-law index in the first quadrant than that in the second Galactic quadrant. 
This suggests that column density decreases weakly while clump size increases. 

\item 
The virial parameter of the Galactic edge clumps ranges from 1.6 to 31.7, with a median of 6.8\,$\pm$\,2.4, although CO-dark H$_2$ gas may lead to an underestimation of their gas masses and thus to overestimated virial parameters. 
This suggests that most Galactic clumps are gravitationally unbound, while bound clumps require external pressures of $\sim$\,10$^4$--$10^6$\,K\,cm$^{-3}$. 
\end{enumerate}

The dynamical environment of molecular clouds at the outskirts of the Galaxy is relatively simple. 
Therefore, further investigations of molecular gas in these regions can reveal systematic variations of molecular cloud properties with Galactocentric distance and have the potential to provide important insights into star formation in the early Universe. 

\begin{acknowledgements}
The authors thank the anonymous referee for helpful comments. 
This work acknowledges the support of the National Key R\&D Program of China under grant Nos.\,2023YFA1608002 and 2022YFA1603103,
the Chinese Academy of Sciences (CAS) “Light of West China” Program under grant No.\,xbzg-zdsys-202212, and the Natural Science Foundation of Xinjiang Uygur Autonomous Region under grant No.\,2026D01D58. 
It was also partially supported by the Tianshan Talent Training Program of Xinjiang Uygur Autonomous Region under grant No.\,2024TSYCTD0013, the National Natural Science Foundation of China under grant Nos.\,12373029, 12403033, and 12463006, 
the Xinjiang Key Laboratory of Radio Astrophysics under grant No.\,2023D04033, the Natural Science Foundation of Xinjiang 
Uygur Autonomous Region under grant No.\,2022D01A359, the National SKA Program of China under grant No.\,2025SKA0140100, 
the Strategic Priority Research Program of the CAS under grant No.\,XDB0800301, and the Youth Innovation Promotion Association CAS.
C.\,Henkel acknowledges support by the Chinese Academy of Sciences President's International Fellowship Initiative under grant No.\,2025PVA0048. 
T.\,Liu, X.\,P.\,Chen, K.\,Wang, J.\,W.\,Wu, and G.\,Wu acknowledge support by the Tianchi Talent Program of Xinjiang Uygur Autonomous Region. 
T.\,Komesh acknowledges support from the Science Committee of the Ministry of Science and Higher Education of the Republic of Kazakhstan under grant No. AP26102915. 
MWISP is sponsored by the National Key R\&D Program of China with grant 2023YFA1608000 and the CAS Key Research Program of Frontier Sciences with grant QYZDJ-SSW-SLH047. 
This research has used NASA's Astrophysical Data System (ADS). 
\end{acknowledgements}
\bibliography{CO32} 
\begin{appendix}

\onecolumn
\section{CO\,(1--0) and CO\,(3--2) spectra}
\label{Sect:Spectra of CO(3--2)}
\setcounter{figure}{0} 
\begin{figure*}[!htbp]
\centering
\includegraphics[width=1.0\textwidth]{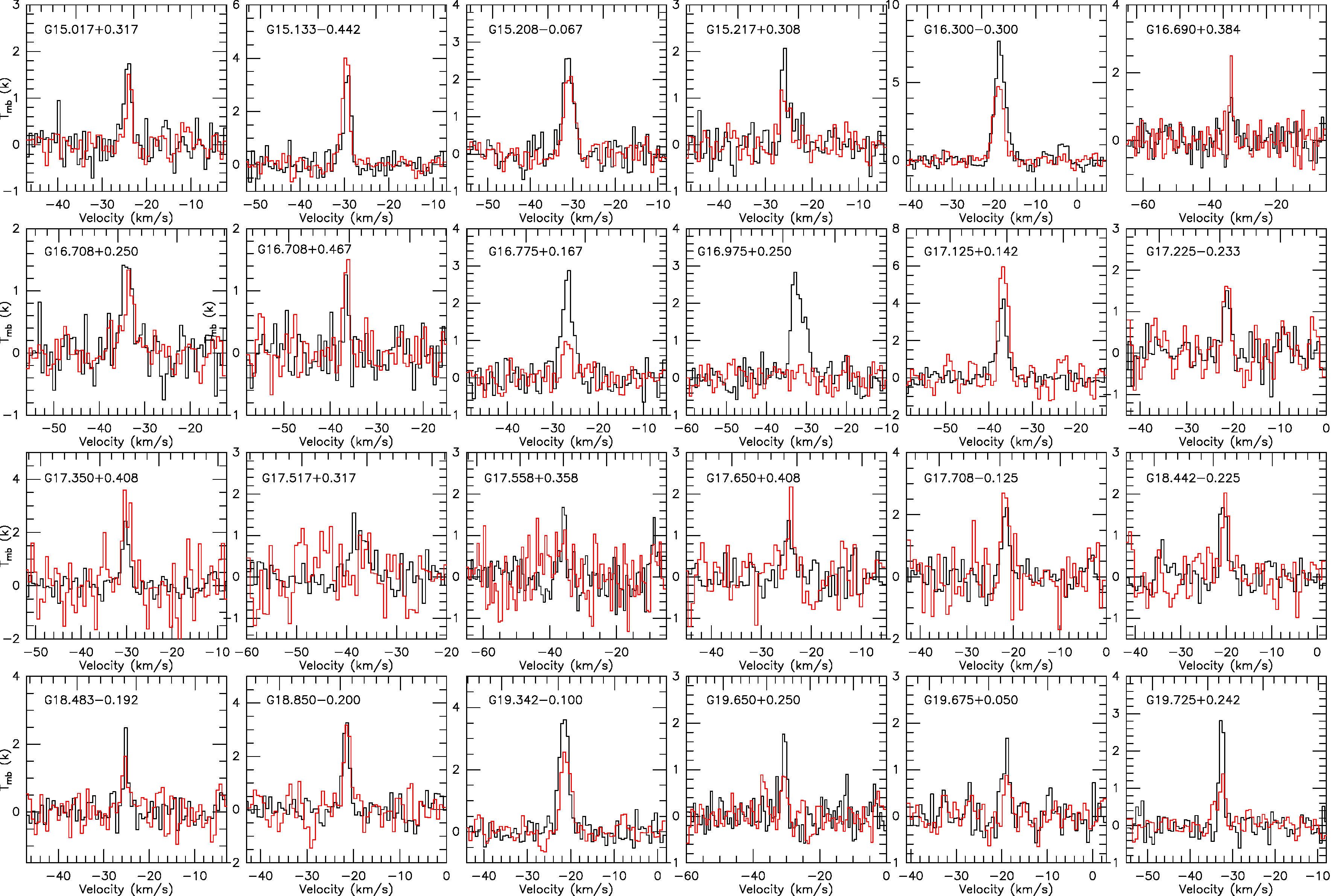}
\includegraphics[width=1.0\textwidth]{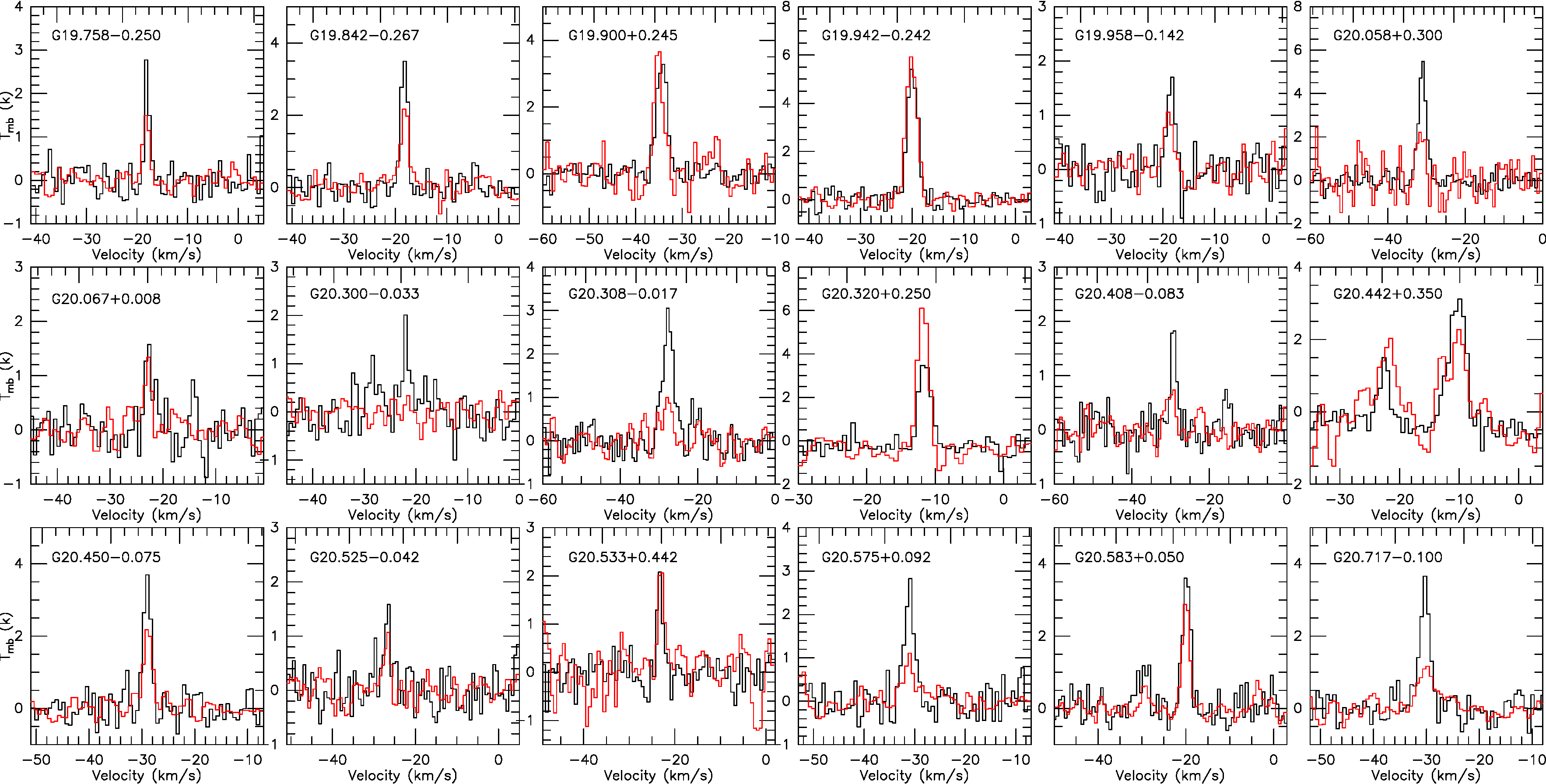}
\caption{CO\,(1--0) ({black line}) and CO\,(3--2) ({red line}) spectra toward 84 Galactic edge clouds. 
The CO\,(3--2) lines were extracted from the CO\,(1--0) emission peaks. 
The CO\,(3--2) line cubes have been smoothed to match the resolution of the CO\,(1--0) line, 50$^{\prime\prime}$, for comparison.}
\label{fig:CO(3-2) spectra}
\end{figure*}

\setcounter{figure}{0}
\begin{figure*}[!htbp]
\centering
\includegraphics[width=1.0\textwidth]{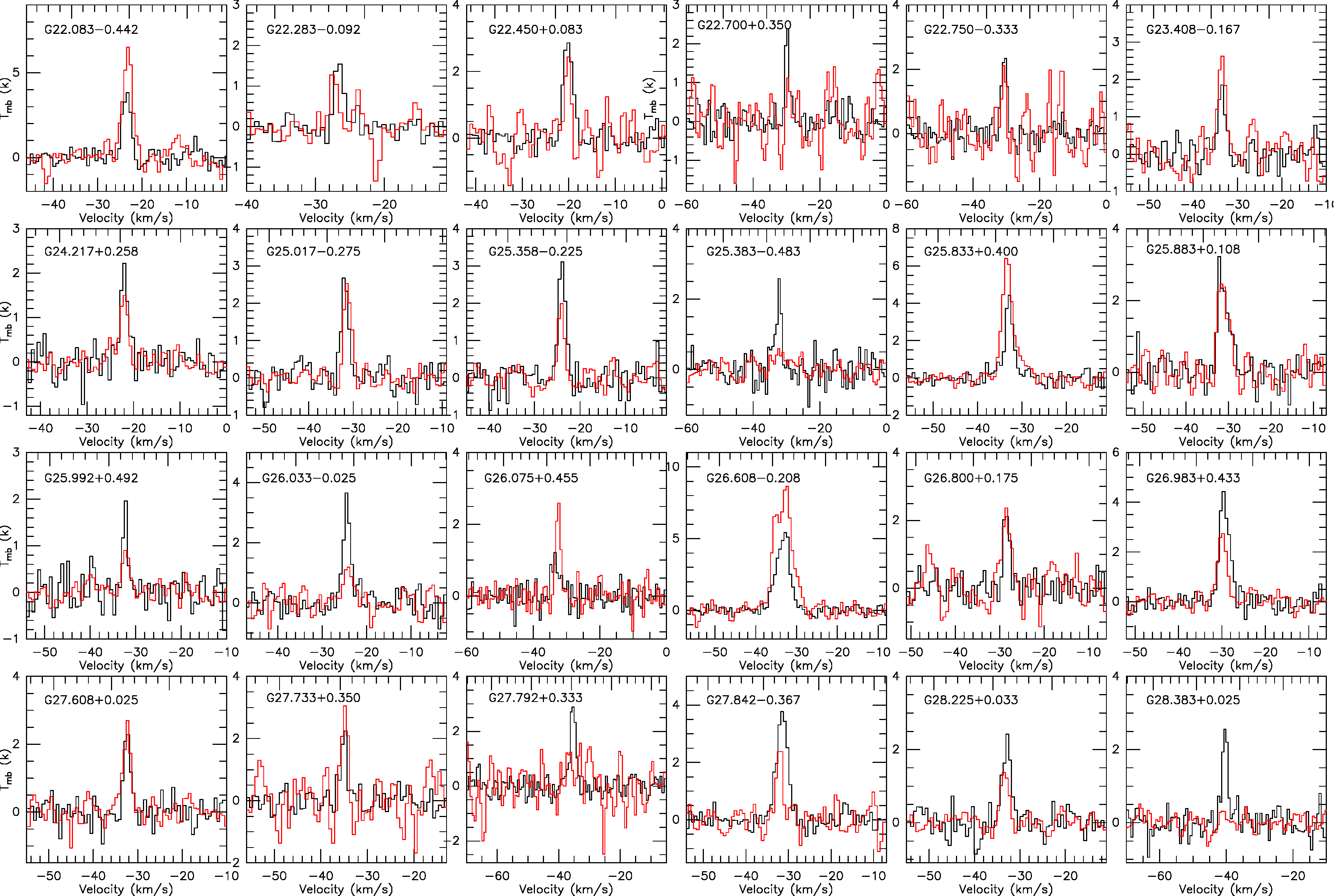}
\includegraphics[width=1.0\textwidth]{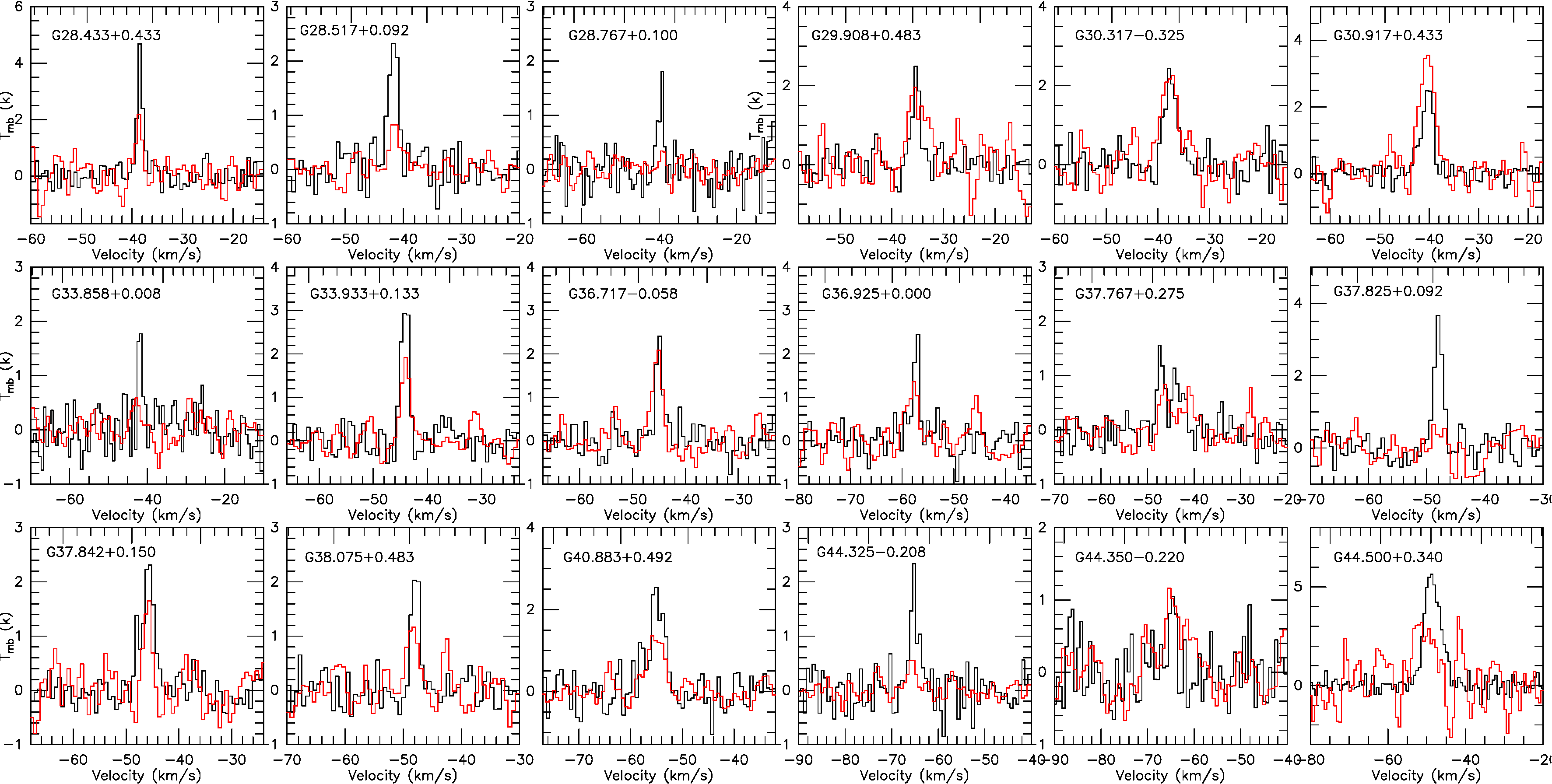}
\caption{Continued.}
\end{figure*}

\onecolumn
\begingroup
\setlength{\tabcolsep}{5.5pt}
\setcounter{table}{0} 
\begin{table*}[!htbp]
\scriptsize
\caption{CO spectral parameters.}
\centering
\begin{tabular}
{ccccccccccccc}
\hline\hline
Cloud name &R.A.(J2000) &Dec.(J2000)  &$V_{\rm LSR}$ &$D_{\rm s}$ &$R_{\rm g}$ &$T_{\rm CO(1-0)}$ &$\Delta V_{\rm CO(1-0)}$ &$I_{\rm CO(1-0)}$ &$T_{\rm CO(3-2)}$ &$\Delta V_{\rm CO(3-2)}$ &$I_{\rm CO(3-2)}$ &$R_{31}$\\
&(h m s) &($\circ$ $\prime$ $\prime\prime$) &km\,s$^{-1}$ &kpc &kpc &K &km\,s$^{-1}$ &K\,km\,s$^{-1}$ &K &km\,s$^{-1}$ &K\,km\,s$^{-1}$ &\\
(1) &(2) &(3) &(4) &(5) &(6) &(7) &(8) &(9) &(10) &(11) &(12) &(13)\\
\hline
G15.017+0.317  &18:16:44.1 &-15:44:16.47   &-24.3  &19.2\,$\pm$\,0.2  &11.6\,$\pm$\,0.3  &1.8\,$\pm$\,0.3  &1.9\,$\pm$\,0.4  &3.7 \,$\pm$\,0.5  &1.6\,$\pm$\,0.2  &1.3\,$\pm$\,0.2  &2.2 \,$\pm$\,0.3  &0.60\,$\pm$\,0.12\\
G15.133-0.442  &18:19:44.6 &-15:59:43.14   &-29.4  &20.5\,$\pm$\,1.0  &12.8\,$\pm$\,1.0  &3.5\,$\pm$\,0.3  &1.7\,$\pm$\,0.2  &6.5 \,$\pm$\,0.5  &4.2\,$\pm$\,0.2  &1.9\,$\pm$\,0.1  &8.5 \,$\pm$\,0.4  &1.31\,$\pm$\,0.12\\
G15.208-0.067  &18:18:31.0 &-15:45:06.94   &-30.9  &21.4\,$\pm$\,0.5  &13.7\,$\pm$\,0.6  &2.6\,$\pm$\,0.3  &2.8\,$\pm$\,0.2  &8.0 \,$\pm$\,0.6  &2.2\,$\pm$\,0.2  &2.8\,$\pm$\,0.2  &6.4 \,$\pm$\,0.4  &0.81\,$\pm$\,0.08\\
G15.217+0.308  &18:17:09.8 &-15:33:58.31   &-25.8  &19.7\,$\pm$\,0.5  &12.0\,$\pm$\,0.5  &2.2\,$\pm$\,0.3  &1.3\,$\pm$\,0.3  &3.0 \,$\pm$\,0.5  &1.1\,$\pm$\,0.2  &2.4\,$\pm$\,0.4  &2.7 \,$\pm$\,0.4  &0.93\,$\pm$\,0.20\\
G16.300-0.300  &18:21:30.4 &-14:53:56.94   &-18.6  &17.5\,$\pm$\,0.4  &9.9 \,$\pm$\,0.4  &7.2\,$\pm$\,0.4  &3.0\,$\pm$\,0.1  &22.9\,$\pm$\,0.8  &5.0\,$\pm$\,0.3  &2.8\,$\pm$\,0.1  &14.8\,$\pm$\,0.5  &0.64\,$\pm$\,0.03\\
G16.690+0.384  &18:19:46.4 &-14:13:58.11   &-35.5  &21.4\,$\pm$\,0.5  &13.8\,$\pm$\,0.6  &1.3\,$\pm$\,0.3  &1.2\,$\pm$\,0.3  &1.7 \,$\pm$\,0.4  &2.5\,$\pm$\,0.5  &1.0\,$\pm$\,0.1  &2.7 \,$\pm$\,0.4  &1.57\,$\pm$\,0.46\\
G16.708+0.250  &18:20:17.7 &-14:16:48.62   &-34.5  &21.4\,$\pm$\,0.4  &13.8\,$\pm$\,0.4  &1.5\,$\pm$\,0.3  &2.6\,$\pm$\,0.4  &4.2 \,$\pm$\,0.6  &1.1\,$\pm$\,0.2  &2.6\,$\pm$\,0.5  &3.1 \,$\pm$\,0.4  &0.73\,$\pm$\,0.15\\
G16.708+0.467  &18:19:30.4 &-14:10:39.88   &-36.3  &22.5\,$\pm$\,0.4  &14.9\,$\pm$\,0.5  &1.4\,$\pm$\,0.3  &0.9\,$\pm$\,0.3  &1.3 \,$\pm$\,0.3  &1.6\,$\pm$\,0.3  &1.4\,$\pm$\,0.2  &2.4 \,$\pm$\,0.4  &1.90\,$\pm$\,0.55\\
G16.775+0.167  &18:20:43.7 &-14:15:36.77   &-26.7  &19.3\,$\pm$\,0.7  &11.7\,$\pm$\,0.8  &2.7\,$\pm$\,0.3  &2.8\,$\pm$\,0.3  &8.0 \,$\pm$\,0.6  &1.1\,$\pm$\,0.2  &2.0\,$\pm$\,0.4  &2.2 \,$\pm$\,0.4  &0.28\,$\pm$\,0.05\\
G16.975+0.250  &18:20:48.9 &-14:02:40.92   &-33.3  &20.5\,$\pm$\,0.4  &12.9\,$\pm$\,0.4  &2.5\,$\pm$\,0.3  &4.1\,$\pm$\,0.3  &11.0\,$\pm$\,0.7  &(0.3)            &...              &<\,1.3            &<\,0.12          \\
G17.125+0.142  &18:21:29.9 &-13:57:47.56   &-36.9  &22.4\,$\pm$\,0.7  &14.8\,$\pm$\,0.7  &3.9\,$\pm$\,0.3  &2.4\,$\pm$\,0.2  &10.1\,$\pm$\,0.6  &6.3\,$\pm$\,0.6  &2.4\,$\pm$\,0.2  &16.0\,$\pm$\,1.1  &1.58\,$\pm$\,0.14\\
G17.225-0.233  &18:23:03.4 &-14:03:03.93   &-21.6  &17.9\,$\pm$\,0.2  &10.4\,$\pm$\,0.2  &1.6\,$\pm$\,0.3  &1.7\,$\pm$\,0.3  &2.8 \,$\pm$\,0.5  &1.9\,$\pm$\,0.4  &1.6\,$\pm$\,0.3  &3.2 \,$\pm$\,0.6  &1.12\,$\pm$\,0.30\\
G17.350+0.408  &18:20:58.1 &-13:38:22.16   &-30.0  &20.0\,$\pm$\,0.3  &12.5\,$\pm$\,0.3  &2.3\,$\pm$\,0.3  &1.7\,$\pm$\,0.2  &4.2 \,$\pm$\,0.5  &3.3\,$\pm$\,0.9  &2.3\,$\pm$\,0.5  &7.9 \,$\pm$\,1.5  &1.87\,$\pm$\,0.42\\
G17.517+0.317  &18:21:37.3 &-13:32:05.90   &-38.4  &22.2\,$\pm$\,0.3  &14.6\,$\pm$\,0.3  &1.1\,$\pm$\,0.3  &4.1\,$\pm$\,0.6  &4.8 \,$\pm$\,0.7  &(0.6)            &...              &<\,2.9            &<\,0.62          \\
G17.558+0.358  &18:21:33.2 &-13:28:46.23   &-35.7  &21.6\,$\pm$\,0.4  &14.1\,$\pm$\,0.4  &2.0\,$\pm$\,0.4  &1.1\,$\pm$\,0.4  &2.3 \,$\pm$\,0.5  &(0.6)            &...              &<\,1.5            &<\,0.63          \\
G17.650+0.408  &18:21:33.0 &-13:22:29.29   &-24.6  &18.6\,$\pm$\,0.4  &11.1\,$\pm$\,0.4  &1.2\,$\pm$\,0.3  &2.2\,$\pm$\,0.5  &2.7 \,$\pm$\,0.5  &2.2\,$\pm$\,0.5  &1.1\,$\pm$\,0.3  &2.5 \,$\pm$\,0.6  &0.92\,$\pm$\,0.29\\
G17.708-0.125  &18:23:35.8 &-13:34:25.78   &-21.6  &17.9\,$\pm$\,0.3  &10.4\,$\pm$\,0.3  &2.3\,$\pm$\,0.4  &1.7\,$\pm$\,0.3  &4.2 \,$\pm$\,0.6  &3.1\,$\pm$\,0.7  &1.5\,$\pm$\,0.3  &4.9 \,$\pm$\,0.9  &1.16\,$\pm$\,0.27\\
G18.442-0.225  &18:25:22.3 &-12:58:18.73   &-20.7  &17.5\,$\pm$\,0.2  &10.1\,$\pm$\,0.2  &1.9\,$\pm$\,0.3  &1.8\,$\pm$\,0.3  &3.5 \,$\pm$\,0.4  &2.1\,$\pm$\,0.4  &1.5\,$\pm$\,0.3  &3.4 \,$\pm$\,0.6  &0.97\,$\pm$\,0.22\\
G18.483-0.192  &18:25:19.8 &-12:55:12.73   &-25.2  &18.4\,$\pm$\,0.3  &11.0\,$\pm$\,0.3  &2.5\,$\pm$\,0.3  &1.0\,$\pm$\,0.1  &2.6 \,$\pm$\,0.3  &1.5\,$\pm$\,0.4  &1.7\,$\pm$\,0.7  &2.8 \,$\pm$\,0.8  &1.06\,$\pm$\,0.33\\
G18.850-0.200  &18:26:03.7 &-12:35:57.73   &-21.6  &17.6\,$\pm$\,0.3  &10.3\,$\pm$\,0.4  &3.4\,$\pm$\,0.4  &1.6\,$\pm$\,0.2  &6.0 \,$\pm$\,0.5  &3.3\,$\pm$\,0.5  &1.6\,$\pm$\,0.3  &5.8 \,$\pm$\,0.8  &0.97\,$\pm$\,0.16\\
G19.342-0.100  &18:26:38.3 &-12:07:02.98   &-21.9  &17.5\,$\pm$\,0.2  &10.1\,$\pm$\,0.2  &3.8\,$\pm$\,0.3  &2.5\,$\pm$\,0.1  &10.2\,$\pm$\,0.5  &2.6\,$\pm$\,0.3  &2.6\,$\pm$\,0.2  &7.2 \,$\pm$\,0.5  &0.71\,$\pm$\,0.06\\
G19.650+0.250  &18:25:57.6 &-11:40:54.59   &-30.9  &19.3\,$\pm$\,0.2  &12.0\,$\pm$\,0.2  &1.9\,$\pm$\,0.3  &1.5\,$\pm$\,0.2  &3.0 \,$\pm$\,0.4  &0.9\,$\pm$\,0.3  &1.8\,$\pm$\,0.4  &1.8 \,$\pm$\,0.4  &0.60\,$\pm$\,0.15\\
G19.675+0.050  &18:26:43.8 &-11:45:10.65   &-18.9  &17.1\,$\pm$\,0.1  &9.8 \,$\pm$\,0.2  &1.4\,$\pm$\,0.3  &1.9\,$\pm$\,0.3  &2.9 \,$\pm$\,0.5  &0.9\,$\pm$\,0.2  &1.6\,$\pm$\,0.4  &1.5 \,$\pm$\,0.3  &0.52\,$\pm$\,0.15\\
G19.725+0.242  &18:26:07.9 &-11:37:09.20   &-32.7  &19.6\,$\pm$\,0.2  &12.3\,$\pm$\,0.3  &3.2\,$\pm$\,0.3  &1.3\,$\pm$\,0.1  &4.3 \,$\pm$\,0.3  &1.1\,$\pm$\,0.2  &2.6\,$\pm$\,0.5  &3.0 \,$\pm$\,0.4  &0.68\,$\pm$\,0.11\\
G19.758-0.250  &18:27:58.4 &-11:49:08.87   &-18.0  &16.9\,$\pm$\,0.2  &9.7 \,$\pm$\,0.2  &3.0\,$\pm$\,0.3  &1.0\,$\pm$\,0.2  &3.1 \,$\pm$\,0.4  &1.6\,$\pm$\,0.2  &1.5\,$\pm$\,0.2  &2.5 \,$\pm$\,0.2  &0.80\,$\pm$\,0.12\\
G19.842-0.267  &18:28:11.7 &-11:45:09.58   &-18.3  &16.9\,$\pm$\,0.3  &9.7 \,$\pm$\,0.3  &3.7\,$\pm$\,0.3  &1.6\,$\pm$\,0.1  &6.2 \,$\pm$\,0.4  &2.2\,$\pm$\,0.2  &1.9\,$\pm$\,0.2  &4.4 \,$\pm$\,0.4  &0.71\,$\pm$\,0.08\\
G19.900+0.245  &18:26:27.2 &-11:27:46.84   &-32.7  &19.8\,$\pm$\,0.3  &12.4\,$\pm$\,0.4  &3.4\,$\pm$\,0.2  &2.7\,$\pm$\,0.1  &9.7 \,$\pm$\,0.4  &3.7\,$\pm$\,0.4  &2.2\,$\pm$\,0.3  &9.1 \,$\pm$\,0.5  &0.94\,$\pm$\,0.07\\
G19.942-0.242  &18:28:17.6 &-11:39:09.00   &-20.1  &16.9\,$\pm$\,0.5  &9.7 \,$\pm$\,0.5  &5.5\,$\pm$\,0.3  &2.4\,$\pm$\,0.1  &13.8\,$\pm$\,0.5  &6.0\,$\pm$\,0.2  &2.4\,$\pm$\,0.1  &15.5\,$\pm$\,0.4  &1.12\,$\pm$\,0.05\\
G19.958-0.142  &18:27:57.8 &-11:35:30.66   &-18.3  &16.9\,$\pm$\,0.2  &9.6 \,$\pm$\,0.2  &1.7\,$\pm$\,0.3  &1.7\,$\pm$\,0.3  &3.1 \,$\pm$\,0.5  &1.1\,$\pm$\,0.2  &2.0\,$\pm$\,0.4  &2.3 \,$\pm$\,0.4  &0.74\,$\pm$\,0.17\\
G20.058+0.300  &18:26:33.4 &-11:17:51.32   &-30.9  &19.1\,$\pm$\,0.2  &11.8\,$\pm$\,0.2  &5.3\,$\pm$\,0.3  &2.4\,$\pm$\,0.1  &13.5\,$\pm$\,0.6  &2.4\,$\pm$\,0.8  &2.4\,$\pm$\,0.5  &5.9 \,$\pm$\,1.3  &0.44\,$\pm$\,0.10\\
G20.067+0.008  &18:27:37.6 &-11:25:32.13   &-22.8  &17.6\,$\pm$\,0.3  &10.3\,$\pm$\,0.4  &1.4\,$\pm$\,0.3  &2.0\,$\pm$\,0.6  &3.0 \,$\pm$\,0.6  &1.3\,$\pm$\,0.2  &1.5\,$\pm$\,0.3  &2.1 \,$\pm$\,0.3  &0.70\,$\pm$\,0.18\\
G20.300-0.033  &18:28:13.1 &-11:14:18.17   &-21.0  &17.4\,$\pm$\,0.6  &10.0\,$\pm$\,0.3  &1.8\,$\pm$\,0.4  &1.5\,$\pm$\,0.7  &2.8 \,$\pm$\,0.9  &(0.4)            &...              &<\,1.1            &<\,0.38          \\
G20.308-0.017  &18:28:10.5 &-11:13:25.90   &-21.9  &17.3\,$\pm$\,0.3  &10.1\,$\pm$\,0.6  &2.7\,$\pm$\,0.3  &3.3\,$\pm$\,0.3  &9.8 \,$\pm$\,0.7  &0.9\,$\pm$\,0.2  &3.2\,$\pm$\,0.7  &3.0 \,$\pm$\,0.6  &0.31\,$\pm$\,0.06\\
G20.320+0.250  &18:27:14.0 &-11:05:20.53   &-27.9  &18.6\,$\pm$\,0.4  &11.3\,$\pm$\,0.4  &1.2\,$\pm$\,0.3  &0.6\,$\pm$\,0.4  &0.8 \,$\pm$\,0.3  &(0.3)            &...              &<\,0.5            &<\,0.62          \\
G20.408-0.083  &18:28:36.2 &-11:09:57.52   &-29.1  &18.7\,$\pm$\,0.3  &11.4\,$\pm$\,0.4  &2.0\,$\pm$\,0.3  &1.7\,$\pm$\,0.3  &3.5 \,$\pm$\,0.4  &0.7\,$\pm$\,0.2  &2.7\,$\pm$\,0.4  &2.1 \,$\pm$\,0.4  &0.59\,$\pm$\,0.13\\
G20.442+0.350  &18:27:06.3 &-10:56:04.24   &-22.5  &17.5\,$\pm$\,0.2  &10.2\,$\pm$\,0.2  &1.7\,$\pm$\,0.3  &1.9\,$\pm$\,0.3  &3.3 \,$\pm$\,0.5  &(0.6)            &...              &<\,2.1            &<\,0.64          \\
G20.450-0.075  &18:28:39.2 &-11:07:30.24   &-29.4  &18.7\,$\pm$\,0.3  &11.4\,$\pm$\,0.3  &3.8\,$\pm$\,0.4  &1.7\,$\pm$\,0.1  &6.7 \,$\pm$\,0.5  &2.3\,$\pm$\,0.2  &1.9\,$\pm$\,0.2  &4.6 \,$\pm$\,0.3  &0.68\,$\pm$\,0.07\\
G20.525-0.042  &18:28:40.6 &-11:02:35.96   &-26.7  &18.2\,$\pm$\,0.3  &10.9\,$\pm$\,0.3  &1.4\,$\pm$\,0.3  &2.0\,$\pm$\,0.9  &3.0 \,$\pm$\,0.8  &1.0\,$\pm$\,0.2  &1.6\,$\pm$\,0.3  &1.8 \,$\pm$\,0.3  &0.61\,$\pm$\,0.20\\
G20.533+0.442  &18:26:56.8 &-10:48:40.11   &-23.4  &17.5\,$\pm$\,0.2  &10.3\,$\pm$\,0.2  &2.1\,$\pm$\,0.3  &1.3\,$\pm$\,0.2  &3.0 \,$\pm$\,0.4  &2.3\,$\pm$\,0.5  &1.6\,$\pm$\,0.4  &3.9 \,$\pm$\,0.7  &1.29\,$\pm$\,0.30\\
G20.575+0.092  &18:28:17.2 &-10:56:12.44   &-30.9  &18.8\,$\pm$\,0.2  &11.5\,$\pm$\,0.3  &2.6\,$\pm$\,0.3  &2.4\,$\pm$\,0.3  &6.4 \,$\pm$\,0.6  &0.9\,$\pm$\,0.2  &3.1\,$\pm$\,0.7  &2.9 \,$\pm$\,0.5  &0.44\,$\pm$\,0.08\\
G20.583+0.050  &18:28:27.2 &-10:56:57.20   &-20.1  &16.9\,$\pm$\,0.3  &9.7 \,$\pm$\,0.3  &3.7\,$\pm$\,0.4  &2.1\,$\pm$\,0.2  &8.2 \,$\pm$\,0.7  &3.0\,$\pm$\,0.2  &1.9\,$\pm$\,0.1  &6.1 \,$\pm$\,0.4  &0.74\,$\pm$\,0.08\\
G20.717-0.100  &18:29:14.9 &-10:54:00.68   &-30.0  &18.9\,$\pm$\,0.4  &11.6\,$\pm$\,0.4  &3.4\,$\pm$\,0.3  &2.2\,$\pm$\,0.2  &8.1 \,$\pm$\,0.6  &1.1\,$\pm$\,0.2  &4.0\,$\pm$\,0.7  &4.6 \,$\pm$\,0.6  &0.56\,$\pm$\,0.08\\
G22.083-0.442  &18:33:03.1 &-9:50:50.96    &-23.4  &17.2\,$\pm$\,0.4  &10.2\,$\pm$\,0.4  &4.0\,$\pm$\,0.4  &2.2\,$\pm$\,0.2  &9.1 \,$\pm$\,0.6  &6.3\,$\pm$\,0.5  &2.6\,$\pm$\,0.2  &17.5\,$\pm$\,1.1  &1.93\,$\pm$\,0.18\\
G22.283-0.092  &18:32:10.0 &-9:30:30.49    &-26.7  &17.5\,$\pm$\,0.2  &10.5\,$\pm$\,0.2  &1.7\,$\pm$\,0.2  &1.5\,$\pm$\,0.2  &2.6 \,$\pm$\,0.3  &1.3\,$\pm$\,0.4  &0.9\,$\pm$\,0.7  &1.3 \,$\pm$\,0.5  &0.48\,$\pm$\,0.19\\
G22.450+0.083  &18:31:51.0 &-9:16:46.19    &-20.4  &16.7\,$\pm$\,0.2  &9.7 \,$\pm$\,0.3  &3.0\,$\pm$\,0.2  &2.4\,$\pm$\,0.1  &7.5 \,$\pm$\,0.4  &2.6\,$\pm$\,0.6  &1.3\,$\pm$\,0.3  &3.7 \,$\pm$\,0.7  &0.49\,$\pm$\,0.10\\
G22.700+0.350  &18:31:21.6 &-8:56:03.72    &-30.0  &18.1\,$\pm$\,0.4  &11.1\,$\pm$\,0.4  &2.5\,$\pm$\,0.3  &1.4\,$\pm$\,0.2  &3.7 \,$\pm$\,0.4  &(0.6)            &...              &<\,1.7            &<\,0.48          \\
G22.750-0.333  &18:33:54.4 &-9:12:19.49    &-30.6  &18.3\,$\pm$\,0.2  &11.2\,$\pm$\,0.2  &2.5\,$\pm$\,0.3  &2.1\,$\pm$\,0.2  &5.4 \,$\pm$\,0.5  &2.3\,$\pm$\,0.7  &1.3\,$\pm$\,0.4  &3.2 \,$\pm$\,1.0  &0.59\,$\pm$\,0.18\\
G23.408-0.167  &18:34:32.3 &-8:32:41.84    &-33.9  &18.6\,$\pm$\,0.4  &11.6\,$\pm$\,0.5  &1.8\,$\pm$\,0.3  &1.9\,$\pm$\,0.3  &3.6 \,$\pm$\,0.5  &2.7\,$\pm$\,0.4  &2.1\,$\pm$\,0.3  &5.9 \,$\pm$\,0.7  &1.64\,$\pm$\,0.30\\
G24.217+0.258  &18:34:31.2 &-7:37:52.06    &-21.9  &16.6\,$\pm$\,0.1  &9.8 \,$\pm$\,0.2  &2.1\,$\pm$\,0.3  &1.9\,$\pm$\,0.3  &4.3 \,$\pm$\,0.5  &1.5\,$\pm$\,0.2  &2.0\,$\pm$\,0.2  &3.2 \,$\pm$\,0.3  &0.75\,$\pm$\,0.12\\
G25.017-0.275  &18:37:54.8 &-7:09:56.83    &-32.4  &17.9\,$\pm$\,0.1  &11.1\,$\pm$\,0.1  &2.7\,$\pm$\,0.3  &1.9\,$\pm$\,0.2  &5.4 \,$\pm$\,0.5  &2.6\,$\pm$\,0.2  &2.0\,$\pm$\,0.1  &5.4 \,$\pm$\,0.3  &0.99\,$\pm$\,0.11\\
G25.358-0.225  &18:38:21.9 &-6:50:23.52    &-24.0  &16.7\,$\pm$\,0.1  &10.0\,$\pm$\,0.3  &3.2\,$\pm$\,0.3  &1.9\,$\pm$\,0.2  &6.4 \,$\pm$\,0.5  &2.1\,$\pm$\,0.2  &1.5\,$\pm$\,0.2  &3.4 \,$\pm$\,0.3  &0.53\,$\pm$\,0.07\\
G25.383-0.483  &18:39:20.1 &-6:56:09.46    &-32.1  &17.9\,$\pm$\,0.3  &11.1\,$\pm$\,0.3  &2.3\,$\pm$\,0.3  &1.9\,$\pm$\,0.4  &4.5 \,$\pm$\,0.7  &(0.2)            &...              &<\,0.7            &<\,0.16          \\
G25.833+0.400  &18:37:00.4 &-6:07:51.44    &-33.0  &17.9\,$\pm$\,0.3  &11.1\,$\pm$\,0.3  &4.3\,$\pm$\,0.3  &2.1\,$\pm$\,0.2  &9.8 \,$\pm$\,0.5  &6.2\,$\pm$\,0.3  &2.8\,$\pm$\,0.1  &18.8\,$\pm$\,0.6  &1.93\,$\pm$\,0.12\\
G25.883+0.108  &18:38:08.6 &-6:13:14.17    &-32.1  &17.6\,$\pm$\,0.2  &10.9\,$\pm$\,0.2  &2.7\,$\pm$\,0.3  &3.0\,$\pm$\,0.3  &8.5 \,$\pm$\,0.7  &2.6\,$\pm$\,0.4  &2.9\,$\pm$\,0.3  &7.8 \,$\pm$\,0.7  &0.91\,$\pm$\,0.11\\
G25.992+0.492  &18:36:58.3 &-5:56:50.88    &-32.4  &17.7\,$\pm$\,0.2  &11.0\,$\pm$\,0.3  &2.0\,$\pm$\,0.3  &1.2\,$\pm$\,0.2  &2.5 \,$\pm$\,0.4  &0.9\,$\pm$\,0.2  &1.9\,$\pm$\,0.3  &1.8 \,$\pm$\,0.3  &0.74\,$\pm$\,0.17\\
G26.033-0.025  &18:38:53.7 &-6:08:53.98    &-24.6  &16.6\,$\pm$\,0.3  &10.0\,$\pm$\,0.3  &3.5\,$\pm$\,0.3  &1.9\,$\pm$\,0.2  &7.1 \,$\pm$\,0.6  &1.2\,$\pm$\,0.4  &2.6\,$\pm$\,0.7  &3.3 \,$\pm$\,0.7  &0.46\,$\pm$\,0.11\\
G26.075+0.455  &18:37:15.5 &-5:53:26.65    &-32.5  &17.6\,$\pm$\,0.2  &10.9\,$\pm$\,0.2  &2.2\,$\pm$\,0.3  &1.2\,$\pm$\,0.2  &2.7 \,$\pm$\,0.4  &2.7\,$\pm$\,0.4  &1.6\,$\pm$\,0.2  &4.5 \,$\pm$\,0.4  &1.70\,$\pm$\,0.30\\
G26.608-0.208  &18:40:36.6 &-5:43:15.99    &-32.4  &17.8\,$\pm$\,0.2  &11.1\,$\pm$\,0.3  &5.3\,$\pm$\,0.2  &3.9\,$\pm$\,0.1  &22.1\,$\pm$\,0.6  &8.1\,$\pm$\,0.4  &5.1\,$\pm$\,0.1  &44.5\,$\pm$\,1.1  &2.02\,$\pm$\,0.07\\
G26.800+0.175  &18:39:35.7 &-5:22:29.95    &-28.8  &17.0\,$\pm$\,0.4  &10.4\,$\pm$\,0.4  &2.4\,$\pm$\,0.3  &1.5\,$\pm$\,0.2  &3.8 \,$\pm$\,0.5  &2.2\,$\pm$\,0.5  &2.5\,$\pm$\,0.5  &5.8 \,$\pm$\,1.0  &1.51\,$\pm$\,0.31\\
G26.983+0.433  &18:39:00.6 &-5:05:38.70    &-29.7  &17.0\,$\pm$\,0.2  &10.5\,$\pm$\,0.3  &4.2\,$\pm$\,0.3  &2.7\,$\pm$\,0.1  &12.2\,$\pm$\,0.6  &2.8\,$\pm$\,0.3  &2.4\,$\pm$\,0.2  &7.3 \,$\pm$\,0.5  &0.60\,$\pm$\,0.05\\
G27.608+0.025  &18:41:37.0 &-4:43:31.33    &-32.4  &17.3\,$\pm$\,0.2  &10.8\,$\pm$\,0.2  &2.4\,$\pm$\,0.3  &1.7\,$\pm$\,0.2  &4.4 \,$\pm$\,0.5  &2.5\,$\pm$\,0.3  &2.1\,$\pm$\,0.3  &5.5 \,$\pm$\,0.5  &1.25\,$\pm$\,0.19\\
G27.733+0.350  &18:40:41.2 &-4:27:55.71    &-35.1  &17.7\,$\pm$\,0.2  &11.2\,$\pm$\,0.3  &2.3\,$\pm$\,0.3  &1.4\,$\pm$\,0.3  &3.5 \,$\pm$\,0.5  &3.2\,$\pm$\,0.7  &1.6\,$\pm$\,0.3  &5.3 \,$\pm$\,1.0  &1.51\,$\pm$\,0.36\\
G27.792+0.333  &18:40:51.3 &-4:25:14.90    &-36.0  &17.8\,$\pm$\,0.2  &11.2\,$\pm$\,0.2  &3.0\,$\pm$\,0.3  &2.1\,$\pm$\,0.2  &6.8 \,$\pm$\,0.5  &(0.8)            &...              &<\,2.7            &<\,0.39          \\
G27.842-0.367  &18:43:26.7 &-4:41:47.50    &-31.5  &17.2\,$\pm$\,0.3  &10.7\,$\pm$\,0.3  &3.8\,$\pm$\,0.2  &3.1\,$\pm$\,0.1  &12.4\,$\pm$\,0.5  &2.7\,$\pm$\,0.4  &1.6\,$\pm$\,0.3  &4.6 \,$\pm$\,0.6  &0.37\,$\pm$\,0.05\\
G28.225+0.033  &18:42:43.2 &-4:10:22.93    &-33.0  &17.2\,$\pm$\,0.1  &10.8\,$\pm$\,0.2  &2.3\,$\pm$\,0.3  &2.3\,$\pm$\,0.3  &5.5 \,$\pm$\,0.6  &1.5\,$\pm$\,0.2  &2.0\,$\pm$\,0.2  &3.1 \,$\pm$\,0.3  &0.56\,$\pm$\,0.08\\
G28.383+0.025  &18:43:02.3 &-4:02:10.23    &-39.9  &18.3\,$\pm$\,0.2  &11.8\,$\pm$\,0.2  &2.6\,$\pm$\,0.3  &1.9\,$\pm$\,0.2  &5.4 \,$\pm$\,0.5  &(0.2)            &...              &<\,0.7            &<\,0.12          \\
G28.433+0.433  &18:41:40.5 &-3:48:18.43    &-38.4  &17.8\,$\pm$\,0.1  &11.4\,$\pm$\,0.1  &4.5\,$\pm$\,0.3  &1.3\,$\pm$\,0.1  &6.4 \,$\pm$\,0.5  &2.1\,$\pm$\,0.5  &1.5\,$\pm$\,0.4  &3.3 \,$\pm$\,0.8  &0.52\,$\pm$\,0.12\\
G28.517+0.092  &18:43:02.7 &-3:53:10.92    &-41.4  &18.5\,$\pm$\,0.1  &12.0\,$\pm$\,0.1  &2.4\,$\pm$\,0.3  &2.0\,$\pm$\,0.2  &5.2 \,$\pm$\,0.5  &0.8\,$\pm$\,0.2  &2.2\,$\pm$\,0.5  &1.9 \,$\pm$\,0.3  &0.37\,$\pm$\,0.07\\
G28.767+0.100  &18:43:28.5 &-3:39:37.28    &-39.0  &18.1\,$\pm$\,0.4  &11.6\,$\pm$\,0.4  &1.5\,$\pm$\,0.3  &1.8\,$\pm$\,0.5  &2.8 \,$\pm$\,0.6  &(0.2)            &...              &<\,0.6            &<\,0.20          \\
G29.908+0.483  &18:44:11.9 &-2:28:13.59    &-35.4  &17.2\,$\pm$\,0.4  &10.9\,$\pm$\,0.4  &2.4\,$\pm$\,0.3  &1.7\,$\pm$\,0.3  &4.3 \,$\pm$\,0.5  &1.8\,$\pm$\,0.5  &4.2\,$\pm$\,0.7  &7.8 \,$\pm$\,1.3  &1.79\,$\pm$\,0.36\\
\hline
\end{tabular}
\label{table:spectral parameters}
\tablefoot{Column 1: cloud name. 
Columns 2 and 3: equatorial coordinates. 
Column 4--6: the local standard of rest velocity, heliocentric distance, and Galactocentric distance taken from \citet{Sun2024a}. 
Column 7: peak brightness temperature of the CO\,$J$=1--0. 
Column 8: line width of the CO\,$J$=1--0. 
Column 9: integrated intensity of the CO\,$J$=1--0. 
Column 10: peak brightness temperature of the CO\,$J$=3--2. 
Column 11: line width of the CO\,$J$=3--2. 
Column 12: integrated intensity of the CO\,$J$=3--2. 
Column 13: the CO\,$J$=3--2/1--0 integrated intensity ratio. 
The $T_{\rm CO(3-2)}$ values with signal-to-noise (S/N) ratios below 3 are presented in parentheses in Column 10. For these sources, the $T_{\rm CO(3-2)}$ values are given by the spectral RMS noise levels ($\sigma_{\rm rms}$) as upper limits. 
And upper limits for the $I_{\rm CO(3-2)}$ values in Column 12 were obtained using $I_{\rm CO(3-2)}$\,<\,3$\sigma_{\rm rms}$\,$\times$\,$\sqrt{\Delta V_{\rm CO(1-0)}\times\Delta v}$, where $\Delta v$ denotes the velocity resolution of 0.64\,km\,s$^{-1}$.} 
\end{table*}

\setlength{\tabcolsep}{5.5pt}
\setcounter{table}{0} 
\begin{table*}[!htbp]
\scriptsize
\caption{Continued.}
\centering
\begin{tabular}
{ccccccccccccc}
\hline\hline
Cloud name &R.A.(J2000) &Dec.(J2000)  &$V_{\rm LSR}$ &$D_{\rm s}$ &$R_{\rm g}$ &$T_{\rm CO(1-0)}$ &$\Delta V_{\rm CO(1-0)}$ &$I_{\rm CO(1-0)}$ &$T_{\rm CO(3-2)}$ &$\Delta V_{\rm CO(3-2)}$ &$I_{\rm CO(3-2)}$  &$R_{31}$\\
&(h m s) &($\circ$ $\prime$ $\prime\prime$) &km\,s$^{-1}$ &kpc &kpc &K &km\,s$^{-1}$ &K\,km\,s$^{-1}$ &K &km\,s$^{-1}$ &K\,km\,s$^{-1}$ &\\
(1) &(2) &(3) &(4) &(5) &(6) &(7) &(8) &(9) &(10) &(11) &(12) &(13)\\
\hline
G30.317-0.325  &18:47:49.4 &-2:28:31.34    &-37.8  &17.4\,$\pm$\,0.3  &11.2\,$\pm$\,0.3  &2.3\,$\pm$\,0.4  &2.8\,$\pm$\,0.3  &6.8 \,$\pm$\,0.7  &2.2\,$\pm$\,0.4  &4.0\,$\pm$\,0.6  &9.5 \,$\pm$\,1.1  &1.40\,$\pm$\,0.22\\
G30.917+0.433  &18:46:13.2 &-1:35:44.18    &-40.2  &17.6\,$\pm$\,0.3  &11.5\,$\pm$\,0.3  &2.5\,$\pm$\,0.2  &2.7\,$\pm$\,0.2  &7.4 \,$\pm$\,0.5  &3.7\,$\pm$\,0.4  &3.5\,$\pm$\,0.3  &13.7\,$\pm$\,0.9  &1.86\,$\pm$\,0.17\\
G33.858+0.008  &18:53:05.9 &+0:49:40.21    &-42.0  &17.1\,$\pm$\,0.3  &11.3\,$\pm$\,0.3  &2.1\,$\pm$\,0.4  &1.1\,$\pm$\,0.3  &2.5 \,$\pm$\,0.5  &(0.2)            &...              &<\,0.6            &<\,0.24          \\
G33.933+0.133  &18:52:47.4 &+0:57:05.71    &-43.8  &17.4\,$\pm$\,0.3  &11.6\,$\pm$\,0.3  &3.3\,$\pm$\,0.3  &1.7\,$\pm$\,0.1  &6.1 \,$\pm$\,0.4  &1.9\,$\pm$\,0.3  &1.9\,$\pm$\,0.2  &3.9 \,$\pm$\,0.4  &0.64\,$\pm$\,0.08\\
G36.717-0.058  &18:58:33.4 &+3:20:30.00    &-45.0  &16.7\,$\pm$\,0.2  &11.3\,$\pm$\,0.2  &2.4\,$\pm$\,0.3  &1.5\,$\pm$\,0.2  &3.8 \,$\pm$\,0.5  &2.1\,$\pm$\,0.3  &2.2\,$\pm$\,0.3  &4.7 \,$\pm$\,0.5  &1.25\,$\pm$\,0.21\\
G36.925+0.000  &18:58:43.8 &+3:33:11.48    &-57.0  &18.4\,$\pm$\,0.3  &12.9\,$\pm$\,0.3  &2.6\,$\pm$\,0.3  &1.2\,$\pm$\,0.2  &3.5 \,$\pm$\,0.4  &1.3\,$\pm$\,0.4  &2.7\,$\pm$\,0.5  &3.5 \,$\pm$\,0.6  &1.02\,$\pm$\,0.21\\
G37.767+0.275  &18:59:17.5 &+4:25:40.10    &-47.1  &16.8\,$\pm$\,0.3  &11.5\,$\pm$\,0.3  &1.6\,$\pm$\,0.2  &1.6\,$\pm$\,0.2  &2.8 \,$\pm$\,0.4  &0.6\,$\pm$\,0.2  &3.1\,$\pm$\,0.9  &1.9 \,$\pm$\,0.5  &0.67\,$\pm$\,0.21\\
G37.825+0.092  &19:00:03.1 &+4:23:44.67    &-48.0  &16.9\,$\pm$\,0.2  &11.6\,$\pm$\,0.2  &3.6\,$\pm$\,0.3  &1.8\,$\pm$\,0.1  &6.8 \,$\pm$\,0.5  &(0.4)            &...              &<\,1.2            &<\,0.18          \\
G37.842+0.150  &18:59:52.5 &+4:26:14.56    &-46.2  &16.7\,$\pm$\,0.2  &11.4\,$\pm$\,0.3  &2.1\,$\pm$\,0.2  &3.0\,$\pm$\,0.3  &6.9 \,$\pm$\,0.5  &1.7\,$\pm$\,0.4  &1.6\,$\pm$\,0.3  &3.0 \,$\pm$\,0.5  &0.44\,$\pm$\,0.08\\
G38.075+0.483  &18:59:06.9 &+4:47:48.51    &-47.7  &16.8\,$\pm$\,0.3  &11.6\,$\pm$\,0.3  &2.3\,$\pm$\,0.3  &1.6\,$\pm$\,0.2  &3.9 \,$\pm$\,0.4  &1.2\,$\pm$\,0.3  &2.3\,$\pm$\,0.5  &3.0 \,$\pm$\,0.6  &0.77\,$\pm$\,0.16\\
G40.883+0.492  &19:04:15.1 &+7:17:50.88    &-54.9  &16.9\,$\pm$\,0.2  &12.0\,$\pm$\,0.2  &2.2\,$\pm$\,0.4  &3.6\,$\pm$\,0.5  &8.6 \,$\pm$\,0.9  &1.4\,$\pm$\,0.2  &3.7\,$\pm$\,0.4  &5.4 \,$\pm$\,0.5  &0.62\,$\pm$\,0.09\\
G44.325-0.208  &19:13:11.2 &+10:01:45.26   &-65.4  &17.4\,$\pm$\,0.3  &12.9\,$\pm$\,0.3  &1.9\,$\pm$\,0.3  &2.1\,$\pm$\,0.4  &4.4 \,$\pm$\,0.6  &0.2\,$\pm$\,0.1  &2.0\,$\pm$\,0.5  &1.2 \,$\pm$\,0.3  &0.28\,$\pm$\,0.07\\
G44.350-0.220  &19:13:16.6 &+10:02:44.96   &-64.5  &17.4\,$\pm$\,0.3  &12.9\,$\pm$\,0.3  &1.2\,$\pm$\,0.4  &1.4\,$\pm$\,0.4  &1.7 \,$\pm$\,0.5  &(0.3)            &...              &<\,0.9            &<\,0.51          \\
G44.500+0.340  &19:11:32.5 &+10:26:17.31   &-48.0  &17.4\,$\pm$\,0.3  &12.9\,$\pm$\,0.3  &5.5\,$\pm$\,0.3  &4.9\,$\pm$\,0.2  &28.9\,$\pm$\,0.9  &(1.2)            &...              &<\,6.2            &<\,0.22          \\
\hline
\end{tabular}
\label{table:spectral parameters}
\end{table*}
\endgroup

\section{CO\,(3--2) integrated intensity maps}
\label{Sect:Integrated intensity map}
\setcounter{figure}{0}
\begin{figure*}[!htbp]
\centering
\includegraphics[width=1.0\textwidth]{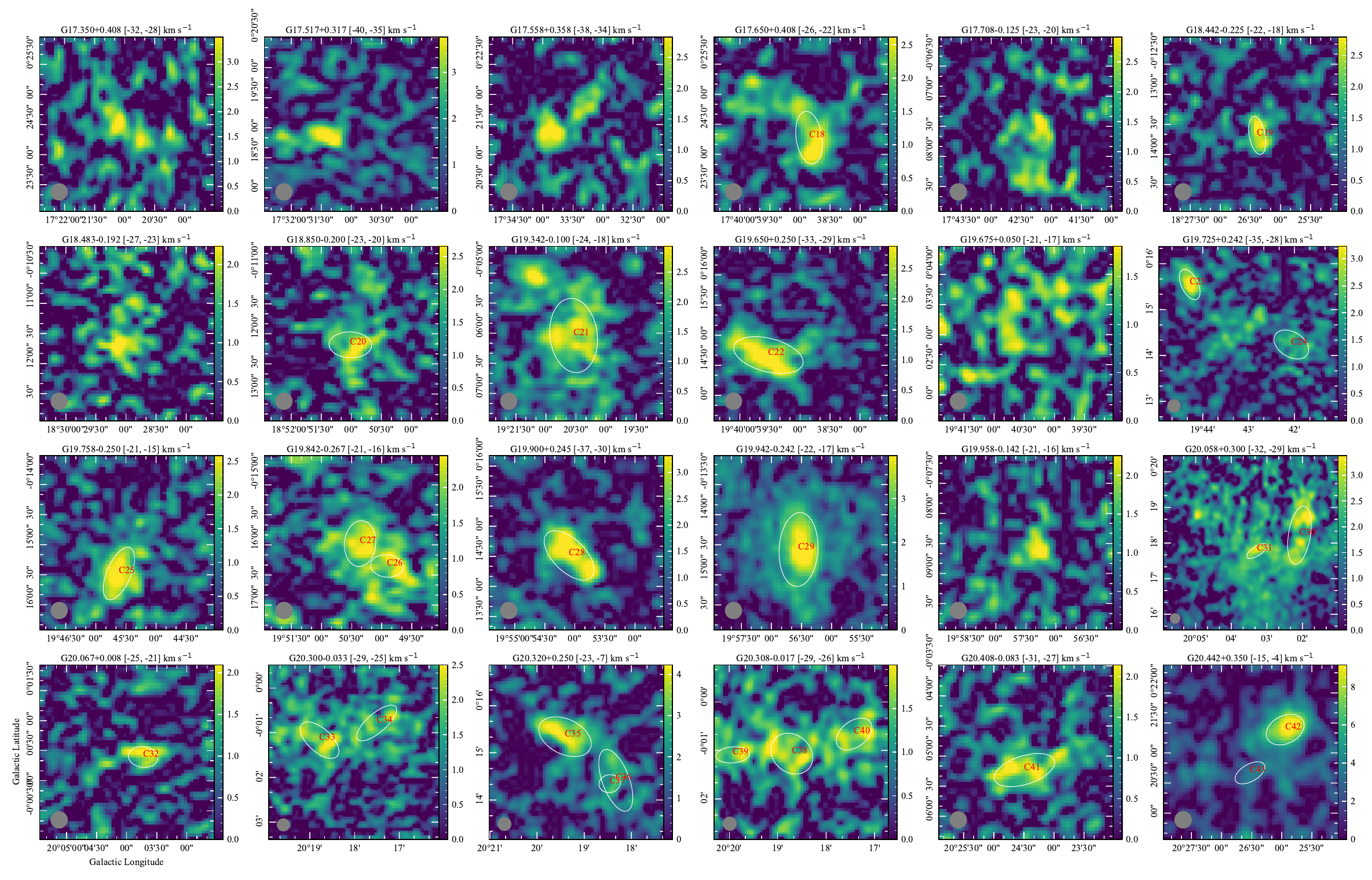}
\caption{CO\,(3--2) velocity-integrated intensity maps of the Galactic edge clouds. 
Detailed descriptions are the same as in Fig.\,\ref{fig:map}. 
} 
\label{fig:map-continue} 
\end{figure*}

\setcounter{figure}{0}
\begin{figure*}[!htbp]
\centering
\includegraphics[width=1.0\textwidth]{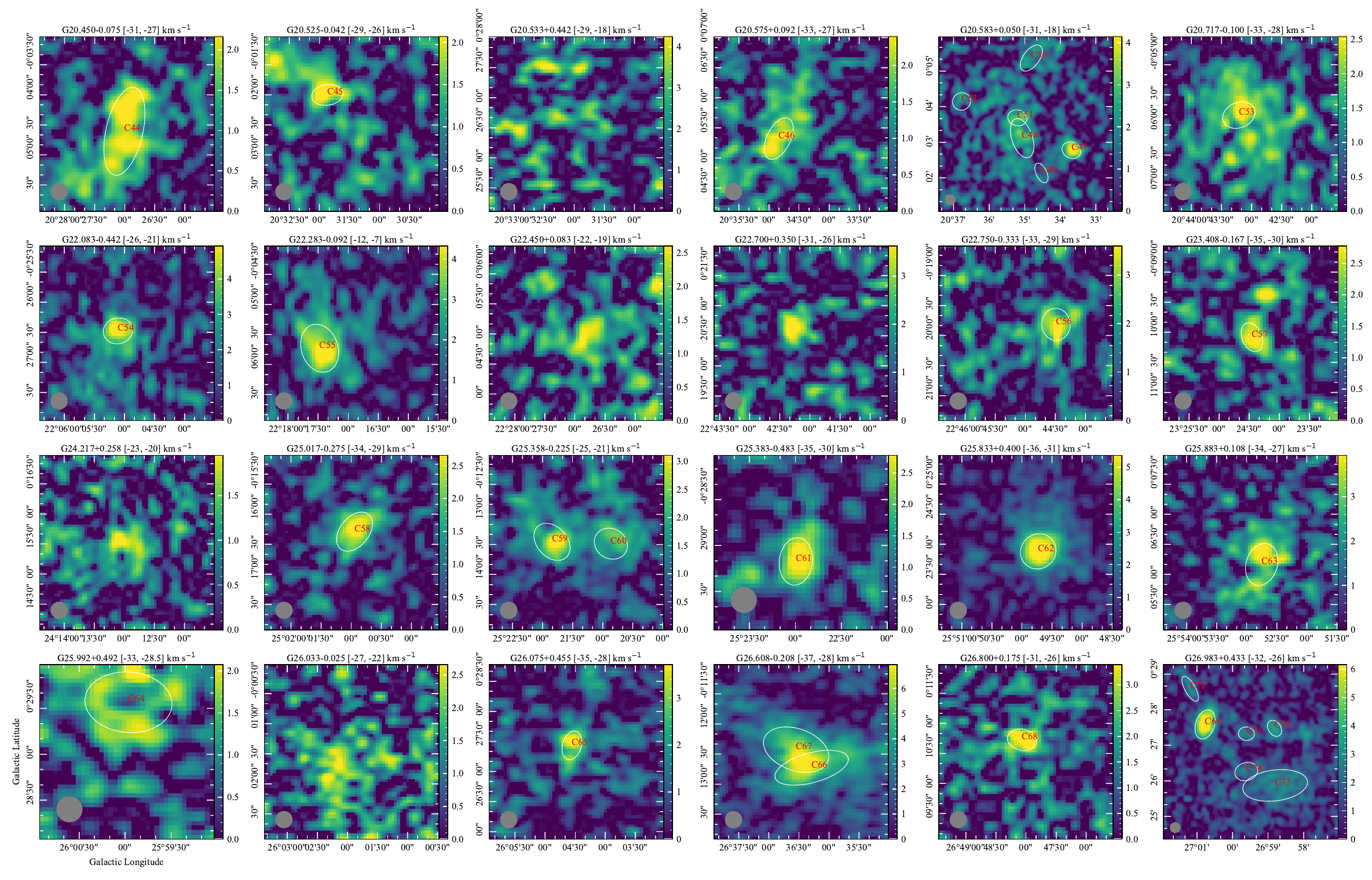}

\vspace{-2mm}

\includegraphics[width=1.0\textwidth]{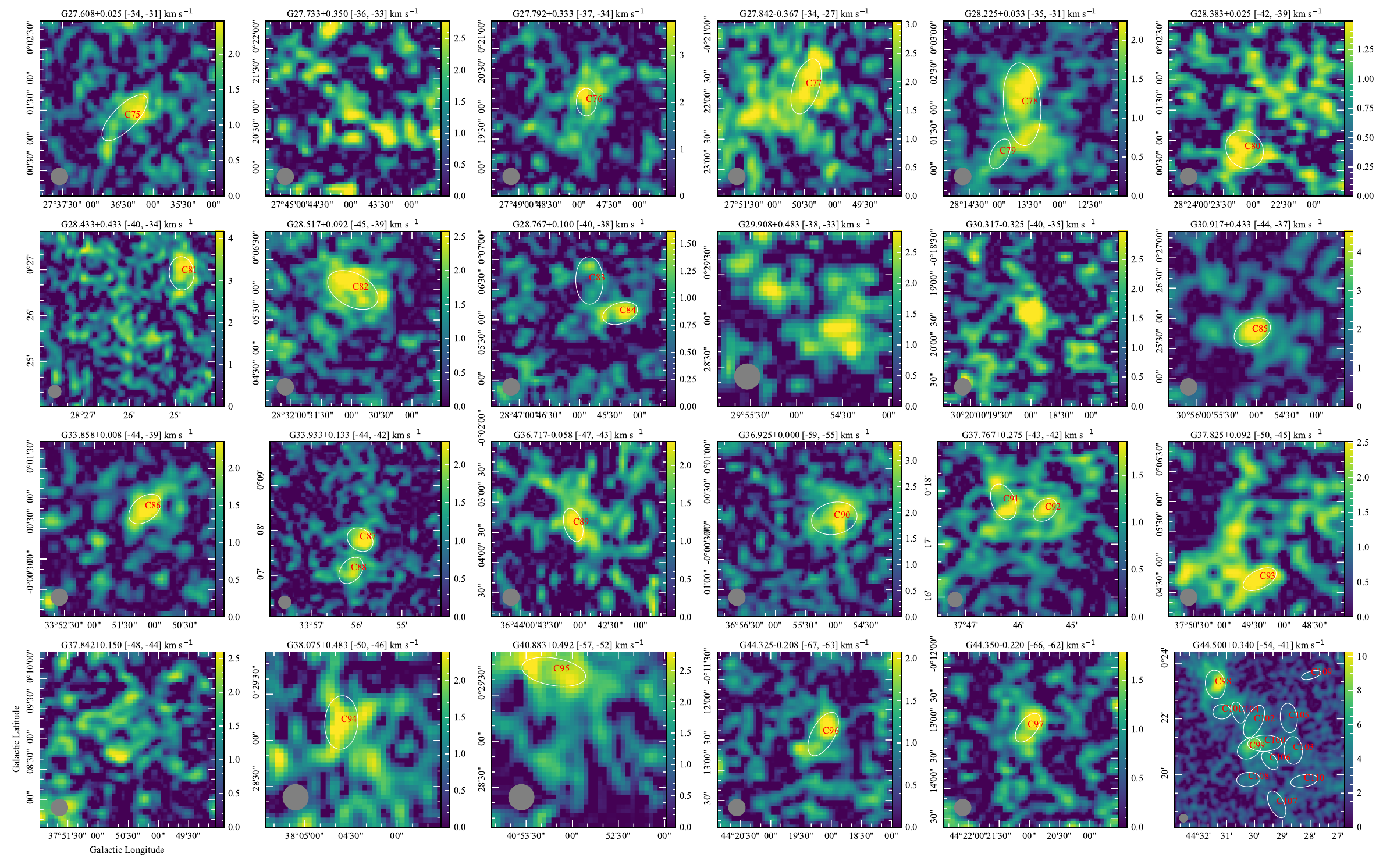}
\caption{continued.}
\end{figure*}

\begingroup
\setlength{\tabcolsep}{2.7pt}
\setcounter{table}{0}
\begin{table*}[!htbp]
\scriptsize
\caption{Physical parameters of CO\,(3--2) clumps}
\centering
\begin{tabular}
{ccccccccccccccc}
\hline\hline
Num &R.A.(J2000) &Dec.(J2000) &$R_{\rm g}$ &$R_{\rm eff}$ &$V_{\rm LSR}$ &$\sigma_{\rm v}$ &$T_{\rm mb}$ &$I_{\rm CO}$ &$L_{\rm CO}$ &$M_{\rm H_2}$ &$N_{\rm H_2}$ &$n_{\rm H_2}$ &$\alpha_{\rm vir}$ &Parent cloud\\
&h m s &$\circ$ $\prime$ $\prime\prime$ & &kpc &pc &km\,s$^{-1}$ &km\,s$^{-1}$ &K &K\,km\,s$^{-1}$ &K\,km\,s$^{-1}$\,pc$^{2}$ &$10^{2}$\,M$_{\odot}$ &$10^{21}$\,cm$^{-2}$ &$10^{2}$\,cm$^{-3}$ & \\
(1) &(2) &(3) &(4) &(5) &(6) &(7) &(8) &(9) &(10) &(11) &(12) &(13) &(14) &(15)\\
\hline
1   &18:16:45.2 &-15:44:10 &11.6\,$\pm$\,0.3 &1.55\,$\pm$\,0.02 &-24.4 &1.15\,$\pm$\,0.05 &2.9\,$\pm$\,0.1 &8.5 \,$\pm$\,0.3 &84.4\,$\pm$\,18.7 & 5.7\,$\pm$\,1.8  &3.5\,$\pm$\,1.1  &5.4\,$\pm$\,1.7   &4.2\,$\pm$\,1.4  &G15.017+0.317\\
2   &18:19:43.9 &-15:59:58 &12.8\,$\pm$\,1.0 &2.57\,$\pm$\,0.14 &-29.7 &1.18\,$\pm$\,0.03 &3.7\,$\pm$\,0.1 &11.3\,$\pm$\,0.3 &114 \,$\pm$\,23   & 8.9\,$\pm$\,2.6  &2.0\,$\pm$\,0.6  &1.9\,$\pm$\,0.6   &4.7\,$\pm$\,1.5  &G15.133-0.442\\
3   &18:19:45.0 &-15:58:23 &12.8\,$\pm$\,1.0 &2.94\,$\pm$\,0.16 &-28.3 &0.59\,$\pm$\,0.03 &2.4\,$\pm$\,0.1 &3.9 \,$\pm$\,0.2 &35.5\,$\pm$\,7.1  & 2.8\,$\pm$\,0.8  &0.5\,$\pm$\,0.1  &0.4\,$\pm$\,0.1   &4.3\,$\pm$\,1.4  &G15.133-0.442\\
4   &18:19:48.3 &-15:58:13 &12.8\,$\pm$\,1.0 &2.02\,$\pm$\,0.11 &-27.7 &0.60\,$\pm$\,0.04 &2.1\,$\pm$\,0.1 &3.4 \,$\pm$\,0.2 &20.4\,$\pm$\,4.1  & 1.6\,$\pm$\,0.5  &0.6\,$\pm$\,0.2  &0.7\,$\pm$\,0.2   &5.3\,$\pm$\,1.8  &G15.133-0.442\\
5   &18:18:30.6 &-15:44:45 &13.7\,$\pm$\,0.6 &2.17\,$\pm$\,0.07 &-31.3 &1.79\,$\pm$\,0.05 &3.6\,$\pm$\,0.1 &16.3\,$\pm$\,0.5 &230 \,$\pm$\,36   &20.2\,$\pm$\,4.9  &6.4\,$\pm$\,1.6  &7.0\,$\pm$\,1.7   &4.0\,$\pm$\,1.0  &G15.208-0.067\\
6   &18:17:08.9 &-15:34:11 &12.0\,$\pm$\,0.5 &1.96\,$\pm$\,0.05 &-26.0 &0.57\,$\pm$\,0.03 &2.3\,$\pm$\,0.1 &3.7 \,$\pm$\,0.2 &21.6\,$\pm$\,5.8  & 1.5\,$\pm$\,0.5  &0.6\,$\pm$\,0.2  &0.7\,$\pm$\,0.3   &4.8\,$\pm$\,1.8  &G15.217+0.308\\
7   &18:21:30.2 &-14:53:29 &9.9 \,$\pm$\,0.4 &5.68\,$\pm$\,0.13 &-18.4 &1.96\,$\pm$\,0.05 &3.8\,$\pm$\,0.1 &19.0\,$\pm$\,0.5 &944 \,$\pm$\,88   &50.9\,$\pm$\,15.7 &2.3\,$\pm$\,0.7  &1.0\,$\pm$\,0.3   &5.0\,$\pm$\,1.6  &G16.300-0.300\\
8   &18:21:32.5 &-14:55:02 &9.9 \,$\pm$\,0.4 &5.91\,$\pm$\,0.14 &-17.1 &2.62\,$\pm$\,0.09 &2.8\,$\pm$\,0.1 &18.5\,$\pm$\,0.7 &874 \,$\pm$\,81   &47.1\,$\pm$\,14.5 &2.0\,$\pm$\,0.6  &0.8\,$\pm$\,0.2   &10.0\,$\pm$\,3.2 &G16.300-0.300\\
9   &18:19:46.2 &-14:13:57 &13.8\,$\pm$\,0.6 &0.72\,$\pm$\,0.03 &-33.2 &0.95\,$\pm$\,0.03 &3.1\,$\pm$\,0.1 &7.6 \,$\pm$\,0.3 &17.2\,$\pm$\,6.1  & 1.5\,$\pm$\,0.6  &4.4\,$\pm$\,1.9  &14.6\,$\pm$\,5.8  &4.9\,$\pm$\,2.0  &G16.690+0.384\\
10  &18:20:16.7 &-14:16:42 &13.8\,$\pm$\,0.4 &1.17\,$\pm$\,0.03 &-33.9 &1.45\,$\pm$\,0.08 &1.8\,$\pm$\,0.1 &6.7 \,$\pm$\,0.4 &41.6\,$\pm$\,10.2 & 3.7\,$\pm$\,1.1  &4.0\,$\pm$\,1.2  &8.2\,$\pm$\,2.5   &7.7\,$\pm$\,2.5  &G16.708+0.250\\
11  &18:19:27.6 &-14:10:20 &14.9\,$\pm$\,0.5 &1.32\,$\pm$\,0.03 &-36.6 &0.55\,$\pm$\,0.03 &3.2\,$\pm$\,0.1 &4.9 \,$\pm$\,0.2 &15.7\,$\pm$\,5.3  & 1.6\,$\pm$\,0.6  &1.4\,$\pm$\,0.5  &2.5\,$\pm$\,0.9   &2.9\,$\pm$\,1.2  &G16.708+0.467\\
12  &18:19:29.0 &-14:10:40 &14.9\,$\pm$\,0.5 &2.33\,$\pm$\,0.06 &-36.4 &0.54\,$\pm$\,0.03 &3.1\,$\pm$\,0.1 &4.7 \,$\pm$\,0.2 &20.1\,$\pm$\,6.8  & 2.1\,$\pm$\,0.8  &0.6\,$\pm$\,0.2  &0.6\,$\pm$\,0.2   &3.8\,$\pm$\,1.5  &G16.708+0.467\\
13  &18:19:27.0 &-14:09:55 &14.9\,$\pm$\,0.5 &1.02\,$\pm$\,0.03 &-35.8 &0.60\,$\pm$\,0.04 &2.9\,$\pm$\,0.1 &4.7 \,$\pm$\,0.3 &11.1\,$\pm$\,3.8  & 1.1\,$\pm$\,0.4  &1.6\,$\pm$\,0.6  &3.8\,$\pm$\,1.4   &3.7\,$\pm$\,1.5  &G16.708+0.467\\
14  &18:19:30.1 &-14:09:49 &14.9\,$\pm$\,0.5 &2.15\,$\pm$\,0.05 &-36.6 &0.61\,$\pm$\,0.04 &2.5\,$\pm$\,0.1 &4.2 \,$\pm$\,0.3 &14.5\,$\pm$\,4.9  & 1.5\,$\pm$\,0.6  &0.5\,$\pm$\,0.2  &0.5\,$\pm$\,0.2   &6.2\,$\pm$\,2.5  &G16.708+0.467\\
15  &18:20:43.6 &-14:15:07 &11.7\,$\pm$\,0.8 &2.24\,$\pm$\,0.10 &-26.8 &0.83\,$\pm$\,0.05 &2.1\,$\pm$\,0.1 &4.6 \,$\pm$\,0.3 &73.5\,$\pm$\,19.3 & 5.0\,$\pm$\,1.8  &1.5\,$\pm$\,0.5  &1.6\,$\pm$\,0.6   &3.6\,$\pm$\,1.4  &G16.775+0.167\\
16  &18:20:48.9 &-14:02:06 &12.9\,$\pm$\,0.4 &2.28\,$\pm$\,0.05 &-31.4 &1.51\,$\pm$\,0.08 &2.2\,$\pm$\,0.1 &8.5 \,$\pm$\,0.5 &61.4\,$\pm$\,13.2 & 4.9\,$\pm$\,1.4  &1.4\,$\pm$\,0.4  &1.5\,$\pm$\,0.4   &12.4\,$\pm$\,3.9 &G16.975+0.250\\
17  &18:21:29.5 &-13:57:36 &14.8\,$\pm$\,0.7 &2.47\,$\pm$\,0.08 &-36.3 &1.60\,$\pm$\,0.07 &6.8\,$\pm$\,0.1 &27.5\,$\pm$\,1.1 &145 \,$\pm$\,22   &14.7\,$\pm$\,3.3  &3.6\,$\pm$\,0.8  &3.5\,$\pm$\,0.8   &5.0\,$\pm$\,1.2  &G17.125+0.142\\
18  &18:21:33.4 &-13:22:43 &11.1\,$\pm$\,0.4 &2.20\,$\pm$\,0.05 &-23.5 &0.62\,$\pm$\,0.05 &3.4\,$\pm$\,0.1 &5.7 \,$\pm$\,0.4 &29.5\,$\pm$\,10.6 & 1.9\,$\pm$\,0.8  &0.6\,$\pm$\,0.2  &0.6\,$\pm$\,0.3   &5.3\,$\pm$\,2.5  &G17.650+0.408\\
19  &18:25:22.7 &-12:58:31 &10.1\,$\pm$\,0.2 &0.81\,$\pm$\,0.01 &-20.4 &0.80\,$\pm$\,0.04 &3.4\,$\pm$\,0.1 &7.3 \,$\pm$\,0.3 &16.1\,$\pm$\,4.1  & 0.9\,$\pm$\,0.3  &2.0\,$\pm$\,0.8  &6.0\,$\pm$\,2.2   &6.7\,$\pm$\,2.7  &G18.442-0.225\\
20  &18:26:04.4 &-12:36:02 &10.3\,$\pm$\,0.4 &1.73\,$\pm$\,0.04 &-21.3 &0.61\,$\pm$\,0.03 &4.8\,$\pm$\,0.1 &7.9 \,$\pm$\,0.3 &29.7\,$\pm$\,6.3  & 1.7\,$\pm$\,0.6  &0.8\,$\pm$\,0.3  &1.1\,$\pm$\,0.4   &4.4\,$\pm$\,1.7  &G18.850-0.200\\
21  &18:26:38.5 &-12:07:03 &10.1\,$\pm$\,0.2 &3.80\,$\pm$\,0.05 &-20.8 &1.45\,$\pm$\,0.06 &2.5\,$\pm$\,0.1 &9.3 \,$\pm$\,0.4 &120 \,$\pm$\,13   & 6.6\,$\pm$\,2.0  &0.7\,$\pm$\,0.2  &0.4\,$\pm$\,0.1   &14.0\,$\pm$\,4.4 &G19.342-0.100\\
22  &18:25:59.8 &-11:40:36 &12.0\,$\pm$\,0.2 &3.31\,$\pm$\,0.04 &-31.4 &1.40\,$\pm$\,0.05 &2.9\,$\pm$\,0.1 &10.5\,$\pm$\,0.4 &161 \,$\pm$\,44   &11.4\,$\pm$\,4.0  &1.5\,$\pm$\,0.5  &1.1\,$\pm$\,0.4   &6.6\,$\pm$\,2.4  &G19.650+0.250\\
23  &18:26:05.7 &-11:35:59 &12.3\,$\pm$\,0.3 &1.66\,$\pm$\,0.03 &-31.0 &1.23\,$\pm$\,0.05 &2.5\,$\pm$\,0.1 &8.0 \,$\pm$\,0.3 &80.4\,$\pm$\,15.5 & 5.9\,$\pm$\,1.7  &3.2\,$\pm$\,0.9  &4.6\,$\pm$\,1.3   &4.9\,$\pm$\,1.5  &G19.725+0.242\\
24  &18:26:06.2 &-11:38:33 &12.3\,$\pm$\,0.3 &2.61\,$\pm$\,0.04 &-32.4 &0.80\,$\pm$\,0.04 &2.4\,$\pm$\,0.1 &5.1 \,$\pm$\,0.2 &51.6\,$\pm$\,10.0 & 3.8\,$\pm$\,1.1  &0.8\,$\pm$\,0.2  &0.8\,$\pm$\,0.2   &5.1\,$\pm$\,1.6  &G19.725+0.242\\
25  &18:28:00.4 &-11:49:15 &9.7 \,$\pm$\,0.2 &1.96\,$\pm$\,0.03 &-18.1 &1.22\,$\pm$\,0.05 &2.1\,$\pm$\,0.1 &6.6 \,$\pm$\,0.3 &45.0\,$\pm$\,7.9  & 2.4\,$\pm$\,0.8  &0.9\,$\pm$\,0.3  &1.1\,$\pm$\,0.4   &14.3\,$\pm$\,5.1 &G19.758-0.250\\
26  &18:28:11.7 &-11:45:51 &9.7 \,$\pm$\,0.3 &1.09\,$\pm$\,0.02 &-18.1 &0.57\,$\pm$\,0.03 &2.6\,$\pm$\,0.1 &4.2 \,$\pm$\,0.2 &14.9\,$\pm$\,2.3  & 0.8\,$\pm$\,0.3  &1.0\,$\pm$\,0.3  &2.1\,$\pm$\,0.7   &5.3\,$\pm$\,1.9  &G19.842-0.267\\
27  &18:28:11.2 &-11:45:17 &9.7 \,$\pm$\,0.3 &2.07\,$\pm$\,0.04 &-18.1 &0.59\,$\pm$\,0.03 &2.7\,$\pm$\,0.1 &4.4 \,$\pm$\,0.2 &35.1\,$\pm$\,5.2  & 1.8\,$\pm$\,0.6  &0.6\,$\pm$\,0.2  &0.7\,$\pm$\,0.2   &4.5\,$\pm$\,1.6  &G19.842-0.267\\
28  &18:26:28.1 &-11:27:47 &12.4\,$\pm$\,0.4 &3.06\,$\pm$\,0.05 &-33.5 &2.09\,$\pm$\,0.12 &2.7\,$\pm$\,0.1 &14.4\,$\pm$\,0.8 &137 \,$\pm$\,18   &10.2\,$\pm$\,2.5  &1.6\,$\pm$\,0.4  &1.3\,$\pm$\,0.3   &15.2\,$\pm$\,4.2 &G19.900+0.245\\
29  &18:28:17.9 &-11:39:09 &9.7 \,$\pm$\,0.5 &3.25\,$\pm$\,0.10 &-19.9 &1.50\,$\pm$\,0.05 &3.3\,$\pm$\,0.1 &12.7\,$\pm$\,0.4 &221 \,$\pm$\,23   &11.6\,$\pm$\,3.7  &1.6\,$\pm$\,0.5  &1.2\,$\pm$\,0.4   &7.3\,$\pm$\,2.5  &G19.942-0.242\\
30  &18:26:29.9 &-11:18:59 &11.8\,$\pm$\,0.2 &4.26\,$\pm$\,0.05 &-30.7 &1.51\,$\pm$\,0.08 &5.4\,$\pm$\,0.1 &21.0\,$\pm$\,1.0 &448 \,$\pm$\,112  &30.8\,$\pm$\,10.3 &2.5\,$\pm$\,0.8  &1.4\,$\pm$\,0.5   &3.6\,$\pm$\,1.3  &G20.058+0.300\\
31  &18:26:33.8 &-11:18:08 &11.8\,$\pm$\,0.2 &0.92\,$\pm$\,0.01 &-30.9 &1.08\,$\pm$\,0.06 &4.9\,$\pm$\,0.1 &13.7\,$\pm$\,0.7 &65.8\,$\pm$\,16.4 & 4.5\,$\pm$\,1.5  &7.9\,$\pm$\,2.7  &20.7\,$\pm$\,6.9  &2.7\,$\pm$\,1.0  &G20.058+0.300\\
32  &18:27:37.4 &-11:25:50 &10.3\,$\pm$\,0.4 &0.83\,$\pm$\,0.02 &-21.9 &0.81\,$\pm$\,0.05 &2.4\,$\pm$\,0.1 &5.1 \,$\pm$\,0.3 &18.3\,$\pm$\,5.6  & 1.0\,$\pm$\,0.4  &2.2\,$\pm$\,0.9  &6.5\,$\pm$\,2.7   &6.1\,$\pm$\,2.7  &G20.067+0.008\\
33  &18:28:11.6 &-11:13:14 &10.1\,$\pm$\,0.6 &2.69\,$\pm$\,0.10 &-27.2 &1.74\,$\pm$\,0.09 &1.9\,$\pm$\,0.1 &8.3 \,$\pm$\,0.4 &53.7\,$\pm$\,13.2 & 3.0\,$\pm$\,1.1  &0.6\,$\pm$\,0.2  &0.5\,$\pm$\,0.2   &31.7\,$\pm$\,12.7&G20.300-0.033\\
34  &18:28:07.8 &-11:14:11 &10.1\,$\pm$\,0.6 &2.43\,$\pm$\,0.09 &-27.6 &1.11\,$\pm$\,0.06 &1.7\,$\pm$\,0.1 &5.0 \,$\pm$\,0.3 &28.2\,$\pm$\,6.9  & 1.6\,$\pm$\,0.6  &0.4\,$\pm$\,0.1  &0.4\,$\pm$\,0.1   &22.3\,$\pm$\,8.9 &G20.300-0.033\\
35  &18:27:13.2 &-11:05:00 &10.0\,$\pm$\,0.3 &3.53\,$\pm$\,0.07 &-10.6 &1.48\,$\pm$\,0.05 &4.0\,$\pm$\,0.1 &15.1\,$\pm$\,0.5 &1034\,$\pm$\,237  &56.4\,$\pm$\,20.7 &6.7\,$\pm$\,2.5  &4.6\,$\pm$\,1.7   &1.6\,$\pm$\,0.6  &G20.320+0.250\\
36  &18:27:14.5 &-11:06:23 &10.0\,$\pm$\,0.3 &3.32\,$\pm$\,0.06 &-11.7 &1.22\,$\pm$\,0.06 &2.2\,$\pm$\,0.1 &6.8 \,$\pm$\,0.4 &323 \,$\pm$\,74   &17.6\,$\pm$\,6.5  &2.4\,$\pm$\,0.9  &1.7\,$\pm$\,0.6   &3.2\,$\pm$\,1.3  &G20.320+0.250\\
37  &18:27:15.0 &-11:06:18 &10.0\,$\pm$\,0.3 &0.84\,$\pm$\,0.02 &-21.0 &1.06\,$\pm$\,0.06 &1.8\,$\pm$\,0.1 &4.8 \,$\pm$\,0.3 &49.7\,$\pm$\,11.4 & 2.7\,$\pm$\,1.0  &5.7\,$\pm$\,2.1  &16.3\,$\pm$\,6.0  &4.0\,$\pm$\,1.6  &G20.320+0.250\\
38  &18:28:11.1 &-11:13:15 &11.3\,$\pm$\,0.4 &3.36\,$\pm$\,0.08 &-27.2 &1.73\,$\pm$\,0.09 &1.8\,$\pm$\,0.1 &8.0 \,$\pm$\,0.4 &63.6\,$\pm$\,14.0 & 4.1\,$\pm$\,1.3  &0.5\,$\pm$\,0.2  &0.4\,$\pm$\,0.1   &28.4\,$\pm$\,9.9 &G20.308-0.017\\
39  &18:28:13.5 &-11:12:11 &11.3\,$\pm$\,0.4 &1.40\,$\pm$\,0.04 &-28.9 &1.05\,$\pm$\,0.05 &1.7\,$\pm$\,0.1 &4.7 \,$\pm$\,0.2 &14.7\,$\pm$\,3.3  & 0.9\,$\pm$\,0.3  &0.7\,$\pm$\,0.2  &1.2\,$\pm$\,0.4   &18.8\,$\pm$\,6.5 &G20.308-0.017\\
40  &18:28:07.2 &-11:14:12 &11.3\,$\pm$\,0.4 &2.46\,$\pm$\,0.06 &-27.6 &1.09\,$\pm$\,0.06 &1.7\,$\pm$\,0.1 &4.6 \,$\pm$\,0.3 &27.5\,$\pm$\,6.1  & 1.8\,$\pm$\,0.6  &0.4\,$\pm$\,0.1  &0.4\,$\pm$\,0.1   &19.1\,$\pm$\,6.7 &G20.308-0.017\\
41  &18:28:37.1 &-11:10:04 &11.4\,$\pm$\,0.4 &2.66\,$\pm$\,0.06 &-29.2 &1.22\,$\pm$\,0.05 &2.0\,$\pm$\,0.1 &6.3 \,$\pm$\,0.3 &57.2\,$\pm$\,15.1 & 3.7\,$\pm$\,1.3  &0.8\,$\pm$\,0.3  &0.7\,$\pm$\,0.3   &12.3\,$\pm$\,4.6 &G20.408-0.083\\
42  &18:27:03.8 &-10:56:24 &10.2\,$\pm$\,0.2 &1.74\,$\pm$\,0.03 &-10.1 &4.51\,$\pm$\,0.15 &6.5\,$\pm$\,0.1 &74.0\,$\pm$\,2.4 &814 \,$\pm$\,171  &45.6\,$\pm$\,15.8 &22.3\,$\pm$\,7.8 &30.8\,$\pm$\,10.7 &9.0\,$\pm$\,3.2  &G20.442+0.350\\
43  &18:27:07.5 &-10:56:13 &10.2\,$\pm$\,0.2 &0.66\,$\pm$\,0.02 &-12.5 &1.28\,$\pm$\,0.06 &4.3\,$\pm$\,0.1 &14.1\,$\pm$\,0.7 &26.7\,$\pm$\,5.7  & 1.5\,$\pm$\,0.5  &5.1\,$\pm$\,1.9  &18.5\,$\pm$\,6.4  &8.4\,$\pm$\,3.1  &G20.442+0.350\\
44  &18:28:39.6 &-11:07:33 &11.4\,$\pm$\,0.3 &3.91\,$\pm$\,0.07 &-28.7 &1.18\,$\pm$\,0.05 &2.1\,$\pm$\,0.1 &6.3 \,$\pm$\,0.3 &127 \,$\pm$\,18   & 8.3\,$\pm$\,2.3  &0.8\,$\pm$\,0.2  &0.5\,$\pm$\,0.1   &7.6\,$\pm$\,2.2  &G20.450-0.075\\
45  &18:28:39.4 &-11:02:02 &10.9\,$\pm$\,0.3 &0.97\,$\pm$\,0.03 &-27.9 &1.69\,$\pm$\,0.08 &1.7\,$\pm$\,0.1 &7.4 \,$\pm$\,0.4 &32.3\,$\pm$\,12.1 & 2.0\,$\pm$\,0.9  &3.1\,$\pm$\,1.4  &7.7\,$\pm$\,3.5   &16.2\,$\pm$\,7.5 &G20.525-0.042\\
46  &18:28:18.6 &-10:56:00 &11.5\,$\pm$\,0.3 &1.79\,$\pm$\,0.03 &-30.2 &1.03\,$\pm$\,0.05 &1.9\,$\pm$\,0.1 &5.1 \,$\pm$\,0.2 &45.7\,$\pm$\,9.8  & 3.0\,$\pm$\,1.0  &1.4\,$\pm$\,0.4  &1.9\,$\pm$\,0.6   &7.3\,$\pm$\,2.5  &G20.575+0.092\\
47  &18:28:25.6 &-10:58:12 &9.7 \,$\pm$\,0.3 &1.36\,$\pm$\,0.04 &-29.1 &1.06\,$\pm$\,0.03 &3.5\,$\pm$\,0.1 &9.7 \,$\pm$\,0.3 &95.3\,$\pm$\,14.9 & 5.0\,$\pm$\,1.7  &4.0\,$\pm$\,1.4  &7.1\,$\pm$\,2.4   &3.5\,$\pm$\,1.3  &G20.583+0.050\\
48  &18:28:27.0 &-10:56:49 &9.7 \,$\pm$\,0.3 &2.92\,$\pm$\,0.07 &-20.2 &1.56\,$\pm$\,0.06 &2.1\,$\pm$\,0.1 &8.2 \,$\pm$\,0.3 &89.9\,$\pm$\,14.0 & 4.7\,$\pm$\,1.6  &0.8\,$\pm$\,0.3  &0.7\,$\pm$\,0.2   &17.4\,$\pm$\,6.1 &G20.583+0.050\\
49  &18:28:29.5 &-10:57:45 &9.7 \,$\pm$\,0.3 &0.88\,$\pm$\,0.03 &-19.3 &0.59\,$\pm$\,0.03 &1.8\,$\pm$\,0.1 &2.9 \,$\pm$\,0.2 &10.3\,$\pm$\,1.6  & 0.5\,$\pm$\,0.2  &1.0\,$\pm$\,0.4  &2.8\,$\pm$\,0.9   &6.6\,$\pm$\,2.4  &G20.583+0.050\\
50  &18:28:18.4 &-10:56:00 &9.7 \,$\pm$\,0.3 &2.02\,$\pm$\,0.05 &-30.2 &1.03\,$\pm$\,0.05 &1.7\,$\pm$\,0.1 &4.6 \,$\pm$\,0.2 &25.9\,$\pm$\,4.1  & 1.4\,$\pm$\,0.5  &0.5\,$\pm$\,0.2  &0.6\,$\pm$\,0.2   &18.2\,$\pm$\,6.4 &G20.583+0.050\\
51  &18:28:25.2 &-10:56:26 &9.7 \,$\pm$\,0.3 &1.33\,$\pm$\,0.04 &-29.5 &1.07\,$\pm$\,0.06 &1.7\,$\pm$\,0.1 &4.7 \,$\pm$\,0.3 &18.6\,$\pm$\,2.9  & 1.0\,$\pm$\,0.3  &0.8\,$\pm$\,0.3  &1.5\,$\pm$\,0.5   &18.1\,$\pm$\,6.5 &G20.583+0.050\\
52  &18:28:26.5 &-10:54:50 &9.7 \,$\pm$\,0.3 &1.38\,$\pm$\,0.04 &-30.2 &0.59\,$\pm$\,0.03 &1.7\,$\pm$\,0.1 &2.8 \,$\pm$\,0.1 &9.4 \,$\pm$\,1.5  & 0.5\,$\pm$\,0.2  &0.4\,$\pm$\,0.1  &0.7\,$\pm$\,0.2   &11.3\,$\pm$\,4.0 &G20.583+0.050\\
53  &18:29:14.7 &-10:53:46 &11.6\,$\pm$\,0.4 &1.43\,$\pm$\,0.04 &-30.4 &1.77\,$\pm$\,0.09 &1.8\,$\pm$\,0.1 &8.2 \,$\pm$\,0.4 &58.3\,$\pm$\,10.8 & 3.9\,$\pm$\,1.2  &2.8\,$\pm$\,0.9  &4.8\,$\pm$\,1.4   &13.3\,$\pm$\,4.3 &G20.717-0.100\\
54  &18:33:03.2 &-09:50:43 &10.2\,$\pm$\,0.4 &1.16\,$\pm$\,0.03 &-23.5 &1.57\,$\pm$\,0.05 &6.6\,$\pm$\,0.1 &26.3\,$\pm$\,0.8 &46.6\,$\pm$\,6.6  & 2.6\,$\pm$\,0.8  &2.9\,$\pm$\,0.9  &6.0\,$\pm$\,1.9   &12.7\,$\pm$\,4.1 &G22.083-0.442\\
55  &18:32:11.7 &-09:30:10 &10.5\,$\pm$\,0.2 &2.50\,$\pm$\,0.03 &-9.3  &1.18\,$\pm$\,0.05 &4.9\,$\pm$\,0.1 &14.9\,$\pm$\,0.6 &200 \,$\pm$\,61   &11.6\,$\pm$\,4.7  &2.8\,$\pm$\,1.1  &2.6\,$\pm$\,1.1   &3.5\,$\pm$\,1.5  &G22.283-0.092\\
56  &18:33:52.9 &-09:12:42 &11.2\,$\pm$\,0.2 &1.51\,$\pm$\,0.02 &-30.6 &0.83\,$\pm$\,0.04 &6.1\,$\pm$\,0.1 &13.3\,$\pm$\,0.6 &70.2\,$\pm$\,23.0 & 4.5\,$\pm$\,1.8  &2.9\,$\pm$\,1.2  &4.6\,$\pm$\,1.9   &2.7\,$\pm$\,1.2  &G22.750-0.333\\
57  &18:34:32.5 &-08:32:45 &11.6\,$\pm$\,0.5 &0.80\,$\pm$\,0.03 &-33.4 &0.59\,$\pm$\,0.03 &3.2\,$\pm$\,0.1 &5.3 \,$\pm$\,0.2 &10.0\,$\pm$\,2.4  & 0.7\,$\pm$\,0.2  &1.6\,$\pm$\,0.6  &4.7\,$\pm$\,1.6   &4.8\,$\pm$\,1.8  &G23.408-0.167\\
58  &18:37:53.9 &-07:09:55 &11.1\,$\pm$\,0.1 &2.01\,$\pm$\,0.02 &-31.2 &1.17\,$\pm$\,0.04 &2.4\,$\pm$\,0.1 &7.3 \,$\pm$\,0.3 &48.8\,$\pm$\,6.0  & 3.1\,$\pm$\,0.8  &1.1\,$\pm$\,0.3  &1.3\,$\pm$\,0.4   &10.4\,$\pm$\,3.0 &G25.017-0.275\\
59  &18:38:22.4 &-06:50:03 &10.0\,$\pm$\,0.3 &1.85\,$\pm$\,0.02 &-23.7 &0.84\,$\pm$\,0.05 &2.7\,$\pm$\,0.1 &5.9 \,$\pm$\,0.3 &94.4\,$\pm$\,13.7 & 5.2\,$\pm$\,1.7  &2.2\,$\pm$\,0.7  &2.9\,$\pm$\,0.9   &2.9\,$\pm$\,1.1  &G25.358-0.225\\
60  &18:38:20.7 &-06:50:56 &10.0\,$\pm$\,0.3 &1.55\,$\pm$\,0.01 &-24.3 &0.81\,$\pm$\,0.04 &2.1\,$\pm$\,0.1 &4.6 \,$\pm$\,0.2 &28.0\,$\pm$\,4.1  & 1.5\,$\pm$\,0.5  &0.9\,$\pm$\,0.3  &1.5\,$\pm$\,0.5   &7.7\,$\pm$\,2.6  &G25.358-0.225\\
61  &18:39:20.8 &-06:56:14 &11.1\,$\pm$\,0.3 &1.01\,$\pm$\,0.02 &-32.5 &1.18\,$\pm$\,0.04 &2.4\,$\pm$\,0.1 &7.2 \,$\pm$\,0.3 &32.3\,$\pm$\,6.8  & 2.0\,$\pm$\,0.7  &3.0\,$\pm$\,1.0  &7.0\,$\pm$\,2.3   &8.0\,$\pm$\,2.7  &G25.383-0.483\\
62  &18:37:00.4 &-06:08:08 &11.1\,$\pm$\,0.3 &1.96\,$\pm$\,0.04 &-32.4 &1.80\,$\pm$\,0.05 &5.4\,$\pm$\,0.1 &24.5\,$\pm$\,0.6 &157 \,$\pm$\,16   & 9.9\,$\pm$\,2.6  &3.8\,$\pm$\,1.0  &4.7\,$\pm$\,1.2   &7.4\,$\pm$\,2.1  &G25.833+0.400\\
63  &18:38:09.3 &-06:13:35 &10.9\,$\pm$\,0.2 &2.04\,$\pm$\,0.03 &-30.5 &1.26\,$\pm$\,0.06 &4.0\,$\pm$\,0.1 &13.0\,$\pm$\,0.6 &82.9\,$\pm$\,11.9 & 5.1\,$\pm$\,1.5  &1.8\,$\pm$\,0.5  &2.1\,$\pm$\,0.6   &7.4\,$\pm$\,2.3  &G25.883+0.108\\
64  &18:36:59.0 &-05:56:26 &11.0\,$\pm$\,0.3 &2.94\,$\pm$\,0.05 &-31.8 &1.29\,$\pm$\,0.07 &1.7\,$\pm$\,0.1 &5.6 \,$\pm$\,0.3 &32.4\,$\pm$\,8.6  & 2.0\,$\pm$\,0.7  &0.3\,$\pm$\,0.1  &0.3\,$\pm$\,0.1   &28.1\,$\pm$\,10.8&G25.992+0.492\\
65  &18:37:15.1 &-05:53:20 &10.9\,$\pm$\,0.2 &0.61\,$\pm$\,0.01 &-32.7 &1.59\,$\pm$\,0.06 &2.9\,$\pm$\,0.1 &11.8\,$\pm$\,0.4 &22.2\,$\pm$\,5.1  & 1.4\,$\pm$\,0.5  &5.4\,$\pm$\,1.9  &21.3\,$\pm$\,7.2  &13.2\,$\pm$\,4.6 &G26.075+0.455\\
66  &18:40:37.1 &-05:43:33 &11.1\,$\pm$\,0.3 &2.95\,$\pm$\,0.05 &-34.7 &2.43\,$\pm$\,0.07 &5.3\,$\pm$\,0.1 &32.7\,$\pm$\,1.0 &200 \,$\pm$\,14   &12.6\,$\pm$\,3.2  &2.1\,$\pm$\,0.6  &1.7\,$\pm$\,0.5   &16.1\,$\pm$\,4.3 &G26.608-0.208\\
67  &18:40:36.5 &-05:43:11 &11.1\,$\pm$\,0.3 &3.40\,$\pm$\,0.06 &-32.1 &2.30\,$\pm$\,0.08 &5.0\,$\pm$\,0.1 &29.1\,$\pm$\,1.0 &188 \,$\pm$\,13   &11.8\,$\pm$\,3.0  &1.5\,$\pm$\,0.4  &1.1\,$\pm$\,0.3   &17.6\,$\pm$\,4.8 &G26.608-0.208\\
68  &18:39:35.0 &-05:22:20 &10.4\,$\pm$\,0.4 &0.81\,$\pm$\,0.02 &-27.6 &1.04\,$\pm$\,0.05 &4.7\,$\pm$\,0.1 &12.7\,$\pm$\,0.6 &27.4\,$\pm$\,7.0  & 1.6\,$\pm$\,0.6  &3.5\,$\pm$\,1.4  &10.5\,$\pm$\,3.9  &6.5\,$\pm$\,2.5  &G26.800+0.175\\
69  &18:38:58.2 &-05:03:19 &10.5\,$\pm$\,0.3 &2.26\,$\pm$\,0.04 &-28.1 &1.60\,$\pm$\,0.03 &7.1\,$\pm$\,0.1 &29.0\,$\pm$\,0.5 &526 \,$\pm$\,63   &30.6\,$\pm$\,9.0  &8.9\,$\pm$\,2.6  &9.4\,$\pm$\,2.8   &2.2\,$\pm$\,0.7  &G26.983+0.433\\
70  &18:38:55.0 &-05:05:06 &10.5\,$\pm$\,0.3 &0.83\,$\pm$\,0.02 &-26.9 &1.79\,$\pm$\,0.06 &3.6\,$\pm$\,0.1 &16.1\,$\pm$\,0.6 &81.6\,$\pm$\,9.7  & 4.8\,$\pm$\,1.4  &10.2\,$\pm$\,3.1 &29.6\,$\pm$\,8.7  &6.5\,$\pm$\,2.0  &G26.983+0.433\\
71  &18:38:57.0 &-05:04:28 &10.5\,$\pm$\,0.3 &0.82\,$\pm$\,0.02 &-30.0 &0.54\,$\pm$\,0.03 &3.4\,$\pm$\,0.1 &5.2 \,$\pm$\,0.2 &21.8\,$\pm$\,2.6  & 1.3\,$\pm$\,0.4  &2.8\,$\pm$\,0.9  &8.2\,$\pm$\,2.4   &2.2\,$\pm$\,0.7  &G26.983+0.433\\
72  &18:38:55.4 &-05:02:29 &10.5\,$\pm$\,0.3 &1.28\,$\pm$\,0.03 &-28.3 &1.04\,$\pm$\,0.05 &3.1\,$\pm$\,0.1 &8.3 \,$\pm$\,0.4 &38.2\,$\pm$\,4.6  & 2.2\,$\pm$\,0.7  &2.0\,$\pm$\,0.6  &3.8\,$\pm$\,1.1   &7.2\,$\pm$\,2.3  &G26.983+0.433\\
73  &18:39:00.7 &-05:05:51 &10.5\,$\pm$\,0.3 &4.84\,$\pm$\,0.09 &-29.0 &1.50\,$\pm$\,0.07 &2.8\,$\pm$\,0.1 &10.5\,$\pm$\,0.5 &204 \,$\pm$\,25   &11.9\,$\pm$\,3.5  &0.7\,$\pm$\,0.2  &0.4\,$\pm$\,0.1   &10.6\,$\pm$\,3.3 &G26.983+0.433\\
74  &18:39:00.8 &-05:04:57 &10.5\,$\pm$\,0.3 &1.79\,$\pm$\,0.04 &-29.5 &0.60\,$\pm$\,0.04 &3.0\,$\pm$\,0.1 &4.9 \,$\pm$\,0.2 &34.7\,$\pm$\,4.2  & 2.0\,$\pm$\,0.6  &0.9\,$\pm$\,0.3  &1.2\,$\pm$\,0.4   &3.7\,$\pm$\,1.2  &G26.983+0.433\\
75  &18:41:37.4 &-04:43:35 &10.8\,$\pm$\,0.2 &2.08\,$\pm$\,0.03 &-31.9 &1.06\,$\pm$\,0.05 &3.0\,$\pm$\,0.1 &8.3 \,$\pm$\,0.4 &24.1\,$\pm$\,4.3  & 1.5\,$\pm$\,0.5  &0.5\,$\pm$\,0.2  &0.6\,$\pm$\,0.2   &18.5\,$\pm$\,6.0 &G27.608+0.025\\
\hline
\end{tabular}
\label{table:clump parameters}
\tablefoot{Column 1: number is consistent with the red number marked in Fig.\,\ref{fig:map}. 
Columns 2--3: equatorial coordinates of the clump centroid derived from elliptical fitting. 
Columns 4--5: the Galactocentric distance ($R_{\rm g}$) and effective radius ($R_{\rm eff}$) of the clump. 
Columns 6--9: local standard of rest velocity ($V_{\rm LSR}$), velocity dispersion ($\sigma_{\rm v}$), peak main-beam temperature ($T_{\rm mb}$), and integrated intensity ($I_{\rm CO}$) obtained from Gaussian fitting from the averaged spectral profile over each clump. 
Columns 10--14: the CO luminosity ($L_{\rm CO}$), luminosity-based H$_2$ masses ($M_{\rm H_2}$), mean column density ($N_{\rm H_2}$), mean volume density ($n_{\rm H_2}$), and virial parameter ($\alpha_{\rm vir}$) of each clump. For detailed derivation of parameters refer to Sect.\,\ref{sect:Clump identification}. 
Column 15: parent cloud name identifying the clumps. 
}
\end{table*}

\setcounter{table}{0}
\begin{table*}[!htbp]
\scriptsize
\caption{Continued.}
\centering
\begin{tabular}
{cccccccccccccccc}
\hline\hline
Num &R.A.(J2000) &Dec.(J2000) &$R_{\rm g}$ &$R_{\rm eff}$ &$V_{\rm LSR}$ &$\sigma_{\rm v}$ &$T_{\rm mb}$ &$I_{\rm CO}$ &$L_{\rm CO}$ &$M_{\rm H_2}$ &$N_{\rm H_2}$ &$n_{\rm H_2}$ &$\alpha_{\rm vir}$ &Parent cloud\\
&h m s &$\circ$ $\prime$ $\prime\prime$ & &kpc &pc &km\,s$^{-1}$ &km\,s$^{-1}$ &K &K\,km\,s$^{-1}$ &K\,km\,s$^{-1}$\,pc$^{2}$ &$10^{2}$\,M$_{\odot}$ &$10^{21}$\,cm$^{-2}$ &$10^{2}$\,cm$^{-3}$ & \\
(1) &(2) &(3) &(4) &(5) &(6) &(7) &(8) &(9) &(10) &(11) &(12) &(13) &(14) &(15)\\
\hline
76  &18:40:51.5 &-04:24:52 &11.2\,$\pm$\,0.2 &0.61\,$\pm$\,0.01 &-35.7 &1.06\,$\pm$\,0.05 &6.5\,$\pm$\,0.1 &17.9\,$\pm$\,0.8 &44.9\,$\pm$\,8.9  & 2.9\,$\pm$\,0.9  &11.4\,$\pm$\,3.7 &44.8\,$\pm$\,14.0 &2.8\,$\pm$\,1.0  &G27.792+0.333\\
77  &18:43:25.2 &-04:41:39 &10.7\,$\pm$\,0.3 &2.16\,$\pm$\,0.04 &-32.0 &1.08\,$\pm$\,0.06 &3.2\,$\pm$\,0.1 &8.9 \,$\pm$\,0.5 &116 \,$\pm$\,20   & 6.9\,$\pm$\,2.2  &2.2\,$\pm$\,0.7  &2.4\,$\pm$\,0.8   &4.2\,$\pm$\,1.5  &G27.842-0.367\\
78  &18:42:43.0 &-04:10:13 &10.8\,$\pm$\,0.2 &3.43\,$\pm$\,0.04 &-32.7 &1.63\,$\pm$\,0.06 &1.7\,$\pm$\,0.1 &7.1 \,$\pm$\,0.3 &141 \,$\pm$\,24   & 8.5\,$\pm$\,2.6  &1.1\,$\pm$\,0.3  &0.8\,$\pm$\,0.2   &12.3\,$\pm$\,3.9 &G28.225+0.033\\
79  &18:42:46.6 &-04:10:16 &10.8\,$\pm$\,0.2 &0.61\,$\pm$\,0.01 &-32.8 &0.81\,$\pm$\,0.04 &1.6\,$\pm$\,0.1 &3.4 \,$\pm$\,0.2 &10.7\,$\pm$\,1.8  & 0.6\,$\pm$\,0.2  &2.6\,$\pm$\,0.8  &10.1\,$\pm$\,3.1  &7.2\,$\pm$\,2.3  &G28.225+0.033\\
80  &18:43:04.9 &-04:02:19 &11.8\,$\pm$\,0.2 &2.19\,$\pm$\,0.04 &-40.1 &0.59\,$\pm$\,0.03 &1.8\,$\pm$\,0.1 &3.0 \,$\pm$\,0.2 &19.8\,$\pm$\,4.2  & 1.4\,$\pm$\,0.4  &0.4\,$\pm$\,0.1  &0.5\,$\pm$\,0.1   &6.5\,$\pm$\,2.1  &G28.383+0.025\\
81  &18:41:35.1 &-03:48:52 &11.4\,$\pm$\,0.1 &2.13\,$\pm$\,0.03 &-36.2 &1.45\,$\pm$\,0.05 &5.7\,$\pm$\,0.1 &20.9\,$\pm$\,0.7 &153 \,$\pm$\,39   &10.0\,$\pm$\,3.4  &3.3\,$\pm$\,1.1  &3.7\,$\pm$\,1.3   &5.2\,$\pm$\,1.9  &G28.433+0.433\\
82  &18:43:00.9 &-03:53:00 &12.0\,$\pm$\,0.1 &2.63\,$\pm$\,0.02 &-42.0 &1.25\,$\pm$\,0.07 &2.1\,$\pm$\,0.1 &6.6 \,$\pm$\,0.4 &151 \,$\pm$\,31   &10.6\,$\pm$\,3.1  &2.3\,$\pm$\,0.7  &2.1\,$\pm$\,0.6   &4.5\,$\pm$\,1.5  &G28.517+0.092\\
83  &18:43:25.8 &-03:39:29 &11.6\,$\pm$\,0.4 &2.03\,$\pm$\,0.05 &-12.0 &1.26\,$\pm$\,0.06 &1.7\,$\pm$\,0.1 &5.7 \,$\pm$\,0.3 &42.6\,$\pm$\,9.4  & 2.9\,$\pm$\,0.9  &1.0\,$\pm$\,0.3  &1.2\,$\pm$\,0.4   &13.1\,$\pm$\,4.5 &G28.767+0.100\\
84  &18:43:26.8 &-03:40:11 &11.6\,$\pm$\,0.4 &1.18\,$\pm$\,0.03 &-38.7 &1.22\,$\pm$\,0.05 &1.6\,$\pm$\,0.1 &5.0 \,$\pm$\,0.2 &17.5\,$\pm$\,3.9  & 1.2\,$\pm$\,0.4  &1.2\,$\pm$\,0.4  &2.5\,$\pm$\,0.8   &17.3\,$\pm$\,5.8 &G28.767+0.100\\
85  &18:46:14.0 &-01:35:47 &11.5\,$\pm$\,0.3 &1.55\,$\pm$\,0.03 &-40.5 &1.59\,$\pm$\,0.06 &4.4\,$\pm$\,0.1 &17.9\,$\pm$\,0.6 &48.4\,$\pm$\,6.1  & 3.2\,$\pm$\,0.9  &2.0\,$\pm$\,0.5  &3.1\,$\pm$\,0.8   &14.1\,$\pm$\,4.0 &G30.917+0.433\\
86  &18:53:04.1 &+00:49:33 &11.3\,$\pm$\,0.3 &1.33\,$\pm$\,0.03 &-41.5 &1.01\,$\pm$\,0.04 &2.2\,$\pm$\,0.1 &5.8 \,$\pm$\,0.3 &27.7\,$\pm$\,5.9  & 1.8\,$\pm$\,0.6  &1.5\,$\pm$\,0.5  &2.7\,$\pm$\,0.9   &8.8\,$\pm$\,3.0  &G33.858+0.008\\
87  &18:52:48.0 &+00:56:58 &11.6\,$\pm$\,0.3 &1.66\,$\pm$\,0.03 &-43.5 &0.76\,$\pm$\,0.04 &2.8\,$\pm$\,0.1 &5.7 \,$\pm$\,0.3 &43.6\,$\pm$\,7.0  & 2.9\,$\pm$\,0.8  &1.6\,$\pm$\,0.4  &2.3\,$\pm$\,0.7   &3.8\,$\pm$\,1.2  &G33.933+0.133\\
88  &18:52:50.8 &+00:56:50 &11.6\,$\pm$\,0.3 &1.69\,$\pm$\,0.03 &-43.9 &1.23\,$\pm$\,0.07 &2.3\,$\pm$\,0.1 &7.3 \,$\pm$\,0.4 &48.9\,$\pm$\,7.8  & 3.3\,$\pm$\,0.9  &1.7\,$\pm$\,0.5  &2.4\,$\pm$\,0.7   &9.0\,$\pm$\,2.8  &G33.933+0.133\\
89  &18:58:33.2 &+03:20:38 &11.3\,$\pm$\,0.2 &0.77\,$\pm$\,0.01 &-44.8 &0.79\,$\pm$\,0.04 &2.9\,$\pm$\,0.1 &6.1 \,$\pm$\,0.3 &8.4 \,$\pm$\,1.7  & 0.5\,$\pm$\,0.2  &1.4\,$\pm$\,0.4  &4.2\,$\pm$\,1.3   &10.3\,$\pm$\,3.4 &G36.717-0.058\\
90  &18:58:42.3 &+03:32:51 &12.9\,$\pm$\,0.3 &2.27\,$\pm$\,0.04 &-56.9 &1.33\,$\pm$\,0.04 &4.2\,$\pm$\,0.1 &14.3\,$\pm$\,0.5 &71.7\,$\pm$\,17.1 & 5.7\,$\pm$\,1.8  &1.6\,$\pm$\,0.5  &1.7\,$\pm$\,0.5   &8.2\,$\pm$\,2.6  &G36.925+0.000\\
91  &18:59:13.4 &+04:26:30 &11.5\,$\pm$\,0.3 &1.67\,$\pm$\,0.03 &-43.2 &1.52\,$\pm$\,0.09 &2.6\,$\pm$\,0.1 &10.1\,$\pm$\,0.6 &41.9\,$\pm$\,14.7 & 2.8\,$\pm$\,1.2  &1.5\,$\pm$\,0.6  &2.1\,$\pm$\,0.9   &16.1\,$\pm$\,7.1 &G37.767+0.275\\
92  &18:59:12.5 &+04:25:44 &11.5\,$\pm$\,0.3 &0.94\,$\pm$\,0.02 &-42.8 &1.48\,$\pm$\,0.08 &2.5\,$\pm$\,0.1 &9.5 \,$\pm$\,0.5 &34.1\,$\pm$\,11.9 & 2.3\,$\pm$\,1.0  &3.8\,$\pm$\,1.6  &9.7\,$\pm$\,4.1   &10.6\,$\pm$\,4.7 &G37.767+0.275\\
93  &19:00:05.8 &+04:23:18 &11.6\,$\pm$\,0.2 &0.90\,$\pm$\,0.02 &-48.1 &0.84\,$\pm$\,0.05 &3.0\,$\pm$\,0.1 &6.6 \,$\pm$\,0.4 &15.3\,$\pm$\,3.1  & 1.0\,$\pm$\,0.3  &1.9\,$\pm$\,0.6  &5.0\,$\pm$\,1.5   &7.1\,$\pm$\,2.4  &G37.825+0.092\\
94  &18:59:06.3 &+04:48:00 &11.6\,$\pm$\,0.3 &1.11\,$\pm$\,0.02 &-47.2 &1.03\,$\pm$\,0.06 &2.8\,$\pm$\,0.1 &7.6 \,$\pm$\,0.4 &21.0\,$\pm$\,5.2  & 1.4\,$\pm$\,0.5  &1.7\,$\pm$\,0.6  &3.7\,$\pm$\,1.2   &9.7\,$\pm$\,3.5  &G38.075+0.483\\
95  &19:04:14.7 &+07:18:09 &12.0\,$\pm$\,0.2 &1.11\,$\pm$\,0.02 &-54.4 &1.78\,$\pm$\,0.09 &2.4\,$\pm$\,0.1 &10.8\,$\pm$\,0.5 &45.7\,$\pm$\,7.8  & 3.2\,$\pm$\,0.9  &3.9\,$\pm$\,1.1  &8.4\,$\pm$\,2.3   &12.6\,$\pm$\,3.8 &G40.883+0.492\\
96  &19:13:10.2 &+10:01:28 &12.9\,$\pm$\,0.3 &1.69\,$\pm$\,0.04 &-65.1 &1.01\,$\pm$\,0.04 &1.7\,$\pm$\,0.1 &4.4 \,$\pm$\,0.2 &65.0\,$\pm$\,19.3 & 5.2\,$\pm$\,1.8  &2.7\,$\pm$\,1.0  &3.8\,$\pm$\,1.4   &3.9\,$\pm$\,1.5  &G44.325-0.208\\
97  &19:13:15.9 &+10:02:52 &12.9\,$\pm$\,0.3 &0.75\,$\pm$\,0.02 &-64.2 &0.81\,$\pm$\,0.04 &1.5\,$\pm$\,0.1 &3.1 \,$\pm$\,0.2 &8.9 \,$\pm$\,1.9  & 0.7\,$\pm$\,0.2  &1.9\,$\pm$\,0.6  &5.9\,$\pm$\,1.7   &8.1\,$\pm$\,2.5  &G44.350-0.220\\
98  &19:11:24.9 &+10:28:51 &12.9\,$\pm$\,0.3 &3.19\,$\pm$\,0.08 &-45.9 &5.44\,$\pm$\,0.15 &7.1\,$\pm$\,0.1 &97.4\,$\pm$\,2.6 &2682\,$\pm$\,595  & 213\,$\pm$\,64   &31.0\,$\pm$\,9.2 &23.3\,$\pm$\,6.9  &5.1\,$\pm$\,1.6  &G44.500+0.340\\
99  &19:11:30.9 &+10:26:42 &12.9\,$\pm$\,0.3 &2.94\,$\pm$\,0.07 &-48.5 &2.48\,$\pm$\,0.06 &7.4\,$\pm$\,0.1 &46.2\,$\pm$\,1.0 &773 \,$\pm$\,172  &61.4\,$\pm$\,18.2 &10.5\,$\pm$\,3.1 &8.6\,$\pm$\,2.5   &3.4\,$\pm$\,1.1  &G44.500+0.340\\
100 &19:11:29.2 &+10:26:18 &12.9\,$\pm$\,0.3 &3.14\,$\pm$\,0.08 &-46.2 &2.11\,$\pm$\,0.06 &6.3\,$\pm$\,0.1 &33.7\,$\pm$\,0.9 &423 \,$\pm$\,94   &33.6\,$\pm$\,10.0 &5.0\,$\pm$\,1.5  &3.9\,$\pm$\,1.1   &4.8\,$\pm$\,1.5  &G44.500+0.340\\
101 &19:11:28.0 &+10:28:11 &12.9\,$\pm$\,0.3 &1.91\,$\pm$\,0.05 &-45.8 &2.21\,$\pm$\,0.07 &5.8\,$\pm$\,0.1 &32.3\,$\pm$\,0.9 &286 \,$\pm$\,64   &22.7\,$\pm$\,6.7  &9.2\,$\pm$\,2.8  &11.6\,$\pm$\,3.4  &4.8\,$\pm$\,1.5  &G44.500+0.340\\
102 &19:11:27.1 &+10:27:00 &12.9\,$\pm$\,0.3 &3.23\,$\pm$\,0.08 &-46.6 &2.90\,$\pm$\,0.10 &5.1\,$\pm$\,0.1 &37.5\,$\pm$\,1.3 &392 \,$\pm$\,87   &31.1\,$\pm$\,9.2  &4.4\,$\pm$\,1.3  &3.3\,$\pm$\,1.0   &10.1\,$\pm$\,3.1 &G44.500+0.340\\
103 &19:11:28.2 &+10:25:16 &12.9\,$\pm$\,0.3 &2.99\,$\pm$\,0.07 &-46.6 &1.80\,$\pm$\,0.05 &5.2\,$\pm$\,0.1 &23.8\,$\pm$\,0.7 &284 \,$\pm$\,63   &22.5\,$\pm$\,6.7  &3.7\,$\pm$\,1.1  &3.0\,$\pm$\,0.9   &5.0\,$\pm$\,1.6  &G44.500+0.340\\
104 &19:11:27.0 &+10:27:39 &12.9\,$\pm$\,0.3 &1.85\,$\pm$\,0.05 &-42.2 &2.04\,$\pm$\,0.07 &5.6\,$\pm$\,0.1 &28.9\,$\pm$\,0.9 &208 \,$\pm$\,47   &16.5\,$\pm$\,4.9  &7.1\,$\pm$\,2.1  &9.2\,$\pm$\,2.8   &5.4\,$\pm$\,1.7  &G44.500+0.340\\
105 &19:11:24.3 &+10:25:57 &12.9\,$\pm$\,0.3 &2.97\,$\pm$\,0.07 &-47.3 &1.04\,$\pm$\,0.04 &5.1\,$\pm$\,0.1 &13.7\,$\pm$\,0.5 &200 \,$\pm$\,45   &15.8\,$\pm$\,4.7  &2.7\,$\pm$\,0.8  &2.1\,$\pm$\,0.6   &2.4\,$\pm$\,0.8  &G44.500+0.340\\
106 &19:11:31.1 &+10:25:51 &12.9\,$\pm$\,0.3 &1.87\,$\pm$\,0.05 &-46.6 &2.25\,$\pm$\,0.07 &5.0\,$\pm$\,0.1 &28.2\,$\pm$\,0.8 &201 \,$\pm$\,45   &16.0\,$\pm$\,4.7  &6.8\,$\pm$\,2.0  &8.7\,$\pm$\,2.6   &6.9\,$\pm$\,2.1  &G44.500+0.340\\
107 &19:11:36.3 &+10:24:54 &12.9\,$\pm$\,0.3 &2.67\,$\pm$\,0.07 &-46.5 &1.67\,$\pm$\,0.06 &4.8\,$\pm$\,0.1 &20.5\,$\pm$\,0.7 &188 \,$\pm$\,42   &14.9\,$\pm$\,4.4  &3.1\,$\pm$\,0.9  &2.8\,$\pm$\,0.8   &5.8\,$\pm$\,1.8  &G44.500+0.340\\
108 &19:11:35.0 &+10:26:13 &12.9\,$\pm$\,0.3 &2.26\,$\pm$\,0.06 &-49.5 &2.60\,$\pm$\,0.10 &4.4\,$\pm$\,0.1 &28.8\,$\pm$\,1.1 &120 \,$\pm$\,27   & 9.5\,$\pm$\,2.8  &2.8\,$\pm$\,0.8  &2.9\,$\pm$\,0.9   &18.6\,$\pm$\,5.8 &G44.500+0.340\\
109 &19:11:17.2 &+10:25:57 &12.9\,$\pm$\,0.3 &1.05\,$\pm$\,0.03 &-47.0 &0.81\,$\pm$\,0.03 &4.7\,$\pm$\,0.1 &10.1\,$\pm$\,0.3 &39.1\,$\pm$\,8.7  & 3.1\,$\pm$\,0.9  &4.2\,$\pm$\,1.3  &9.5\,$\pm$\,2.8   &2.6\,$\pm$\,0.8  &G44.500+0.340\\
110 &19:11:31.4 &+10:24:25 &12.9\,$\pm$\,0.3 &2.24\,$\pm$\,0.06 &-48.6 &1.56\,$\pm$\,0.07 &4.2\,$\pm$\,0.1 &16.5\,$\pm$\,0.7 &89.6\,$\pm$\,19.9 & 7.1\,$\pm$\,2.1  &2.1\,$\pm$\,0.6  &2.2\,$\pm$\,0.7   &8.9\,$\pm$\,2.8  &G44.500+0.340\\
\hline
\end{tabular}
\end{table*}
\endgroup

\begin{table*}[!htbp]
\caption{Fitted results of the scaling relations.}
\centering
\begin{tabular}
{lccccccccc}
\hline\hline
Type &Number &$v_0$ &$a$ &$r_v$ &$p_v$ &$m_0$ &$b$ &$r_m$ &$p_m$\\
\hline
First quadrant   &110 &0.98\,$\pm$\,0.06 &0.35\,$\pm$\,0.08 &0.39 &<\,10$^{-3}$ &165\,$\pm$\,22 &1.60\,$\pm$\,0.18 &0.67 &<\,10$^{-3}$\\
Second quadrant  &112 &0.69\,$\pm$\,0.03 &0.36\,$\pm$\,0.10 &0.34 &<\,10$^{-3}$ &196\,$\pm$\,16 &2.18\,$\pm$\,0.26 &0.63 &<\,10$^{-3}$\\  
Both             &222 &0.77\,$\pm$\,0.03 &0.52\,$\pm$\,0.06 &0.53 &<\,10$^{-3}$ &188\,$\pm$\,13 &1.60\,$\pm$\,0.13 &0.65 &<\,10$^{-3}$\\
\hline
\end{tabular}
\label{table:Fit results for the scaling relations}
\tablefoot{Column 1: Types of the Galactic edge clumps. Clumps located in the first Galactic quadrant (\emph{this work}), the second Galactic quadrant \citep{Luo2026}, or combined data from both samples (see Figs.\,\ref{fig:Scaling-Relation} and \ref{fig:mass-size ralation}). 
Column 2: The number of samples of each fitted clump type. 
Columns 3--6: The fitted parameters of the $\sigma_v$--$R$ relation, including the normalization coefficient $v_0$, the power-law index $a$, the Pearson correlation coefficient $r_v$, and the statistical significance $p_v$. 
Columns 7--10: The fitted parameters of the $M$--$R$ relation, including the normalization coefficient $m_0$, the power-law index $b$, the Pearson correlation coefficient $r_m$, and the statistical significance $p_m$. 
Power-law fitting and the Pearson correlation coefficients (with statistical significance $p$-values) were calculated using the {\tt linmix} module from \citet{Kelly2007} and the {\tt pearsonr} function of the {\tt Python SciPy} package, respectively. 
The fitting was performed in log--log space using a linear regression method with uncertainties on both axes included.} 
\end{table*}

\end{appendix}
\end{document}